\documentclass{aastex61}
\usepackage{amsmath}
\received{\today}
\revised{}
\accepted{}
\submitjournal{ApJ}

\shorttitle{Chemistry of Glycine}
\shortauthors{Suzuki et al.}

\begin{document}

\title{An Expanded Gas-Grain Model for Interstellar Glycine}

\author[0000-0003-3278-2513]{Taiki Suzuki}
\affil{Department of Astronomy, the Graduate University for Advanced Studies (SOKENDAI), Osawa 2-21-1, Mitaka, Tokyo 181-8588, Japan}
\affil{Astrobiology Center, Osawa 2-21-1, Mitaka, Tokyo 181-8588, Japan}
\email{taiki.suzuki@nao.ac.jp}

\author[0000-0001-7031-8039]{Liton Majumdar}
\affiliation{Jet Propulsion Laboratory, California Institute of Technology, 4800 Oak Grove Drive, Pasadena, CA 91109, USA}
\email{liton.majumdar@jpl.nasa.gov}

\author[0000-0003-2775-7487]{Masatoshi Ohishi}
\affil{Department of Astronomy, the Graduate University for Advanced Studies (SOKENDAI), Osawa 2-21-1, Mitaka, Tokyo 181-8588, Japan}
\affil{National Astronomical Observatory of Japan, Osawa 2-21-1, Mitaka, Tokyo 181-8588, Japan}

\author[0000-0003-0769-8627]{Masao Saito}
\affil{Department of Astronomy, the Graduate University for Advanced Studies (SOKENDAI), Osawa 2-21-1, Mitaka, Tokyo 181-8588, Japan}
\affil{National Astronomical Observatory of Japan, Osawa 2-21-1, Mitaka, Tokyo 181-8588, Japan}

\author[0000-0003-1659-095X]{Tomoya Hirota}
\affil{Department of Astronomy, the Graduate University for Advanced Studies (SOKENDAI), Osawa 2-21-1, Mitaka, Tokyo 181-8588, Japan}
\affil{National Astronomical Observatory of Japan, Osawa 2-21-1, Mitaka, Tokyo 181-8588, Japan}

\author[0000-0001-9676-2605]{Valentine Wakelam}
\affil{Laboratoire d'astrophysique de Bordeaux, Univ. Bordeaux, CNRS, B18N, all\'{e}e Geoffroy Saint-Hilaire, 33615 Pessac, France}



\begin{abstract}
The study of the chemical evolution of glycine in the interstellar medium is one of challenging topics in astrochemistry.
Here, we present the chemical modeling of glycine in hot cores using the state-of-the-art three-phase chemical model NAUTILUS, which is focused on the latest glycine chemistry.
For the formation process of glycine on the grain surface, we obtained consistent results with previous studies that glycine would be formed via the reactions of COOH with CH$_2$NH$_2$.
However, we will report three important findings regarding the chemical evolution and the detectability of interstellar glycine.
First, with the experimentally obtained binding energy from the temperature programmed thermal desorption (TPD) experiment, a large proportion of glycine was destroyed through the grain surface reactions with NH or CH$_3$O radicals before it fully evaporates.
As a result, the formation process in the gas phase is more important than thermal evaporation from grains.
If this is the case, NH$_2$OH and CH$_3$COOH rather than CH$_3$NH$_2$ and CH$_2$NH would be the essential precursors to the gas phase glycine.
Secondly, since the gas phase glycine will be quickly destroyed by positive ions or radicals, early evolutionary phase of the hot cores would be the preferable target for the future glycine surveys.
Thirdly, we suggest the possibility that the suprathermal hydrogen atoms can strongly accelerate the formation of COOH radicals from CO$_2$, resulting in the dramatic increase of formation rate of glycine on grains.
The efficiency of this process should be investigated in detail by theoretical and experimental studies in the future.
\end{abstract}




\section{Introduction}
An attempt to understand the origin of life must necessarily begin with detailed studies of the formation and evolution of complex organic molecules (COMs), products of a complex chemistry that most likely starts in molecular clouds and continues within protoplanetary disks.
However, the synthesis and evolution of organic molecules which form the building blocks of more complex biotic molecules is not well understood.
There are two views regarding how COMs were formed on the early Earth: formation on planetary surface or exogenous delivery \citep{Ehrenfreund02}. 

In the interstellar medium (ISM), almost 200 molecules ranging from simple linear molecules to COMs were detected mainly towards dark clouds, low-mass and high-mass star-forming regions (as listed by CDMS\footnote{\url{https://www.astro.uni-koeln.de/cdms/molecules}}).
\cite{Ehrenfreund02} argued that the exogenous delivery of COMs to the early Earth by comets and/or asteroids could be as high as 1$\times$10$^{11}$~kg~year$^{-1}$, making it more important than their terrestrial formation by three orders of magnitude; thus organic molecules delivered by extraterrestrial processes might have played a crucial role in early Earth chemistry.
If this is the case, understanding the interstellar chemistry will enable us to better understand the first stage of chemical evolution regarding the origin of life: from atoms to very simple prebiotic species.

Since amino acids are the building blocks of life, the search for amino acids and their complex organic precursors at different stages of star and planet formation is one of the exciting topics in modern astronomy. 
Glycine (NH$_2$CH$_2$COOH), which is the simplest and the only non-chiral member out of 20 standard amino acids, was recently detected in the coma of comet 67P/Churyumov-Gerasimenko by the ROSINA (Rosetta Orbiter Spectrometer for Ion and Neutral Analysis) mass spectrometer \citep{Altwegg16}, giving clues about glycine's interstellar origin.
Several attempts have been made to detect glycine towards Sgr B2(N) and other high mass star forming regions, but none of them has been successful \cite[e.g.,][]{Ceccarelli00}.
Although \cite{Kuan03} claimed the first detections of glycine towards high-mass star-forming regions, their detections were refuted by \cite{Snyder05}.
While it is assumed that cometary glycine would be the remnant of the chemistry in Solar nebula \citep{Walsh14}, its formation process in the ISM is still not well understood.

Many authors have discussed the formation processes to interstellar glycine.
\cite{Blagojevic03} assumed that glycine would be synthesized via ion-molecule reactions in the gas phase.
They prepared some combinations of positive ions and neutral species, and found that the reaction of ``NH$_{2}$OH$^+$ + CH$_3$COOH $\longrightarrow$ NH$_{2}$CH$_2$COOH$^+$ + H$_2$O'' and ``NH$_{3}$OH$^+$ + CH$_3$COOH $\longrightarrow$ NH$_{3}$CH$_2$COOH$^+$ + H$_2$O'' can efficiently form protonated glycine, which would be converted to the neutral form via dissociative recombination processes with an electron.
Although NH$_2$OH (hydroxylamine) has not been detected in the ISM, it would be a plausible interstellar molecule since the ultraviolet laser irradiation of ice mixtures simulating interstellar grains produces hydroxylamine \citep{Nishi84}.
NH$_2$OH would be easily protonated in the gas phase chemistry due to its high proton affinity \citep{Blagojevic03}. 
\cite{Charnley06} performed chemical modeling calculation under the environment of hot cores with the constant density (10$^7$~cm$^{-3}$) and temperatures (100~K and 300~K) through the simulations.
Their simulations started from the observed or predicted interstellar ice abundances of CO, N$_2$, CH$_4$, H$_2$CO, C$_2$H$_5$OH, H$_2$S, C$_2$H$_6$, H$_2$O, CH$_3$OH, NH$_3$, HCOOH, and NH$_2$OH, as is found in \cite{Rodgers01} in detail.
They found that above ion-neutral processes in the gas phase can form glycine as much as 10$^{-10}$ relative to molecular hydrogen.

The formation processes of glycine on interstellar grain surface have also been explored.
\cite{Woon02} suggested that glycine would be built on the interstellar ice from COOH and CH$_2$NH$_2$ radicals.
He assumed that the UV irradiation on H$_2$O/CO ice would lead to abundant OH radical, which can react with CO to form COOH radical.
CH$_2$NH$_2$ can be formed via hydrogenation process to HCN.
\cite{Woon02} conducted quantum chemical calculations to show the the activation barriers associated with ``s-CO + s-OH''  (here ``s'' is to represent species on the surface), and successive hydrogenation processes to HCN, which finally lead to CH$_3$NH$_2$. Later, \cite{Theule11} confirmed CH$_3$NH$_2$ formation via UV 
irradiation experiment on ice containing HCN and H atoms, implying that these processes would be valid in the ISM conditions.
%
%
Then, \cite{Majumdar13} used their hydro-chemical model at 10~K to exclude the importance of CH$_2$NH$_2$ + COOH reaction in such low temperature.

Recently, \cite{Garrod13} investigated an effective formation route to glycine through chemical modeling.
He assumed that grain surface radical-radical reactions would form glycine, for instance,

s-COOH + s-CH$_2$NH$_2$ $\longrightarrow$ s-NH$_2$CH$_2$COOH, or

s-NH$_2$ + s-CH$_2$COOH $\longrightarrow$ s-NH$_2$CH$_2$COOH

\cite{Garrod13} also considered the gas phase formation of protonated glycine from CH$_3$COOH and NH$_3$OH$^+$ or NH$_2$OH$^+$, which were suggested by \cite{Blagojevic03}:

NH$_2$OH$^+$ + CH$_3$COOH $\longrightarrow$ NH$_2$CH$_2$COOH$^+$ + H$_2$O

NH$_3$OH$^+$ + CH$_3$COOH $\longrightarrow$ NH$_2$CH$_2$COOH$_2^+$ + H$_2$O

The dissociative recombination process of NH$_2$CH$_2$COOH$_2^+$ with an electron in the gas phase will form glycine.

Under the fast warm-up model, which assumed the environment of high-mass stars where its timescale for the warm-up phase was assumed to be 7.12$\times$10$^4$~years, he found that the grain surface reaction, ``s-COOH + s-CH$_2$NH$_2$ $\longrightarrow$ s-NH$_2$CH$_2$COOH'', was the most dominant formation route to glycine, and the effect of gas phase formation routes is negligible.
He claimed that s-COOH was formed via a photodissociation process of or a chemical hydrogen subtraction from s-HCOOH.
In this case, the origin of s-COOH was not ``s-CO + s-OH'' as previously suggested by \cite{Woon02}.
s-CH$_2$NH$_2$ originates from the photodissociation of s-CH$_3$NH$_2$ in the solid phase, or hydrogen subtraction processes from s-CH$_3$NH$_2$ in the solid phase.

The first detection of interstellar glycine is the holy grail for astronomers. With the current capability of ALMA, which provides greater sensitivity and spatial resolution, glycine will be the prime target across different molecule rich sources in the ISM. It is therefore invaluable to extend the state-of-the-art astrochemical models with latest theoretical and experimental studies to simulate the glycine formation and its precursors in the ISM.

In this work, we investigate the importance of different glycine formation mechanisms and its detectability in hot cores using a newly developed chemical network implemented in the three phase (gas-surface-mantle) NAUTILUS chemical model.
The detailed descriptions of our chemical and physical models will be discussed in section~2; our results will be described in section~3;
Finally, We will summarize our work in section 4.

\section{Chemical Model and Network }

\subsection{The NAUTILUS chemical model}

To investigate the chemistry of glycine in hot cores, we have used the state-of-the-art chemical code NAUTILUS described
in \cite{Ruaud16}. NAUTILUS allows us to compute the time evolution of chemical abundances for a given set of physical 
and chemical parameters. It simulates chemistry in three phases i.e. gas-phase, grain surface and grain mantle. It also considers various 
possible exchanges among the different phases via adsorption of gas phase species on to grain surfaces, the thermal and non-thermal desorption of species 
from grain surface into the gas phase and the surface-mantle and mantle-surface exchange of species. 
The gas-to-dust ratio of 100 by mass, the grain density of 3~g~cm$^{-3}$ and the grain radius of 10$^{-5}$~cm was employed.
Then, the total surface area was calculated assuming compact spherical grains.
We modeled the standard interstellar UV radiation field and its destructions of molecules as was presented by \cite{Ruaud16}.
Competition between reaction, diffusion and evaporation is also taken 
into account in the model by following \cite{Chang07} and \cite{Garrod07}. NAUTILUS computes diffusion energies of each species as a fraction of their binding energies and assumes a ratio of 0.4 for the surface whereas 0.8 for the ice mantle. 
(see \cite{Ruaud16} for more discussions).  
The thickness of the assumed barrier width is 1~$\AA$ that a surface species need to cross while undergoing quantum tunneling to diffuse from one surface site to another.
We have also updated binding energy of each species in our model from \cite{Wakelam17}.
The binding energy of a species is typically obtained by TPD experiments where either the temperature of the substrate is kept constant while the species of interest is deposited or the temperature is linearly increased until the species are desorbed from the substrate. Deposited species on the substrate can be either in the monolayer or multilayer regime.  In the multilayer regime, desorption occurs from the species itself, and the impact of the underlying substrate becomes negligible as pointed out by \cite{Green09}.
For glycine, the binding energy is estimated to be 13000~K from \cite{Tzvetkov04} (private communication with Jean-Christophe Loison).
This experiment was performed in multilayer regime, and so the effect of the substrate will have a minimal impact. Experiments in the monolayer regime are needed in order quantify the mixing of the deposited glycine molecule with the water ice substrate. 
Nevertheless, we used higher binding energy obtained by TPD experiment as the extreme case to see the effect of binding energy on the gas phase abundance of glycine.

To simulate the physical conditions in hot cores, we have used the two stage physical model: free-fall collapse, followed by a dynamically-static warm up by following \cite{Garrod13}. The cold collapse phase starts from 
n$_{H_{2}}$ =$ 3\times10^3$ cm$^{-3}$ to final post collapse density of n$_{H_{2}}$ =$ 2\times10^7$ cm$^{-3}$. The increase in visual extinction during collapse leads to the minimum dust-grain temperature of 8 K 
followed by a warm-up from 8 to 400 K; during this phase, the gas and dust temperatures are assumed to be well coupled and the gas density is fixed. 
We assumed fast and slow warm-up models described in \cite{Garrod13}, whose timescale for the warm-up phases are, respectively, 7.12$\times$10$^{4}$ and 1.43 $\times$10$^{6}$~years.
%


\subsection{Developed chemical network for glycine}
To develop a state-of-the-art chemical network for glycine, we have started from the gas-phase chemical network kida.uva.2014\footnote{\url{http://kida.obs.u-bordeaux1.fr/}} \citep{Wakelam15}. 
It includes 489 species composed of 13 elements (H, He, C, N, O, Si, S, Fe, Na, Mg, Cl, P, F) linked with 7509 reactions. \cite{Garrod13} has adapted osu.2005 as an initial network to include the chemistry of glycine.
We used the similar elemental abundances from \cite{Garrod13}, as listed here in Table~\ref{table:initial_abundance}.
Also, we listed the binding energies used in our model in Table~\ref{table:binding_energy}.
The binding energies for some species were obtained by theoretical quantum chemical computation presented in \cite{Wakelam17}, but there are still some species whose binding energies were not measured or calculated.
Our starting network includes various updates as compared to osu.2005 used in \cite{Garrod13}. It considers the updates of HCN/HNC chemistry by \cite{Loison14b}, carbon chemistry by \cite{Loison14a}, branching ratios for reactions forming C$_{n=2-10}^{(0,+)}$, C$_{n = 2- 4}$H$^{(0, +)}$, and C$_3$H$_2^{(0,+)}$ from \cite{Chabot13} and also various new data sheets available in KIDA database.
Our network for surface reactions and gas-grain interactions is based on the one from \cite{Garrod07} with several additional 
processes from \cite{Ruaud15}. We have added following updates into our initial gas-grain chemical network: 

\begin{itemize}

\item [1.] We have included the chemistry of glycine and other related complex organic molecules which are not available in the KIDA database from \cite{Garrod13} \footnote{\url{http://www.astro.cornell.edu/~rgarrod/resources/}}. It includes 
mainly four radical-addition reactions on the grains for glycine formation: 

\begin{equation}
\mbox{s-NH}\mbox{CH}_2\mbox{COOH} + \mbox{s-H} \rightarrow  \mbox{s-NH}_2\mbox{CH}_2\mbox{COOH}
\label{Eq1}
\end{equation}
\begin{equation}
\mbox{s-NH}_2 + \mbox{s-CH}_2\mbox{COOH}  \rightarrow  \mbox{s-NH}_2\mbox{CH}_2\mbox{COOH}
\label{Eq2}
\end{equation}
\begin{equation}
\mbox{s-CH}_2\mbox{NH}_2 + \mbox{s-COOH}  \rightarrow  \mbox{s-NH}_2\mbox{CH}_2\mbox{COOH}
\label{Eq3}
\end{equation}
\begin{equation}
\mbox{s-NH}_2\mbox{CH}_2\mbox{CO} + \mbox{s-OH} \rightarrow  \mbox{s-NH}_2\mbox{CH}_2\mbox{COOH}
\label{Eq4}
\end{equation}
These reactant radicals are mainly formed either through grain surface reactions or photodissociation of stable molecules. It also includes gas-phase destruction routes for glycine via UV photodissociation, either by the cosmic ray-induced or standard 
interstellar radiation field, or by ion-molecular reactions.

The gas phase reactions of CH$_3$COOH and protonated NH$_2$OH to form protonated glycine were also included from \cite{Garrod13}, but the reaction rate coefficients of NH$_2$OH and its protonated species were updated from the theoretical calculation describe in \cite{Barrientos12}.

\item [2.]  We have updated the chemistry of CH$_2$NH and CH$_3$NH$_2$ from \cite{Suzuki16}. 
These reactions are series of hydrogenation processes to HCN on grains, as summarized in Table~\ref{table:HCN_hydrogenation}.

\item [3.] We have included the glycine formation on the grains from \cite{Singh13} via sequences of reactions with simple species, which are abundant in the ISM. They showed following possible routes to form glycine via quantum chemical calculations: 

\begin{equation}
\mbox{s-CH}_2+ \mbox{s-NH}_2 \rightarrow  \mbox{s-CH}_2\mbox{NH}_2
\label{Eq5}
\end{equation}
\begin{equation}
\mbox{s-CH}_2\mbox{NH}_2  + \mbox{s-CO} \rightarrow  \mbox{s-NH}_2\mbox{CH}_2\mbox{CO}
\label{Eq6}
\end{equation}
\begin{equation}
\mbox{s-NH}_2\mbox{CH}_2\mbox{CO} + \mbox{s-OH}  \rightarrow  \mbox{s-NH}_2\mbox{CH}_2\mbox{COOH}
\tag{\ref{Eq4}}
\end{equation}
While the reactions (2) was employed in \cite{Garrod13}, \cite{Singh13} suggested the new reaction sequences to finally form glycine involving CH$_2$CO, NH$_2$CH$_2$CO, and NH$_2$CHCO. 
We tested the importance of these reaction sequence for the first time.
\cite{Singh13} concluded that Reaction 5 and 7 are barrier-less reactions, while the Reaction 6 has an activation barrier of 7100~K. 

Another sequences of reactions start from a combination of NH$_2$ and CH, as below:

\begin{equation}
\mbox{s-NH}_2+ \mbox{s-CH} \rightarrow  \mbox{s-NH}_2\mbox{CH}
\label{Eq7}
\end{equation}
\begin{equation}
  \mbox{s-NH}_2\mbox{CH} + \mbox{s-CO} \rightarrow  \mbox{s-NH}_2\mbox{CH}\mbox{CO}
\label{Eq8}
\end{equation}
\begin{equation}
\mbox{s-NH}_2\mbox{CH}\mbox{CO} + \mbox{s-OH}  \rightarrow  \mbox{s-NH}_2\mbox{CH}\mbox{COOH}
\label{Eq9}
\end{equation}
\begin{equation}
\mbox{s-NH}_2\mbox{CH}\mbox{COOH} + \mbox{s-H}  \rightarrow  \mbox{s-NH}_2\mbox{CH}_2\mbox{COOH}
\label{Eq10}
\end{equation}
There are no activation barriers for Reaction 7, 8 and 9. However, Reaction 10 possesses an activation barrier of as high as $\sim$37000~K.

The third path starts from CH$_2$ and CO:

\begin{equation}
\mbox{s-CH}_2+ \mbox{s-CO} \rightarrow  \mbox{s-CH}_2\mbox{CO}
\label{Eq11}
\end{equation}
\begin{equation}
\mbox{s-CH}_2\mbox{CO}  + \mbox{s-OH} \rightarrow  \mbox{s-CH}_2\mbox{COOH}
\label{Eq12}
\end{equation}
\begin{equation}
\mbox{s-NH}_2 + \mbox{s-CH}_2\mbox{COOH} \rightarrow  \mbox{s-NH}_2\mbox{CH}_2\mbox{COOH}
\tag{\ref{Eq2}}
\end{equation}
Reaction 11, 12  and 2 are barrier-less reactions.

\item [4.] Based on the latest high level quantum chemical computation from \cite{Barrientos12}, we have also included the following reactions into the gas phase: NH$_2$OH$_2^+$ + CH$_3$COOH with a barrier of 1150~K; NH$_2$OH$^+$ + CH$_3$COOH with a barrier of 12180 K and NH$_3$OH$^+$ + CH$_3$COOH with a barrier of 13600~K respectively.  
The formation of the positive ion of glycine, NH$_2$CH$_2$COOH$_2^+$, will be followed by the dissociative recombination processes with an electron, to form glycine ($\sim2\%$) and other fragmented species ($\sim98\%$), such as NH$_2$, CH$_2$, and CH$_2$NH$_2$.
All products and the reaction coefficients of these dissociative recombination processes were obtained from \cite{Garrod13}, as summarized in Table~5 of \cite{Garrod13}.
\end{itemize}

Finally\footnote{The full chemical network is available on request to taiki.suzuki@nao.ac.jp and liton.majumdar@jpl.nasa.gov}, we have nearly 4500 reactions on grain surface linked with the 9500 reactions in gas phase.

\section{Results and Discussion}
\subsection{The Chemical Compositions of Grain Mantle during the Collapsing Phase}
\begin{figure}
 \begin{tabular}{ll}
\includegraphics[scale=.4]{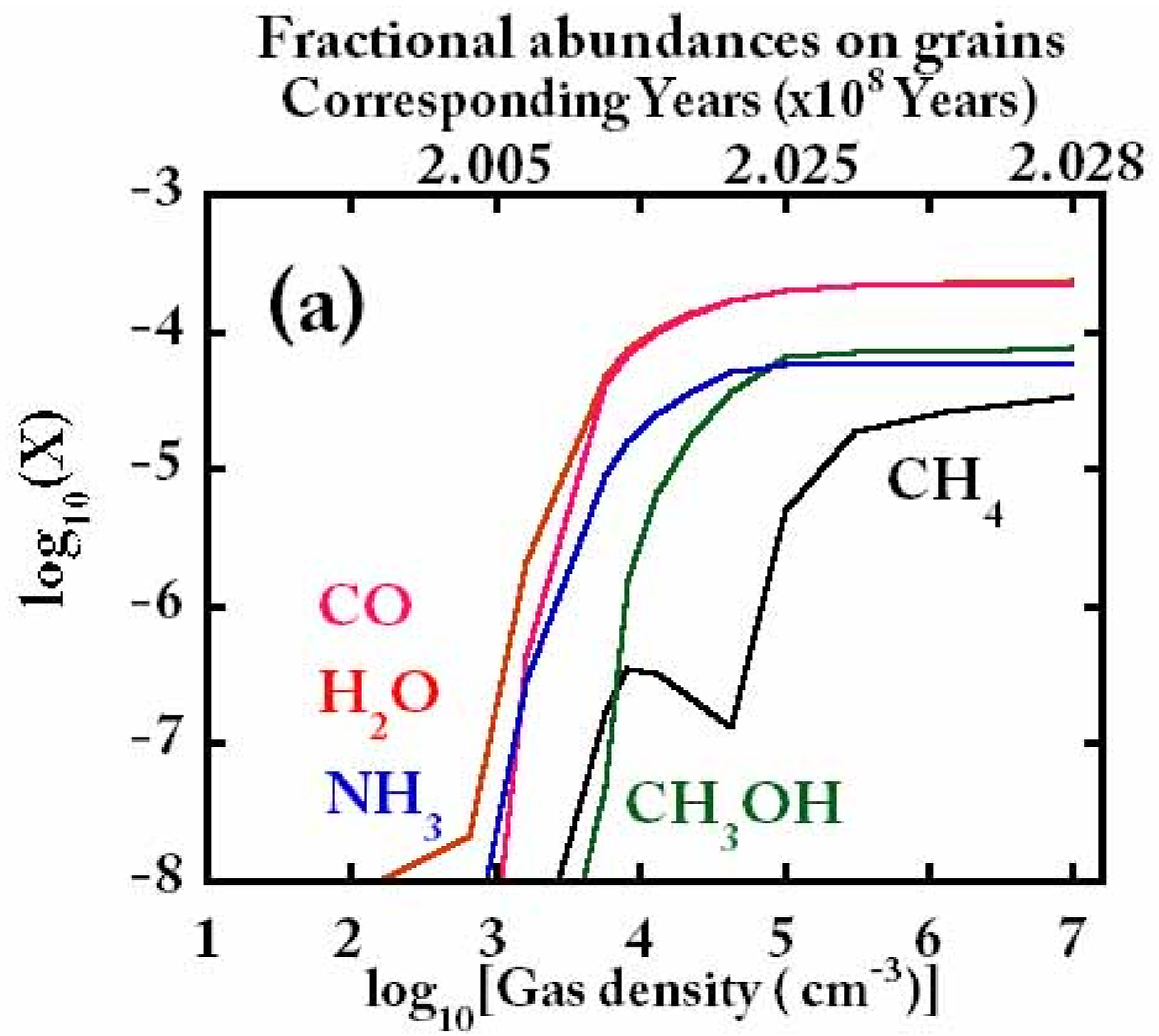}&
\includegraphics[scale=.4]{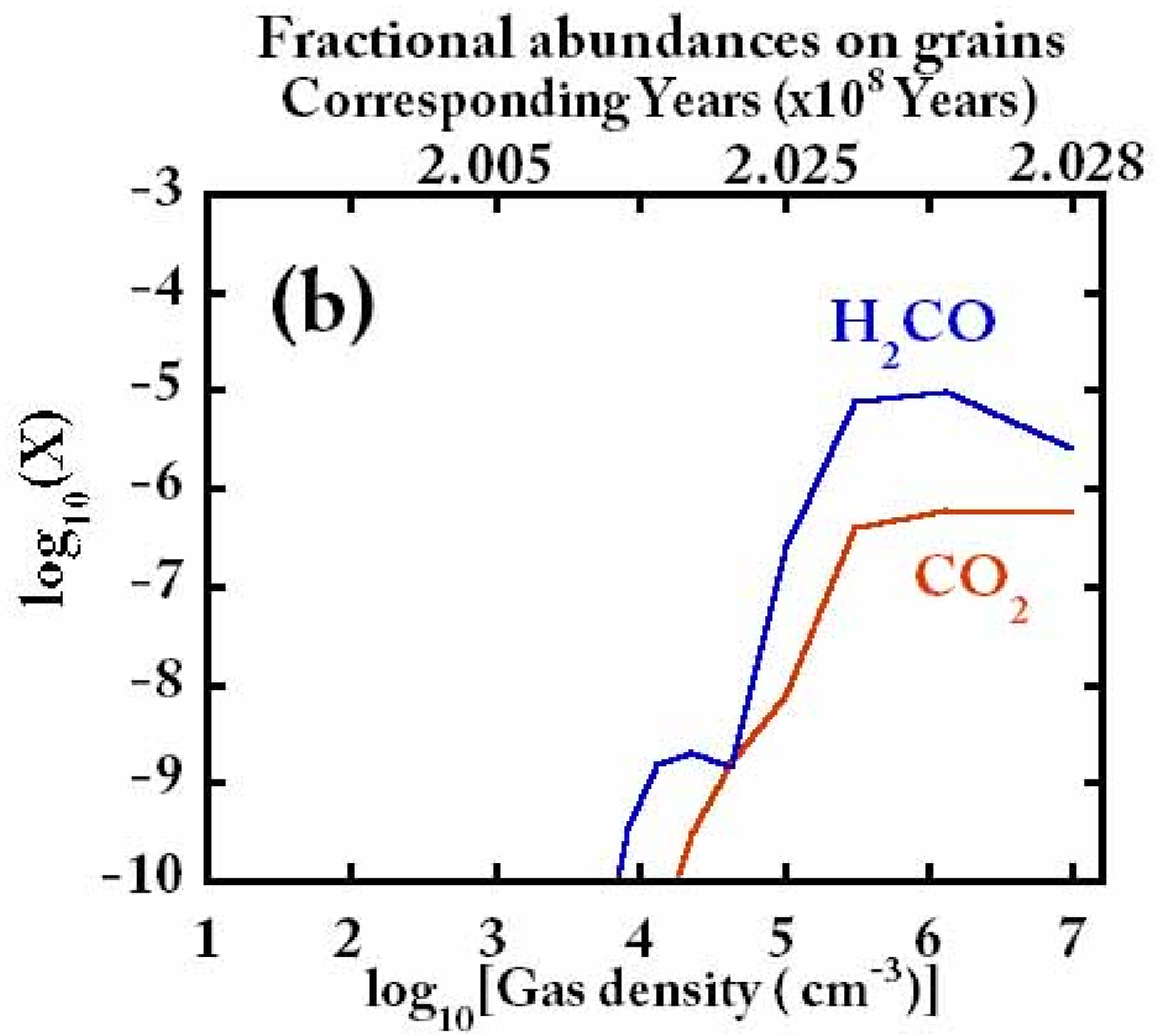}\\
\includegraphics[scale=.4]{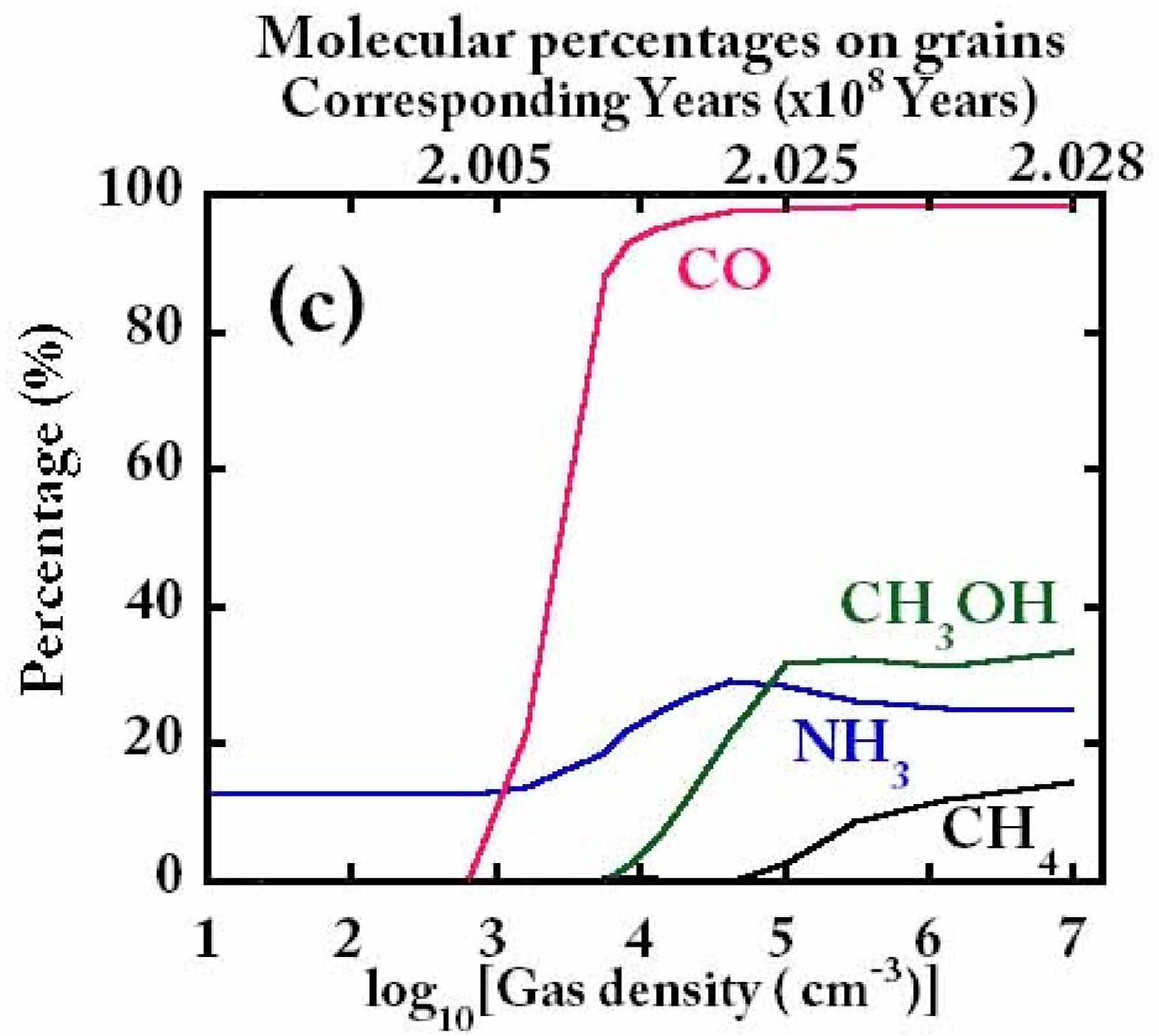}&
\includegraphics[scale=.4]{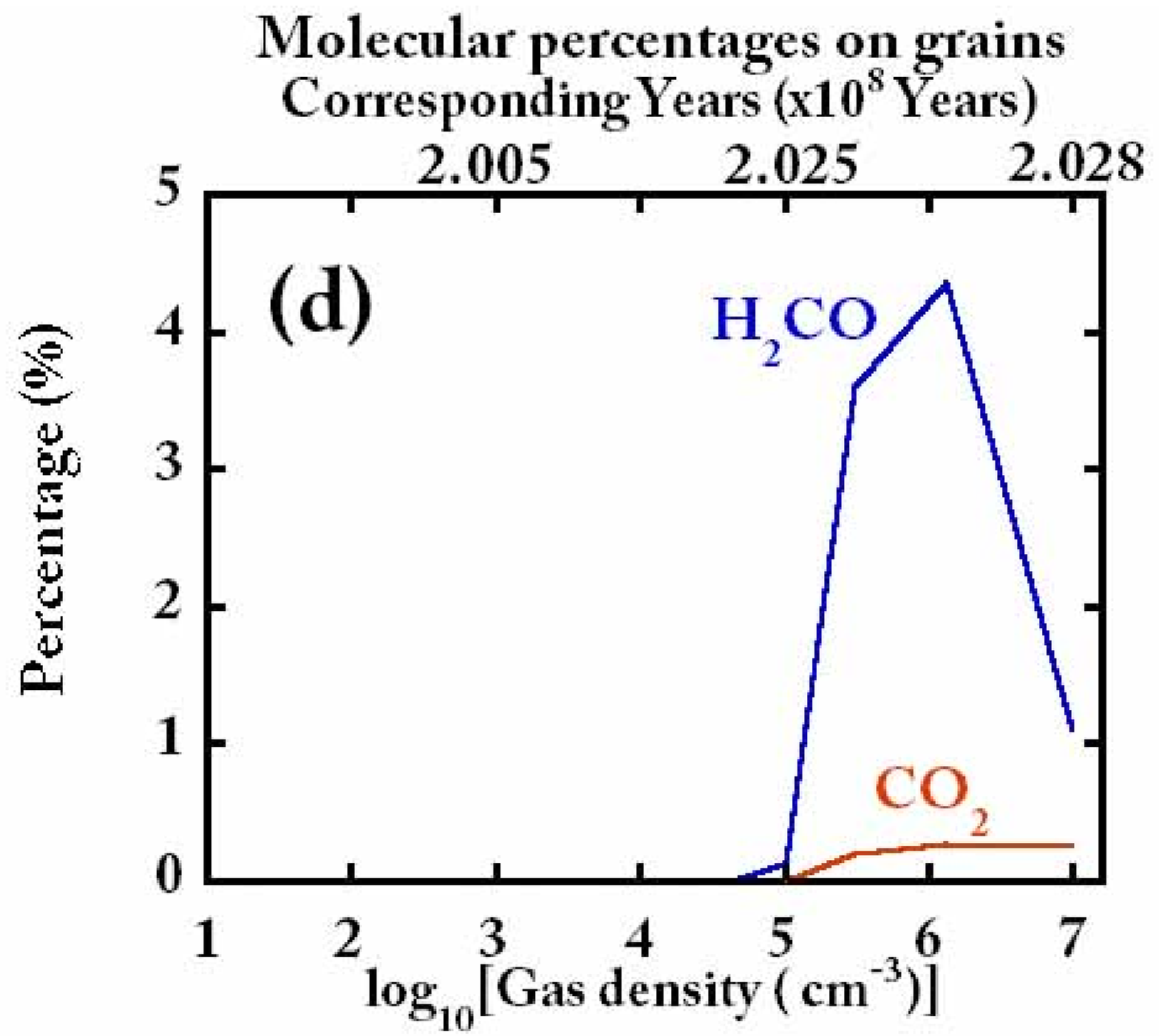}\\
  \end{tabular}
\caption{
 (a) and (b): The fractional abundance of H$_2$O, CO, CO$_2$, CH$_3$OH, NH$_3$, CH$_4$, and H$_2$CO in the solid phase are shown during the collapsing phase.
(c) and (d): The molecular percentages compared to H$_2$O on grains.
\label{fig:abundances_on_grains}
}
\end{figure}

With our model, we calculate the evolution of fractional abundances compared to the total proton density (hereafter presented as X).
In Figure~\ref{fig:abundances_on_grains}~(a) and (b), the fractional ice abundances of H$_2$O, CO, CO$_2$, CH$_3$OH, NH$_3$, CH$_4$, and H$_2$CO are shown.
CO$_2$  is mainly formed from CO and O on grains.
On the other hand, CO is formed in the gas phase and accrete on grains.
The formation processes of H$_2$O, NH$_3$, CH$_4$, H$_2$CO and CH$_3$OH are hydrogenation processes to O, N, C and CO, respectively, on the grain surface.
In our model, water ice starts to form at ~1$\times$10$^3$ years and shows its peak in between 2.005$\times$10$^8$ and 2.025$\times$10$^8$ years.
According to \cite{Ruaud16}, water ice reaches its peak in roughly 1$\times$10$^6$~years. 
This timescale is much shorter than our timescale of 1$\times$10$^8$ years. 
The main reason behind the different time scales is \cite{Ruaud16} assume a constant density of 1$\times$10$^4$ cm$^{-3}$, whereas our model assumed the time evolution during collapsing phase.
In our model, water ice shows its peak after 2$\times$10$^8$ years when density exceeds 1$\times$10$^4$~cm$^{-3}$.

The molecular percentages compared to H$_2$O ice are plotted in Figure~\ref{fig:abundances_on_grains}~(c) and (d).
We compare these molecular percentages in our model with \cite{Boogert15} in Table~\ref{table:ice_abundances}.
It is to be noted that observed ices towards massive YSOs are known to be slight to strongly affect by thermal processing \cite{Boogert15}.
As a result, comparison of abundances of key ice species at the end of collapse may not be directly relevant for simulations run under collapse model, i.e., where the dust temperature is as low as 16~K at the initial visual extinction of Av=2, and during collapse this falls to a minimum temperature of 8 K, whereas gas temperature remains constant at 10 K.
The best-fit age is 2.026$\times$10$^8$~years during the collapsing phase when the gas density gets high enough to form sufficient hydrogenated species via grain-surface reactions.

\subsection{Main Formation Routes to Glycine and its Precursors}
\subsubsection{Fast Warm-up Model}
\begin{figure}
 \begin{tabular}{ll}
\includegraphics[scale=.4]{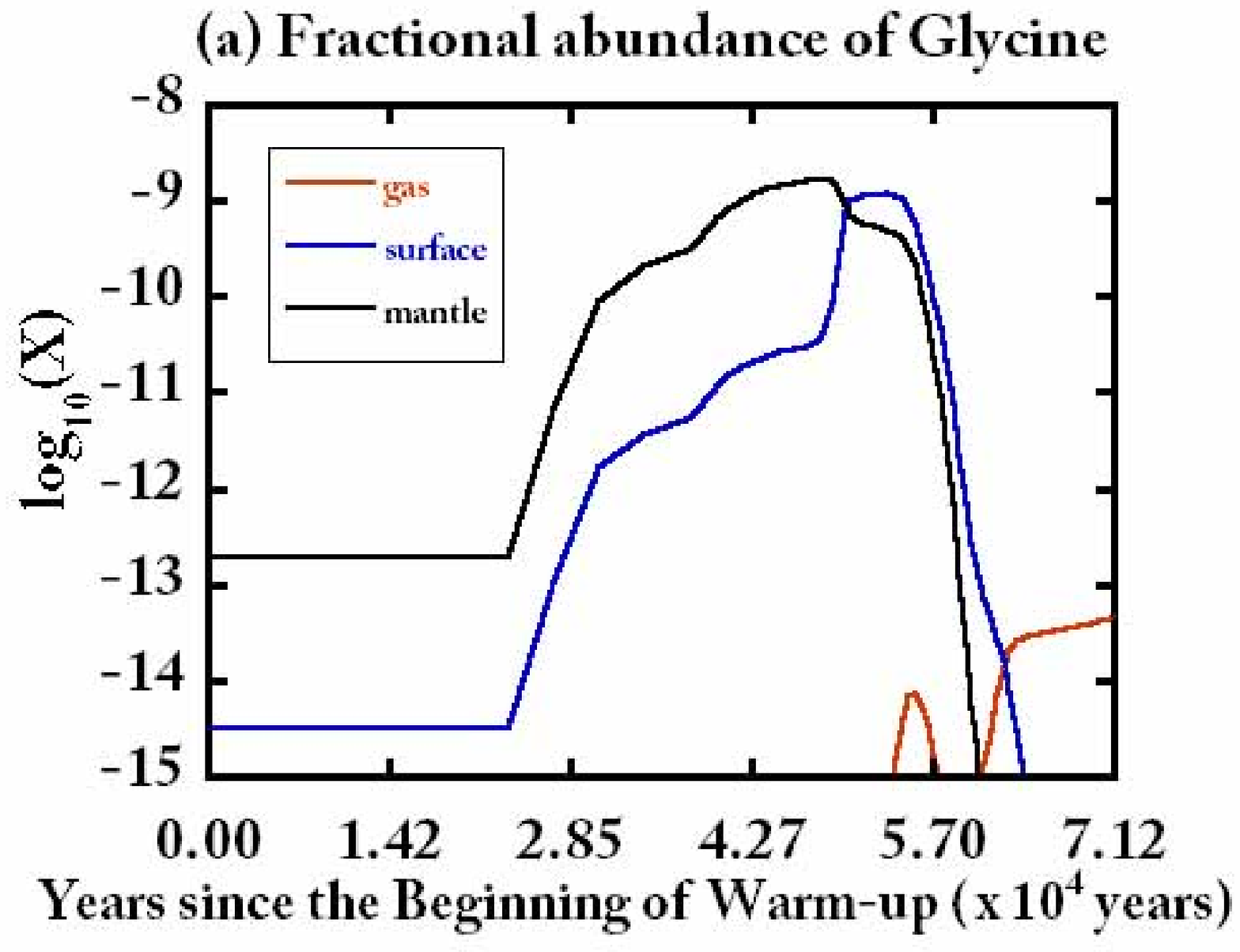}&
\includegraphics[scale=.4]{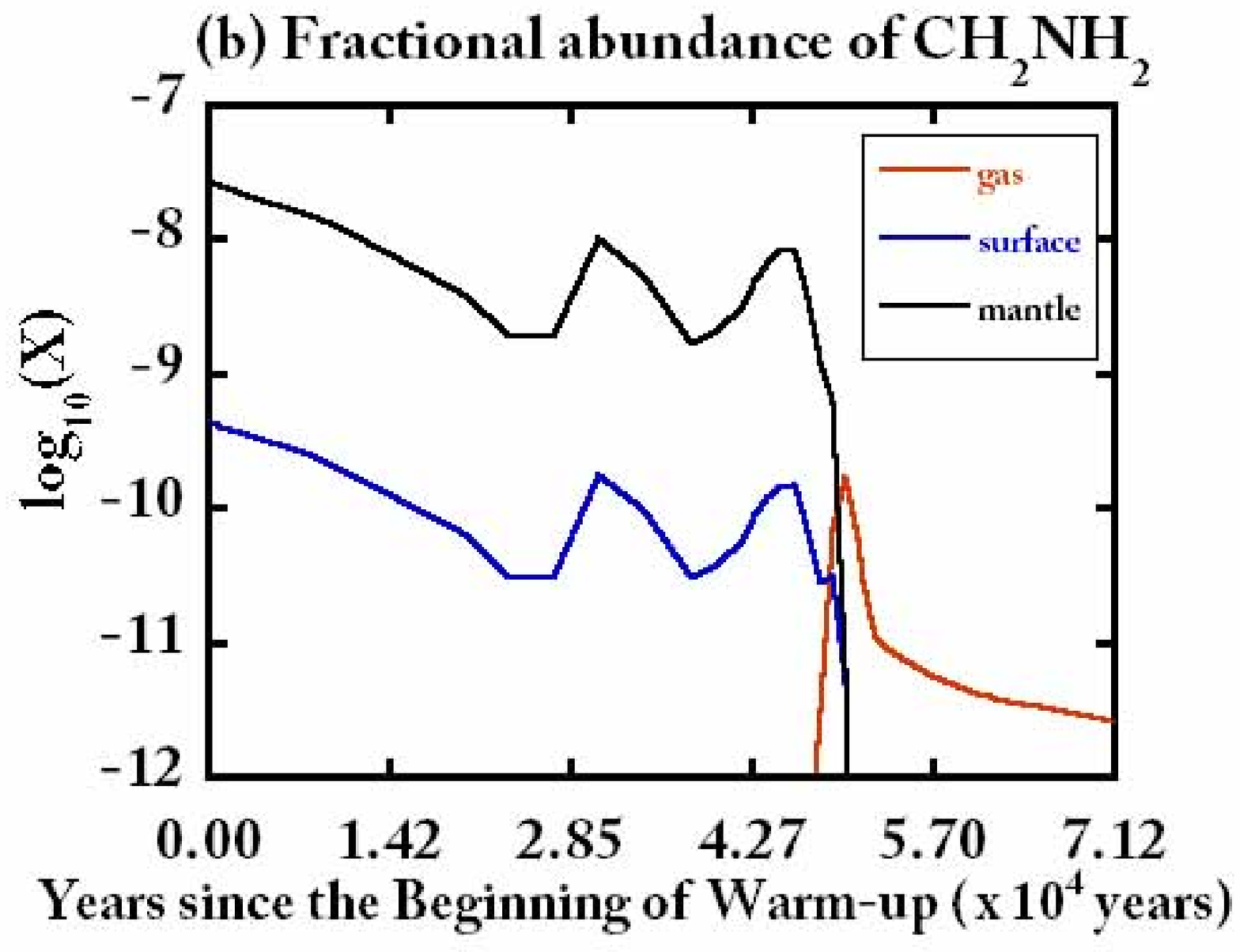}\\
\includegraphics[scale=.4]{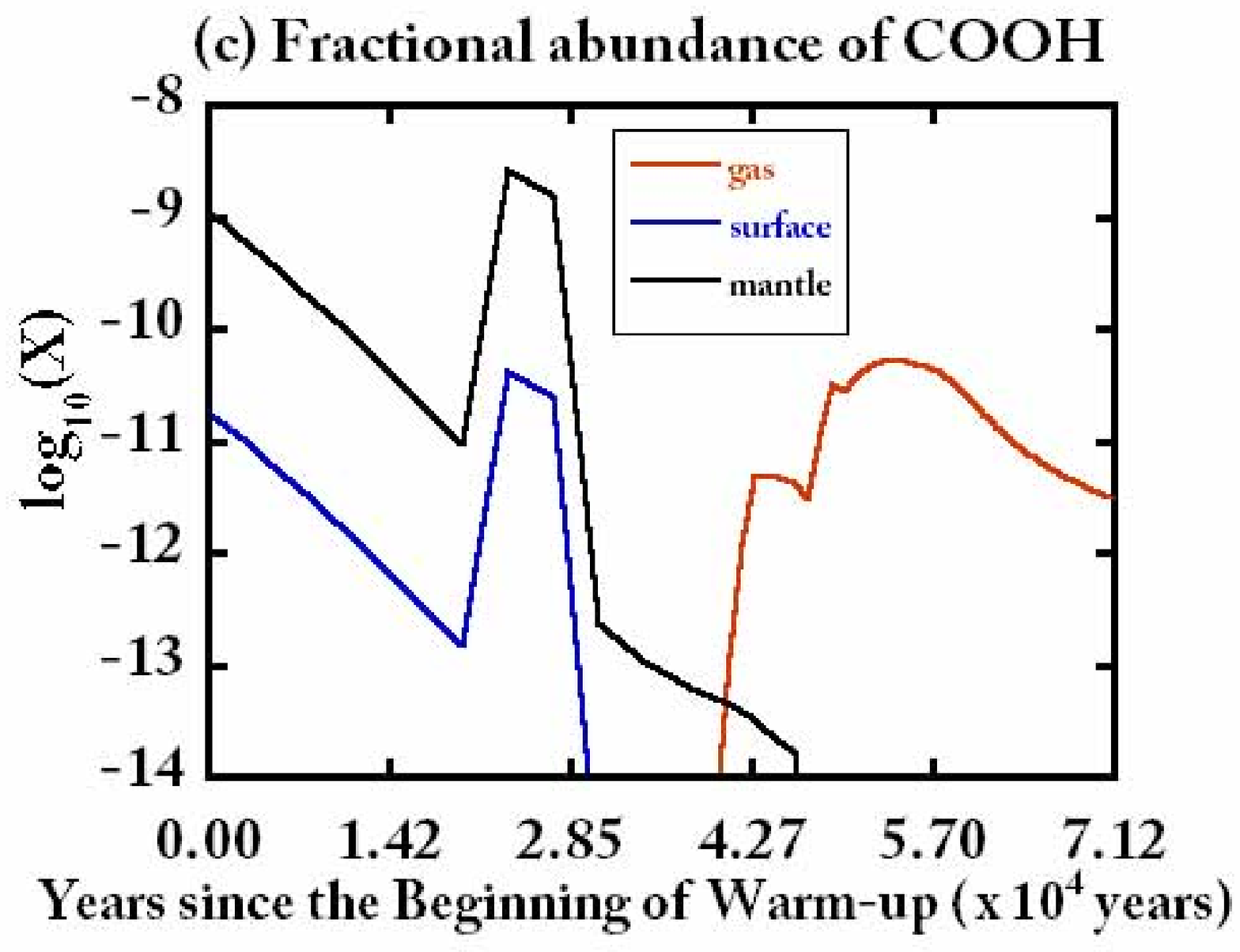}&
\includegraphics[scale=.4]{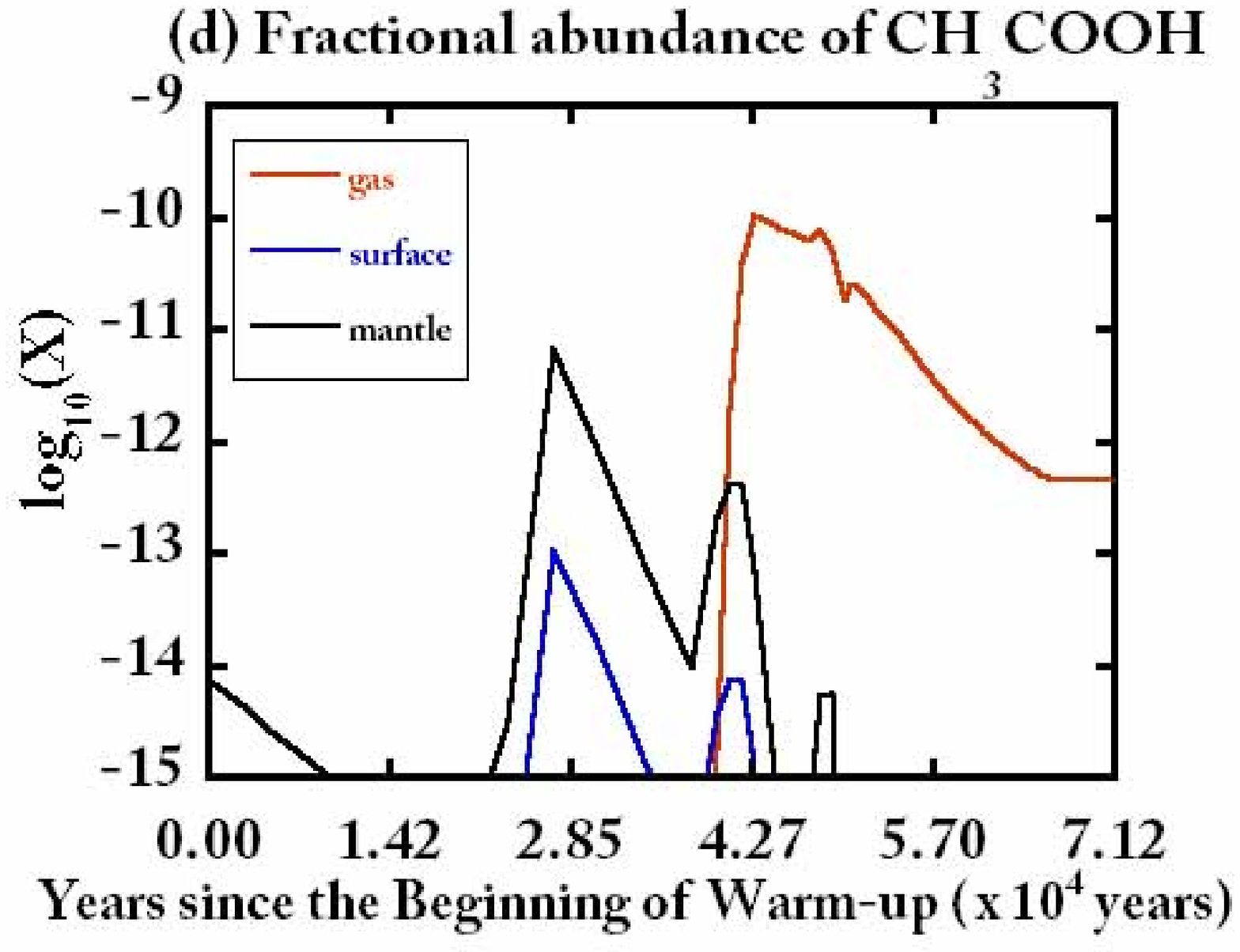}\\
\includegraphics[scale=.4]{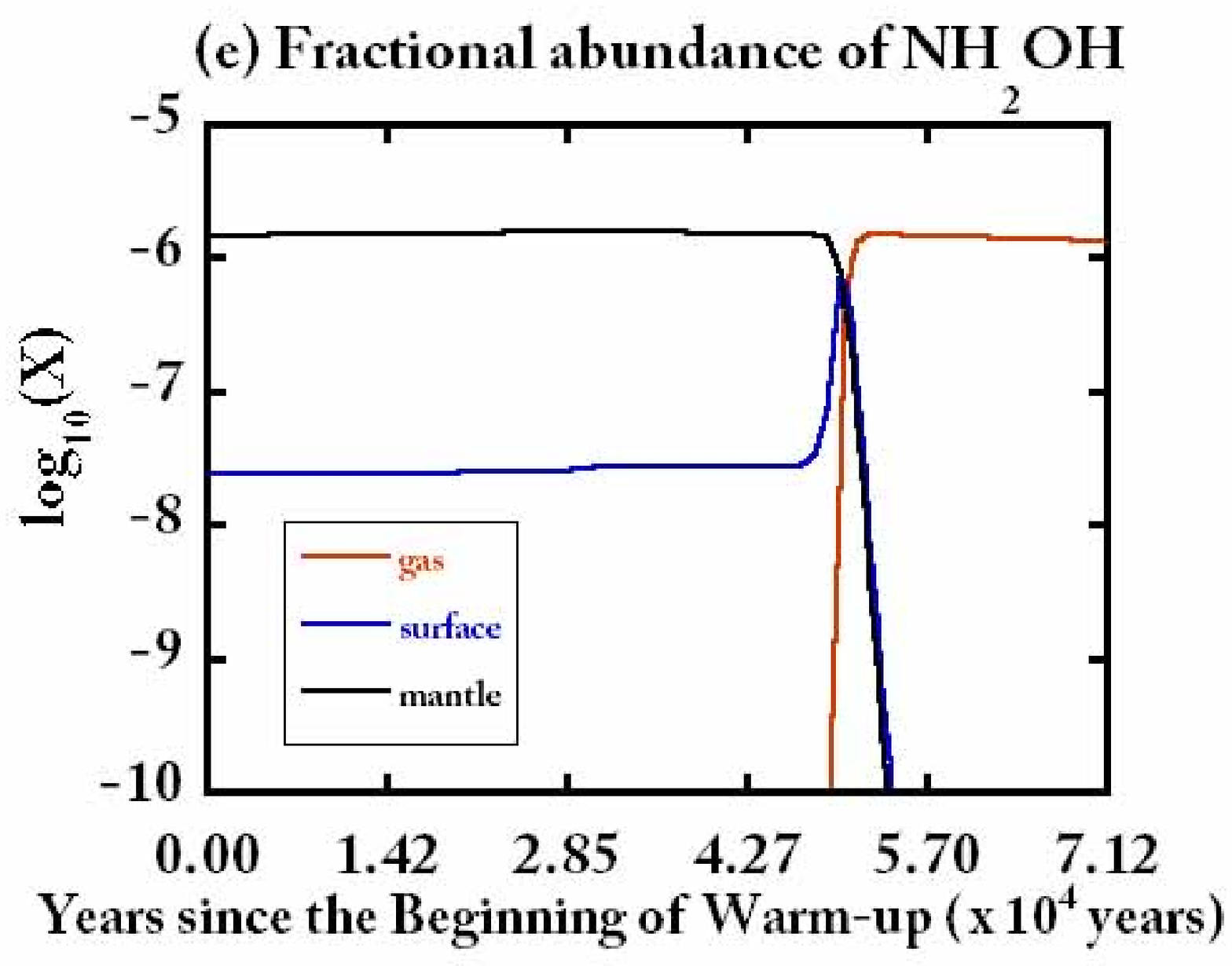}&\\
  \end{tabular}
\caption{
The simulated abundances of glycine and its precursors were compared using the Fast Model.
The red, blue, and black lines, respectively, represent the abundances in the gas phase, on the grain surface, and in the grain mantle.
\label{fig:Glycine_Abundances_fast}
}
\end{figure}

In Figure~\ref{fig:Glycine_Abundances_fast}~(a), we show the fractional abundances of glycine calculated with the fast warm-up model (hereafter we refer it as Fast Model).
The abundances in the gas phase, on grain surface, and in the grain mantle are presented using the red, blue and black lines.
Once the peak abundance of glycine is achieved on grains, it starts to be destroyed through the grain surface reactions before glycine is thermally desorbed.

Figure~\ref{fig:Glycine_Abundances_fast}~(a) suggests that the peak abundance of glycine in the grain mantle is $\sim$1$\times$10$^{-9}$ at 4-5$\times$10$^{4}$ years since the beginning of the warm-up phase.
In other words, the formation of glycine is completed on the grains at that time.
In Figure~\ref{fig:Formation_Rate1_fast}~(a), the formation rates (cm$^{-3}$s$^{-1}$) of glycine on grain surface are compared.
The most dominant formation path on the grains is ``s-CH$_2$NH$_2$ + s-COOH", whose peak is achieved when the temperature is 60-120~K at 4-5$\times$10$^{4}$ years.
In the peak of black lines, HNCH$_2$COOH radical, which is produced via the destruction process of glycine, is used to reproduce glycine again via the hydrogenation process.
Considering the fact that the abundance of glycine in the mantle achieved its peak at 5$\times$10$^{4}$ years, when ``s-CH$_2$NH$_2$ + s-COOH" is the most efficient, this process would be the major path on the grain surface.

We plot the abundances of the glycine precursors, CH$_2$NH$_2$ and COOH in Figures~\ref{fig:Glycine_Abundances_fast}~(b) and (c).
The COOH radical is easily lost from grain surface (blue dotted line) when the temperature is high ($\sim$60~K).
The formation rate of glycine on grain surface via ``s-CH$_2$NH$_2$ + s-COOH" process decreases at $\sim$5$\times$10$^{4}$~years, when COOH radical is completely lost from grain surface.
The formation rates of CH$_2$NH$_2$ and COOH radicals are compared in Figures~\ref{fig:Formation_Rate1_fast}~(b) and (c), in the similar manner with glycine.

In Figures~\ref{fig:Formation_Rate1_fast}~(b), we compare the formation rates of CH$_2$NH$_2$ via ``s-CH$_2$NH + s-H $\longrightarrow$ s-CH$_2$NH$_2$", ``s-CH$_3$O + s-CH$_3$NH$_2$ $\longrightarrow$ s-CH$_3$OH + s-CH$_2$NH$_2$" and the sum of the formation rates via other processes with green, red and black lines. 
The process of ``s-CH$_2$NH + s-H $\longrightarrow$ s-CH$_2$NH$_2$" is efficient at the beginning of the warm-up phase, when the low temperature enabled the hydrogen atoms to be stored on grains.
The other process ,``s-CH$_3$O + s-CH$_3$NH$_2$ $\longrightarrow$ s-CH$_3$OH + s-CH$_2$NH$_2$", is efficient when the temperature is more than $\sim$60~K, so that heavier radicals can move on the grain surface by thermal hopping.
Since glycine is formed on the grains before 5$\times$10$^{4}$ years, the above two processes would be key origins of CH$_2$NH$_2$ radicals for the formation of glycine.
The formation rates of COOH radical are compared in Figures~\ref{fig:Formation_Rate1_fast}~(c).
The red line represents the formation rate of COOH radical via the hydrogen subtraction process by H atoms from HCOOH: ``s-H + s-HCOOH $\longrightarrow$ s-H$_2$ + s-COOH''.
The green and blue lines, respectively, represent the sums of the formation rates of COOH radical via destruction processes by NH$_2$ and OH radicals, respectively, of carboxyl groups, HCOOH, CH$_3$OCOOH, and CH$_2$OHCOOH.
The former process was dominant when the temperature is low, while the latter processes are important when the grain surface temperature gets warmer and radicals can move on the grains.
Figures~\ref{fig:Glycine_Abundances_fast}~(c) shows that the peak of COOH radical on the grain surface is achieved at 3$\times$10$^{4}$~years, when the grain surface temperature is $\sim$30~K and the efficiency of ``s-H + s-HCOOH $\longrightarrow$ s-H$_2$ + s-COOH'' decreases.
Despite the efficient conversion process of HCOOH to COOH on grains at this age, the reverse hydrogenation process of ``s-H + s-COOH $\longrightarrow$ s-HCOOH'' suppresses the abundance of COOH radical on the grains.
Once hydrogen atoms are liberated from grains at 3$\times$10$^{4}$ years, the hydrogenation rate to converte COOH radical to HCOOH decreases.
As a result, the destruction processes of carboxyl groups by NH$_2$ and OH radicals increases COOH radicals on the grain surface.
The above formation process of glycine on grains agrees with \cite{Garrod13}.

Despite the agreement on the formation process of glycine on the grains with \cite{Garrod13}, our chemical model suggests that the major formation process of gas phase glycine is gas phase reactions rather than the thermal evaporation of grains with its binding energy of 13000~K, contrary to \cite{Garrod13}.
In Figure~\ref{fig:Formation_Rate1_fast}~(d), we plot the subtraction of the accretion rate of gas phase glycine from the evaporation rate of glycine on grains.
The accretion rate always overwhelms the evaporation rate, suggesting that the thermal evaporation is not efficient.
The comparison of the gas phase formation rates of glycine in Figure~\ref{fig:Formation_Rate1_fast}~(e) shows that ``NH$_2$CH$_2$COOH$_2^+$ + e$^-$" is the major formation path to glycine at $\sim$5$\times$10$^5$~years since the beginning of the warm-up.
NH$_2$CH$_2$COOH$_2^+$ is produced via the reaction of CH$_3$COOH and NH$_2$OH$_2^+$.
NH$_2$OH$_2^+$ is formed from NH$_2$OH with the reactions of positive ions such as H$_3^+$ and H$_3$O$^+$, as shown in Figure~\ref{fig:Formation_Rate1_fast}~(f).
In our model, both CH$_3$COOH and NH$_2$OH are efficiently built on grains (Figures~\ref{fig:Formation_Rate1_fast}~(g) and (h)).
The formation of NH$_2$OH is almost completed during the collapsing phase.
Once HNO is formed in the gas phase, it accretes on grains and was converted to NH$_2$OH via the hydrogenation processes.
CH$_3$COOH is built on grains from CH$_3$CO and OH radicals.
We conclude that the series of above reactions are the most essential process to form gas phase glycine in this model.
We note that the predicted abundances and the formation mechanisms discussed here are subject to the binding energies, reaction rates and physical evolution of the core, and should be updated based on the latest progress of these studies.
Especially, the precise value of binding energy of glycine should be explored in detail in the future works as the essential parameter to change the predicted glycine abundances, major formation processes, and key precursors.

\subsubsection{Slow Warm-up Model}
In this subsection, we will present our result under slow warm-up model, where the timescale of warm-up is 1.43$\times$10$^{6}$.
We show the simulated abundances and the formation route to glycine in the Figure~\ref{fig:Glycine_Abundances_slow} and Figure~\ref{fig:Formation_Rate1_slow}~(a) in the same manner as the Fast Model.
In this model (referred as Slow Model), the gas phase peak abundance of glycine is much smaller than those on the grain surface and in the grain mantle.
The peak abundance of glycine in the gas phase is lower than in the fast warm-up model, due to the destruction process on grains of glycine by radicals such as OH and NH before thermal evaporation.

In Figure~\ref{fig:Formation_Rate1_slow}~(a), the formation rates (cm$^{-3}$~s$^{-1}$) of glycine on grain surface are compared.
The contributions of other reactions is mainly due to ``s-H + s-NHCH$_2$COOH'', where NHCH$_2$COOH is formed via destruction processes of glycine.
However, similar to the fast warm-up model, the most dominant process is ``s-CH$_2$NH$_2$ + s-COOH'', and its peak is achieved when the temperature is between 60-120~K at 8-10$\times$10$^{5}$ years.

The abundances and the formation rates of the glycine precursors on grains, CH$_2$NH$_2$ and COOH, are compared in Figures~\ref{fig:Glycine_Abundances_slow} and \ref{fig:Formation_Rate1_slow}~(b) and (c) with the same way as the fast warm-up model.
CH$_2$NH$_2$ is produced via the destruction processes of CH$_3$NH$_2$ by NH, OH, and CH$_3$O radicals.
The formation process of COOH was the destruction of HCOOH, which is consistent with the fast warm-up model, while the destruction by UV photons are more important than the fast warm-up model due to the longer timescale.
For the overall trend, the formation process of glycine on the grains does not depend on the warm-up speed.

However, similar to the Fast Model, the fact that the accretion rate of glycine is higher than its evaporation rate (Figure~\ref{fig:Formation_Rate1_slow} (d)) suggests an efficient formation of this molecule in the gas-phase.
In Slow Model, glycine is formed from the reaction of NH$_2$OH$_2^+$ with CH$_3$COOH (Figure~\ref{fig:Formation_Rate1_slow}~(e)).
NH$_2$OH$_2^+$ is formed from the gas phase reactions of NH$_2$OH with positive ions, mainly H$_3$O$^+$ (Figure~\ref{fig:Formation_Rate1_slow}~(f)).
Both NH$_2$OH and CH$_3$COOH thermally evaporate during the warm-up phase (Figure~\ref{fig:Formation_Rate1_slow}~(g) and (h)).

\subsection{Comparison with \cite{Garrod13}}
We report in Table~\ref{table:glycine_abundances}, the gas-phase glycine abundance computed with our model and the one from \cite[Table~8]{Garrod13}.
Similarly, the peak abundance (in the gas, in the surface and in the mantle) of glycine precursors, NH$_2$OH, CH$_2$NH, CH$_3$NH$_2$, HCOOH, and CH$_3$COOH in \cite{Garrod13} are shown on Tables~\ref{table:Garrod_vs_Model1} and \ref{table:Garrod_vs_Model2} together with the temperature corresponding to the peak.
We note that we have only gas-phase abundances for \cite{Garrod13} results.
It is notable that the peak abundance of glycine was decreased by about a factor of 100 in our model as compared to Garrod.
This discrepancy is due to the high desorption energy of glycine in our model, determined by the result of the TPD experiment by \cite{Tzvetkov04}.
With the binding energy of 13000~K, glycine is destroyed through grain surface reactions with NH or CH$_3$O radicals before it fully evaporates.
The dramatical increase of NH$_2$OH anf CH$_3$NH$_2$ in our model are due to the inclusion of a series of hydrogenation processes to NO and HCN, respectively.
The increase of CH$_2$NH would be due to the update of gas phase chemistry in kida.uva.2014, which is described in detail in \cite{Suzuki16}.
In Slow Model, CH$_3$COOH is depleted compared to \cite{Garrod13}.
Considering that this depletion is not seen in Fast Model, the depletion is due to the longer timescale of collapsing phase and different amount of radical species to destroy CH$_3$COOH before its evaporation.

If the binding energy is 13000~K, formation processes of glycine in the gas phase are more important than thermal evaporation from grains.
The higher accretion rate than the evaporation rate shows the gas phase origin of glycine (Figure~\ref{fig:Formation_Rate1_slow}~(d)).
In this case, glycine is formed via ``NH$_2$CH$_2$COOH$_2^+$ + e$^-$", and the precursors are CH$_3$COOH and NH$_2$OH.
These species are built on grains, and then thermally evaporate into gas phase (Figures~\ref{fig:Glycine_Abundances_slow}~(d) and (e)).

We compare the abundances of glycine in the gas phase in Figure~\ref{fig:Abundances_Glycine_ED}, changing the desorption energies of glycine to be 10100~K, corresponding to the one in \cite{Garrod13}.
If we employ the same desorption energy as \cite{Garrod13} in Fast Model~(b), we obtain the peak abundance of gas phase glycine to be 3.1$\times$10$^{-11}$ for the fast warm-up model.
This value agrees with \cite{Garrod13}, where the peak abundance of glycine was reported to be 8.4$\times$10$^{-11}$.
These results clearly suggest that the discrepancy of the formation process of gas phase glycine is due to the different binding energy of glycine, being the essential parameter to affect the abundance and the detectability of gas phase glycine.
With the binding energy of 10100~K for glycine, the gas-phase abundance is mostly due to thermal evaporation.

These results strongly suggest that the binding energy of glycine is a key parameter that affects the abundance of gas phase glycine.
The further study of TPD experiments under monolayer regime is required to improve our understanding further.

\subsection{Evaluation of the Simulated Abundance of Glycine}
In this subsection, we will discuss if the actual abundances of precursors of glycine can be explained with our Fast Model.
\cite{Favre17} detected CH$_3$COOH towards the edge of Orion~KL Hot Core, and reported two velocity components with their CH$_3$COOH column densities of 1.2$\times$10$^{16}$ and 3.3$\times$10$^{15}$~cm$^{-2}$, respectively.
Since the peak position of CH$_3$COOH is close to the source called HKKH7 reported in \cite{Hirota15}, we use the hydrogen column density of 1.1$\times$10$^{25}$~cm$^{-2}$ based on the estimation by \cite{Hirota15}.
Then, the fractional abundances of CH$_3$COOH in these components are, respectively, 1.0$\times$10$^{-9}$ and 3.0$\times$10$^{-10}$.
This value is close to 0.9-7$\times$10$^{-10}$, corresponding to that of Sgr~B2 \citep{Mehringer97}.
With our peak fractional abundance of 1.0$\times$10$^{-10}$, these values can be explained within a factor of 10.
On the other hand, for NH$_2$OH, the upper limits of NH$_2$OH towards Orion~KL and Sgr~B2 are, respectively, 3$\times$10$^{-11}$ and 8$\times$10$^{-12}$ \citep{Pulliam12}.
These upper limits of NH$_2$OH suggest that we strongly overestimated the abundance of NH$_2$OH.
Recently, \cite{Jonusas16} has experimentally shown that heating of NH$_2$OH-H$_2$O ices leads to a decomposition of NH$_2$OH into HNO, NH$_3$ and O$_2$ at 120~K, and it is possible that lack of this effect may have led to strong overestimation of NH$_2$OH in our model.
Since glycine is formed in the gas phase using the positive ion of NH$_2$OH, the gas phase abundance of glycine would be overestimated and gives us only the upper limit.
This result also suggests us the importance of thermal decomposition processes of not only NH$_2$OH, but also other species for the high-temperature chemistry.

On the grain surface, the important precursors are CH$_2$NH, CH$_3$NH$_2$ and HCOOH.
As \cite{Suzuki16} claimed that gas phase CH$_2$NH would originate from gas phase reaction of ``NH + CH$_3$'', the gas phase abundance of CH$_2$NH does not represent the grain surface origin of CH$_2$NH.
Therefore, we will focus on HCOOH and CH$_3$NH$_2$ abundances in the gas phase for the benchmark of our calculation.
\cite{Liu02} reported the fractional abundance of HCOOH to be $\sim$3$\times$10$^{-9}$ towards Orion~KL, which agrees well with our peak value of 3.4$\times$10$^{-9}$.
However, our HCOOH abundance is higher than that of Sgr~B2 \citep{Ikeda01}, probably because the physical condition of Sgr~B2 is different from our model.
\cite{Pagani17} reported the actual observation of CH$_3$NH$_2$ towards Orion~KL to be 1$\times$10$^{16}$~cm$^{-2}$, corresponding to the fractional abundance of 1$\times$10$^{-9}$ assuming the hydrogen column density of 1.1$\times$10$^{25}$cm$^{-2}$ \citep{Hirota15}.
\cite{Halfen13} reported similar CH$_3$NH$_2$ abundance of 1.7$\times$10$^{-9}$ towards Sgr~B2.
Since our peak abundance of CH$_3$NH$_2$ is 7.6$\times$10$^{-6}$, we overestimated the abundance of CH$_3$NH$_2$ in the gas phase.
This disagreement may be due to (1) hydrogenation processes to HCN and CH$_2$NH are not as efficient as we assume, and/or (2) the actual source age is $\sim$10$^5$~years after the completion of the warm up and gas phase CH$_3$NH$_2$ have already destroyed in Orion~KL.
We will also assess the latter possibility in the subsequent section.

\subsection{Future of Glycine Surveys in the era of ALMA}
The first detection of glycine is one of the biggest challenges in the field of radio astronomy.
In this subsection, we will discuss the future detectability of glycine.

In \cite{Suzuki17}, we performed the chemical modeling study for high-mass star-forming regions, where we assumed that the temperature gradient inside hot cores would be represented by hot and warm temperature.
%
%
While the temperature in the inner part of the hot core was set to be 200~K, the temperature of outer region of the hot core was taken as a free parameter.
We looked for the best combination of parameters in our modeling to explain the observed abundance of COMs, changing the age of the core, the temperature of warm region, and the volume ratio of hot and warm region inside the hot core, using ``the degree of proximity (DoP)'' from \cite{Wakelam06} as a criteria to compare the modeling results with observed abundances.
%
Through the comparison, we suggested the age of G10.47+0.03 and NGC6334F were, respectively, 8$\times$10$^{5}$ and 6.5$\times$10$^{5}$~years after the birth of the star, where chemical compositions were altered via the gas phase reactions after the thermal evaporation.
Hence, these ages would be reasonable to predict the abundances of glycine or other COMs in G10.47+0.03 and NGC6334F, although \cite{Garrod13} stopped the chemical evolution just after the completion of the warm-up phase.
%
%
In this subsection, we will simulate the chemical evolution up to 1$\times$10$^6$~years to compare with \cite{Suzuki17}, fixing the temperature and the density once the warm-up phase is completed.

We will use Fast Model with the binding energy of 13000~K for glycine.
In Figure~\ref{fig:Glycine_Suzuki17} we plotted the time evolution of gas phase abundance of glycine with the peak temperature of 400~K, along with the gas phase abundances of CH$_2$NH, CH$_3$NH$_2$, NH$_2$OH, and CH$_3$COOH.
The temperature of 400~K corresponds to the environment of the closer region to the central star and susceptible to stellar radiation than  200~K region in \cite{Suzuki17}.
We note that the gas phase abundance of glycine is higher than that of Table~\ref{table:Garrod_vs_Model1}, since the longer timescale enabled gas phase reaction of ``NH$_2$OH$_2^+$ + CH$_3$COOH" to form protonated glycine, followed by the formation of glycine through the dissociative recombination with an electron.
We plotted the time evolutions of glycine, CH$_3$NH$_2$, CH$_2$NH, NH$_2$OH, and CH$_3$COOH in Figure~\ref{fig:Glycine_Suzuki17}. 
Figure~\ref{fig:Glycine_Suzuki17} implies that the abundances of glycine, CH$_3$NH$_2$, NH$_2$OH, and CH$_3$COOH will decrease as the passage of time due to the destruction processes by positive ions and radicals.
The gas phase abundance of glycine between 6.5$\times$10$^{5}$ and 8$\times$10$^{5}$~years was $\sim$10$^{-14}$, lowered by one orders of magnitude compared to its peak abundance.

Our modeling results suggests that the gas phase abundance of glycine will be lowered by almost one order of magnitude after 5$\times$10$^{5}$~years since the beginning of the warm-up.
Hence, the very early phase of the hot cores, where gas phase destruction processes are less efficient, would be the ideal target sources to observe gas phase glycine.
%
%
%
%
We show the time evolutions of glycine and its precursors, NH$_2$OH and CH$_3$COOH, in Figure~\ref{fig:Glycine_Suzuki17}.
Figure~\ref{fig:Glycine_Suzuki17} tells us that NH$_2$OH, and CH$_3$COOH can be used as the direct precursor to search for glycine rich sources since they contribute directly in chemical kinetics of glycine in the gas phase and also follows similar evolutionary paths.
We note that the number of sources of CH$_3$COOH is still limited.
NH$_2$OH was not detected in the ISM yet.
Future survey observations of these species in the hot components will be helpful to constrain our chemical modeling.
In addition, molecules built on grain surface, such as CH$_3$NH$_2$, can be used as the indicator of the early phase hot core and hence potentially glycine rich sources.
They decrease with time via the destruction processes in the gas phase.
By contrast, molecules originate in the gas phase reactions, such as CH$_2$NH  \citep{Suzuki16}, do not show this trend.
These species would also give us useful information regarding the age of the hot cores.
%

\subsection{Comparison with Chemical Composition of Comet 67P from ROSETTA Mission}
The comets are believed to be formed via the aggregation of interstellar dust particles within the protoplanetary disks, and the link between interstellar chemistry and the chemical compositions of comet is an interesting problem.
Rosetta mission reported many organic compounds in the coma of comet 67P/Churyumov-Gerasimenko.
In the past, \cite{Biver15} have found a good agreement in cometary species and those in warm molecular clouds.
This motivated us to compare our model predictions for glycine with the recent detection in comet 67P. 

The recent detection of glycine by ROSETTA mission in 67P/Churyumov-Gerasimenko \citep{Altwegg16} implies that glycine would possibly be inherited from the parental cloud of our Solar System. 
\cite{Altwegg16} reported the detections of glycine in several positions, with its highest abundance ratio to water, ``Glycine/H$_2$O'', of 1.6$\times10^{-3}$.
Our fast and slow warm-up models developed for the high- and low-mass stars would give us some insights regarding the pristine chemical compositions preserved in the comets.

Comet are believed to have been formed via the accretion of interstellar dust particles.
Considering that the large part of molecules would be frozen in the grain mantle rather than on the grain surface, the chemical composition of the grain mantle would be inherited by the cometary materials.
We plotted the abundance ratio in the grain mantle of ``Glycine/H$_2$O'' in Figure~\ref{fig:Glycine_Comet} with green lines.
While the evaporation rate for glycine is very low below 300~K due to its high binding energy, H$_2$O is quickly lost compared to glycine.
As a result, the ratio of ``Glycine/H$_2$O'' has increased and ``Glycine/H$_2$O $<$ 1.6$\times10^{-3}$'' is achieved when the temperature was less than 127~K.
At that moment, the fractional abundances of glycine and water in the solid phase are, respectively, 9.4$\times10^{-10}$ and 6.9$\times10^{-7}$.

%

\subsection{Special Case: Glycine Chemistry with Suprathermal H*}
\subsubsection{The Background of Suprathermal H*}
\cite{Munoz02} conducted laboratory experiments and confirmed 16 amino acids, including glycine, after UV irradiation on ice containing H$_2$O, CH$_3$OH, NH$_3$, CO and CO$_2$.
Their experiments suggested that the formation of glycine is possible without HCN.
However, the detailed formation mechanism was unclear at that time.
To deepen the understanding of formation path to glycine in this scheme, \cite{Holtom05} conducted experiments assuming cosmic ray inducing environment on the interstellar ice analogue containing CH$_3$NH$_2$ and CO$_2$, under the assumption that CH$_3$NH$_2$ would have formed from CH$_3$ and NH$_2$ radicals during the experiments by \cite{Munoz02}.
As a result, they reported the detection of glycine.
They claimed that this process starts from photo cleavages of C-H and N-H bonds in CH$_3$NH$_2$, and creates CH$_2$NH$_2$ and H.
Although a grain surface process of ``s-CO$_2$ + s-H $\longrightarrow$ s-COOH'' is not likely to occur due to its high activation barrier of 65.6~kJ~mol$^{-1}$ ($\sim$7900~K) \citep{Zhu01}, the nascent H atom by strong UV photons would have extra energy to overcome this barrier \citep{Holtom05,Lee09}.
The following reaction ``s-CH$_2$NH$_2$ + s-COOH $\longrightarrow$ s-NH$_2$CH$_2$COOH'' has no entrance barrier \citep{Holtom05}.
Furthermore, \cite{Lee09} reported the formation of glycine from CH$_3$NH$_2$ and CO$_2$ on interstellar ice analog films under UV irradiation at 7.3-10~eV, corresponding to the effective energy range for interstellar UV radiation. 
We note that \cite{Garrod13} also claimed the importance of the reaction of ``s-CH$_2$NH$_2$ + s-COOH $\longrightarrow$ s-NH$_2$CH$_2$COOH''.
However, he assumed that COOH was formed from the subtraction of H atom from HCOOH by radicals such as OH, rather then hydrogenation process to CO$_2$.

Two sources of UV photons would be available to form H* in actual star-forming regions.
The first possibility would be the black body emission from the central star.
However, we deduce that this process is important in only limited region through the following calculation with the assumption of spherical symmetric sphere.

It is known that Av is proportional to the hydrogen column density: N[H] (cm$^{-2}$) =2.21$\times$10$^{21}$ Av (mag) \citep{Guver09}.
According to the density profile based on the observations and theoretical studies of UCHI\hspace*{-1pt}I regions by \cite{Nomura04}, the hydrogen atom number densities (n$_H$) are not so different within 0.1~pc.
With a peak density of 1$\times$10$^7$~cm$^{-3}$, we are able to calculate Av by using the distance from the star ``r" as the following:

\begin{equation}
Av=1.4 \times 10^{4} \times r (pc),
\end{equation}

In NAUTILUS, the photodissociation rates are presented by the formula below with Av:

\begin{equation}
k = A\exp(-C Av)~s^{-1}.
\end{equation}

The coefficients A and C can be theoretically and/or experimentally obtained.
In this case, the photodissociation rate will decrease by a factor of $\sim$8.3$\times$10$^{-7}$ by every 0.01~pc.
Since the UV photons from the star would be available only in a limited region, we do not employ the star as the source of UV photons.

The other possibility for the source of UV photons is related to the cosmic rays. 
\cite{Prasad83} suggested that the cosmic rays are an important source of UV flux in dense regions with large Av, where interstellar UV photons cannot penetrate.
These mechanisms are included in kida.uva.2014.
For the modeling of H*, we simply assumed that H* is created by the dissociation process of any H-bearing species by UV field by the Prasad-Tarafdar mechanism instead of H.

For instance,

s-CH$_3$NH$_2$ + $h\nu$ $\longrightarrow$ s-H* + s-CH$_2$NH$_2$, or

s-NH$_3$ + $h\nu$ $\longrightarrow$ s-H* + s-NH$_2$

The activation energy for the grain surface reaction, ``s-CO$_2$ + s-H $\longrightarrow$ s-COOH'', is $\sim$7900~K (0.68~eV).
When UV photons with their energy of above 5~eV destroy CH$_3$NH$_2$, $\sim$4.3~eV is used to dissociate C-H bond, and extra energy is given to the dissociated hydrogen atoms that can sufficiently overcome the potential barrier associated with the ``s-CO$_2$ + s-H'' process.
Since all grain surface reactions in our model have lower activation barriers than ``s-CO$_2$ + s-H $\longrightarrow$ s-COOH'' (E$_A$$\sim$7900~K), H* would be able to overcome any activation barriers associated with grain surface reactions in our model.  
Therefore, we develop a new model referred as ``Fast + H*~Model", where H* can lead to any hydrogenation process by penetrating activation barriers  (E$_A$=0~K).
For instance, not only ``s-CO$_2$ + s-H* $\longrightarrow$ s-COOH'', but also other grain surface hydrogenation  processes such as ``s-CO + s-H* $\longrightarrow$ s-HCO'', are treated as barrierless reactions.
We presented a diagram of formation paths of H* and COOH in the above scenario in Figure~\ref{fig:suprathermal_hydrogen}.

For simplification, other chemical properties for H* were assumed to be the same as the usual H atoms.
It was assumed that H* acts as an usual H atom in the gas phase chemistry after the evaporation process.

The inclusion of H* changes the reaction rates of hydrogenation processes dramatically.
Recall that reaction rates on grain is proportional to $\kappa_{ij}$, as is given in the below formula using activation energy E$_A$:

\begin{equation}
\kappa_{ij}=\exp(-2(a/\hbar)(2\mu E_A)^{1/2}),
\end{equation}

where ``a'' is the rectangle barrier of thickness (1~\AA) and $\mu$ is the reduced mass.
$\kappa_{ij}$ is 2.7$\times$10$^{-16}$ for usual hydrogenation process to CO$_2$ (s-H + s-CO$_2$ $\longrightarrow$ s-COOH), with its activation energy of $\sim$7900~K.
With such a low reaction rate, usual hydrogenation process to CO$_2$ would be negligible even if CO$_2$ is abundant.
However, $\kappa_{ij}$ becomes unity for hydrogenation processes of H*, where E$_A$=0.
The conversion of CO$_2$ to COOH by the hydrogenation process of H* may strongly change the abundances of COOH on grains.

We also added the conversion processes of H* to usual hydrogen atoms:

s-H* + s-H* $\longrightarrow$ s-H$_2$,

s-H$_2$ + s-H* $\longrightarrow$ s-H$_2$ + s-H, and

s-H + s-H* $\longrightarrow$ s-H$_2$

Finally, we note that the we neglect the thermalization process of H* in our model.
According to the theoretical study of \cite{Andersson08}, the H atom following the UV photo-dissociation can move tens of Angstrom's before the thermalization due to interactions with H$_2$O ice.
Although the timescale of thermalization and chemical reaction of H* is not well known, it is thought that the timescale chemical reaction should be longer than this thermalization process \citep{Cuppen17}.
Therefore, our model may overestimate the effect of H* and the impact of thermalization process for the chemical evolution has to be deeply investigated in the following studies.

\subsubsection{Impacts of Suprathermal Hydrogen Atoms on the Abundance of Glycine}
We compared the formation rate of glycine in the same way as Fast Model.
In Slow Model, glycine is mainly built via the reaction of ``s-CH$_2$NH$_2$ + s-COOH'', same as in Fast Model.
The formation rate of glycine in the ``Fast + H*~Model" was highest in 4-5$\times$10$^{4}$ years since the beginning of the warm-up phase, with its value of between $\sim$10$^{-11}$ and 10$^{-12}$~cm$^{-3}$ s$^{-1}$ .
This formation rate was much higher than that of Fast Model if H* is available.

``Fast + H*~Model" showed that the abundance of glycine on grains was as high as $\sim$1$\times$10$^{-8}$.
If we excluded the reaction of ``s-H* + s-CO$_2$'', the simulated fractional abundance of glycine was almost the same as Fast Model.
This dramatical increase of glycine in ``Fast + H*~Model" is due to the efficient formation of COOH radical on the grains via ``s-H* + s-CO$_2$''.
Although we have included super thermal hydrogen atoms in the very simple way, it is apparent that H* would have a strong impact on the hot core chemistry.
The suprathermal hydrogen atoms may be potentially important for glycine chemistry and we suggest to investigate the efficiency of suprathermal hydrogen atoms via theoretical and/or experimental studies.
We will discuss the impact of the suprathermal hydrogen atoms on other COMs in the following subsection.

\subsubsection{Implication of the H* Chemistry on the Other COMs}
We showed the chemical composition for the selected COMs in Figure~\ref{fig:COM_abundances}, for the Fast Model and ``Fast + H*~Model".
The time of zero corresponds to the beginning of the warm-up phase.
The solid and dotted lines, respectively, represent the abundances in gas phase and on grains (surface and mantle).
%

Figure~\ref{fig:COM_abundances} suggests that the suprathermal hydrogen atoms do not change the abundance of simple molecules, such as CH$_4$, CO, CO$_2$, and H$_2$O, and complex organic molecules, such as CH$_3$OH, HCOOCH$_3$, CH$_3$OH, and CH$_3$NH$_2$.
However, with the reaction of ``s-CO$_2$ + s-H*'', the abundance of COOH radical has increased by more than a factor of 10.
As a result, the abundance of molecules containing carboxyl group, (i.e., HCOOH, CH$_3$COOH, CH$_3$CH$_2$COOH, NH$_2$CH$_2$COOH) has increased.
We note that the molecules discussed in \cite{Suzuki17}, CH$_3$OH, HCOOCH$_3$, CH$_3$OCH$_3$, (CH$_3$)$_2$CO, NH$_2$CHO, NH$_2$CHO, CH$_2$CHCN and CH$_2$NH, are not sensitive to the addition of H*, suggesting that ``Fast + H*~Model" is able to reproduce the observed abundance in the same manner as our previous model \citep{Suzuki17}.
We summarized the peak abundances of important species in Table~\ref{table:Model1_vs_Model3}, which showed the trend that the abundances of carboxyl groups were enhanced compared the other species.
The species with carboxyl group were mainly formed on grains via the radical-radical reactions with COOH radical.
We suggest that the observations of these species towards various star-forming regions would be a key to discuss the importance of suprathermal hydrogen atoms.

%

\subsubsection{Dependence of the Binding Energy of H*}
One of the essential parameters to determine the significance of H* would be the binding energy.
Lowering binding energy will allow H* atom to diffuse rapidly on grains, while the timescale to reside on grain surface will become short.
In the above section, we assumed that the binding energy of H* is the same with that of usual H atom, however, its extra kinetic energy would increase the opportunity for H* to move, making the binding energy much smaller than usual H atom.
Despite of this importance, the value of best binding energy of H* is not known.
It has to be carefully determined considering the frequency dependency of UV flux inside the cores. 
In this subsection, we show the effect of binding energy on the abundance of glycine.

In Figure~\ref{fig:abundance_glycine_HotH_binding_energy}, we showed the abundances of glycine in the gas phase and on grains (surface and mantle), using the solid and dotted lines, respectively.
In these modeling, we used the different binding energy of 650, 300, and 100~K, and its diffusion energy was given as 40$\%$ of its binding energy.
With the binding energy of 650~K, the peak abundance of glycine on grains is 2.0$\times$10$^{-8}$, while they are 1.3$\times$10$^{-8}$ and 6.5$\times$10$^{-9}$, respectively, with the binding energies of 300 and 100~K.
The peak gas phase abundance of glycine are, respectively, 5.5$\times$10$^{-9}$, 1.5$\times$10$^{-9}$, and 2.1$\times$10$^{-11}$, respectively, with the binding energies of 650, 300, and 100~K.
This result suggests that the smaller binding energy of H* will decrease the abundance of glycine on grains due to more efficient thermal evaporation than usual hydrogen atom.
The detailed studies regarding kinetics of H* would be required to further promote the knowledge of the formation of glycine and to predict its reliable abundance by chemical modeling studies.

\section{Conclusion}

The main results of this paper can be summarized as follows:

\begin{enumerate}
\item With the updated gas-grain chemical network and binding energies of the related species along with glycine, our model predicts a peak gas phase glycine abundance of the order of 10$^{-14}$ which is almost a factor 100 lower than \cite{Garrod13}. 
The non detection of glycine towards Sgr B2(N) and other high mass star forming regions \citep{Kuan03,Snyder05} could then be explained by non complete desorption of this species in these sources.

\item We investigated the importance of newly suggested formation processes of glycine.
We found that on the grains, glycine is mostly formed by the reaction s-CH$_2$NH$_2$ + s-COOH as previously suggested by Garrod (2013) with his fast warm-up model. 
However, with glycine binding energy of 13000 K, glycine is quickly destroyed on the grains by radicals and the gas phase reactions play a major role.
The formation process of glycine and its precursors strongly depend on its binding energy that should be explored in detail in the future work.
In addition, chemistry of NH$_2$OH should be investigated to understand the validity of gas phase formation of glycine.

\item Once glycine is thermally evaporated, it is destroyed in the gas phase by positive ions and radicals.
Therefore, the hot cores in the early evolutionary phase would be preferable targets for future glycine surveys.
%

\item We developed a very simple model including suprathermal hydrogen atoms, which is expected to be formed via photodissociation process of UV photons and/or cosmic rays.
We found that these suprathermal hydrogen atoms can increase the abundance of COOH radical by more than a factor of ten, penetrating activation barrier associated with the reaction of ``s-H + s-CO$_2$" on grain surface.
Although COOH radical was believed to be formed via the destruction of HCOOH, we suggest that the hydrogenation of CO$_2$ is the dominant source for COOH radical.

\item The addition of suprathermal hydrogen atoms has strong impact on the abundance of molecules containing the carboxyl group.
Our model predicted that not only glycine, but also other molecules with carboxyl group, such as HCOOH, CH$_3$COOH, and CH$_3$CH$_2$COOH would increase with the reactions of suprathermal hydrogens.
Astronomical observations of these species would be keys to discuss the effectiveness of the reactions with suprathermal hydrogens.
\end{enumerate}

\acknowledgments
We thank to Dr.Hideko Nomura, and staffs in Bordeaux University for fruitful discussions.
This study was supported by the Astrobiology Program of National Institutes of Natural Sciences (NINS) and by the JSPS Kakenhi Grant  Numbers 15H03646 and 14J03618.
Valentine Wakelam acknowledge the European Research Council (3DICE, grant agreement 336474) and the CNRS program ``Physique et Chimie du Milieu Interstellaire'' (PCMI) co-funded by the Centre National d'Etudes Spatiales (CNES).  
Liton Majumdar acknowledges support from the NASA postdoctoral program. 
A portion of this research was carried out at the Jet Propulsion Laboratory, California Institute of Technology, under a
contract with the National Aeronautics and Space Administration.






\appendix

\clearpage
\clearpage

\clearpage

\begin{figure}
 \begin{tabular}{ll}
\includegraphics[scale=.4]{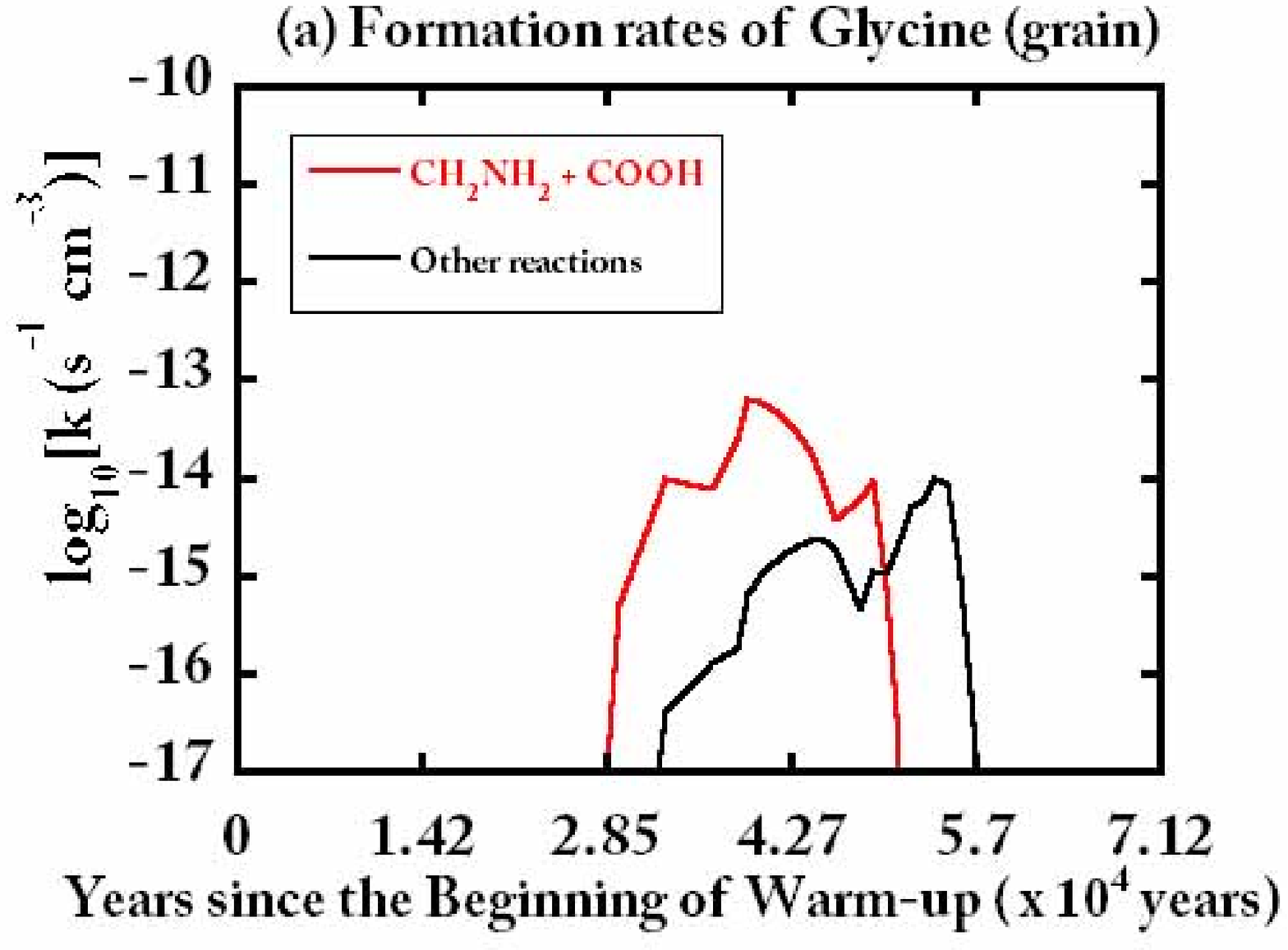}&
\includegraphics[scale=.4]{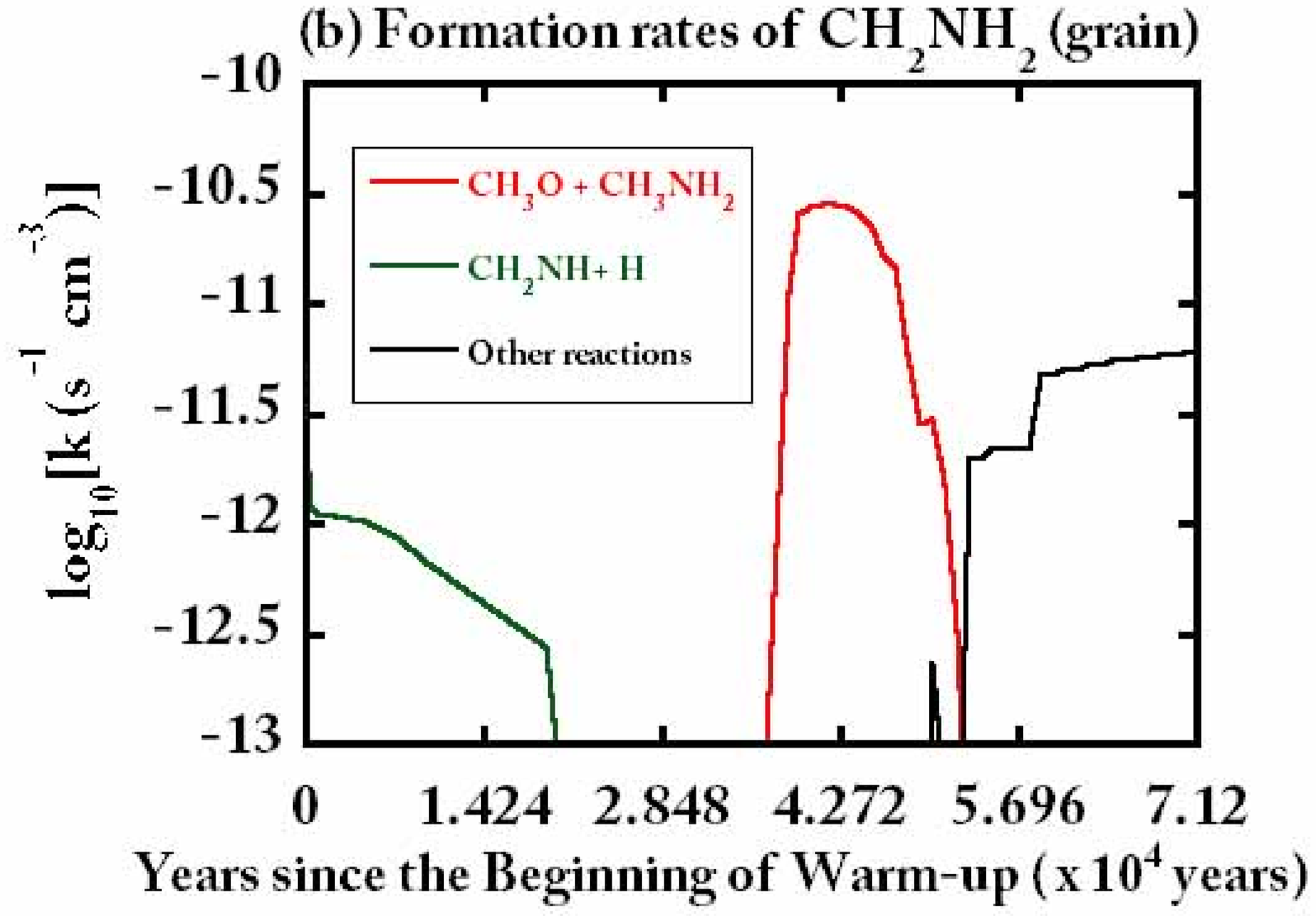}\\
\includegraphics[scale=.4]{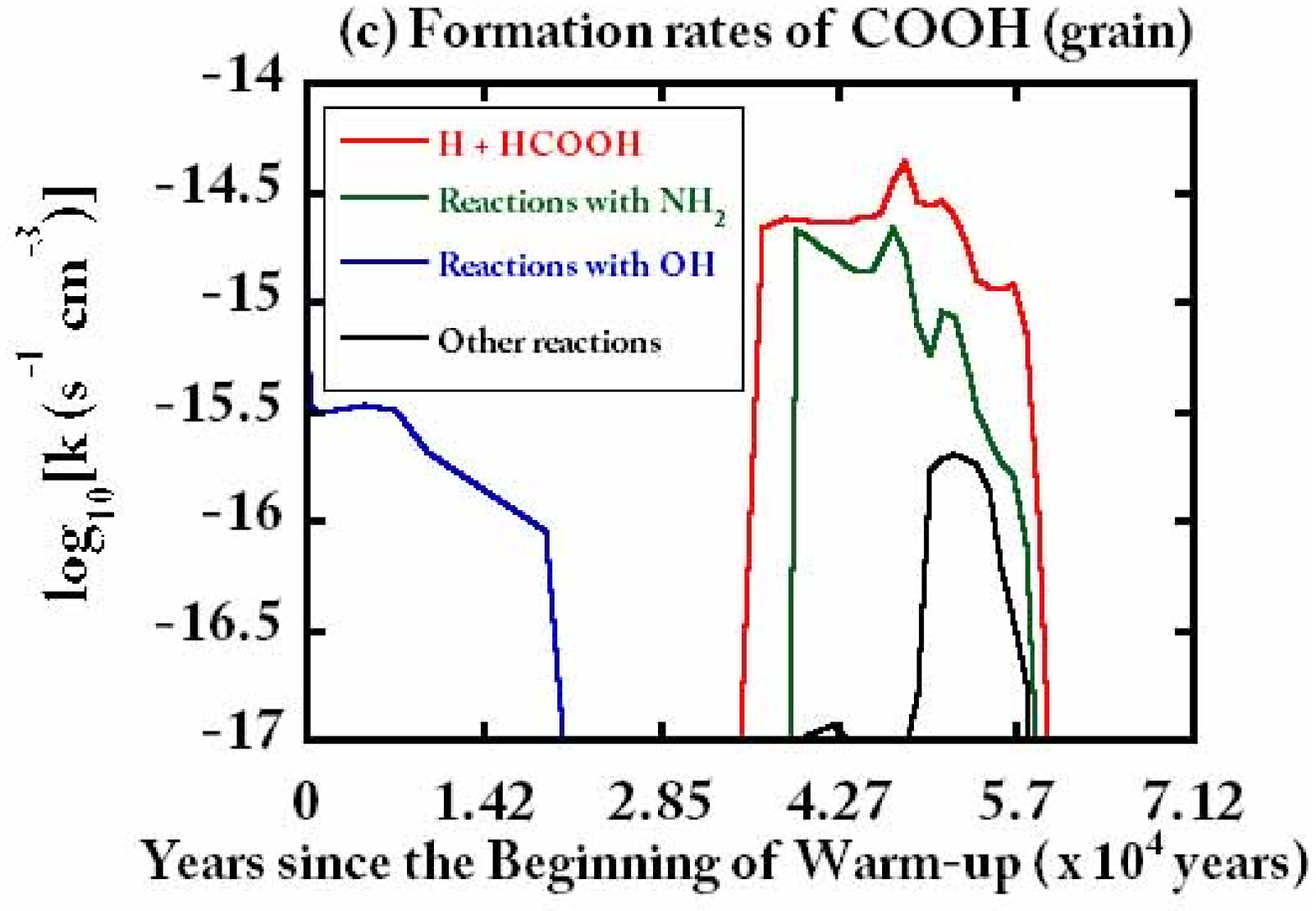}&\\
  \end{tabular}
\caption{
The formation rates (cm$^{-3}$ s$^{-1}$) of grain surface reactions for glycine and its precursors were compared using Fast Model.
(a) The formation rates for glycine on the grain surface were compared.
The red line are corresponding to ``s-CH$_2$NH$_2$ + s-COOH'' process, while the black line represent the sum of reaction rates for the other processes.
(b) The formation rates for CH$_2$NH$_2$ radical on the grain surface were compared.
The red and green lines are corresponding to ``s-CH$_3$O + s-CH$_2$NH$_2$  $\longrightarrow$ s-CH$_3$OH + s-CH$_3$NH$_2$'' and ``s-CH$_2$NH$_2$ + s-H  $\longrightarrow$ s-CH$_2$NH$_2$'' process, while the black lines represent the sum of reaction rates for the other processes.
(c) The formation rates for COOH radical on the grain surface were compared.
The red lines are corresponding to ``s-H + s-HCOOH  $\longrightarrow$ s-H$_2$ + s-COOH'' process.
The green and blue lines were, respectively, represents the sum of the formation rates of COOH through hydrogen subtraction processes from HCOOH, CH$_2$(OH)COOH, and CH$_3$OCOOH by NH$_2$ and OH radicals.
For instance, these processes are ``HCOOH + NH$_2$ $\longrightarrow$ NH$_3$ + COOH'' or ``CH$_2$(OH)COOH + NH$_2$ $\longrightarrow$ NH$_3$ + H$_2$CO + COOH'', and ``CH$_3$OCOOH + NH$_2$ $\longrightarrow$ NH$_3$ + H$_2$CO + COOH''.
It was assumed that they are the only products for the destruction process of HCOOH, CH$_2$(OH)COOH, and CH$_3$OCOOH.
The black lines represent the sum of reaction rates for the other processes.
\label{fig:Formation_Rate1_fast}
}
\end{figure}
\clearpage
\addtocounter{figure}{-1}
\begin{figure}
 \begin{tabular}{ll}
\includegraphics[scale=.4]{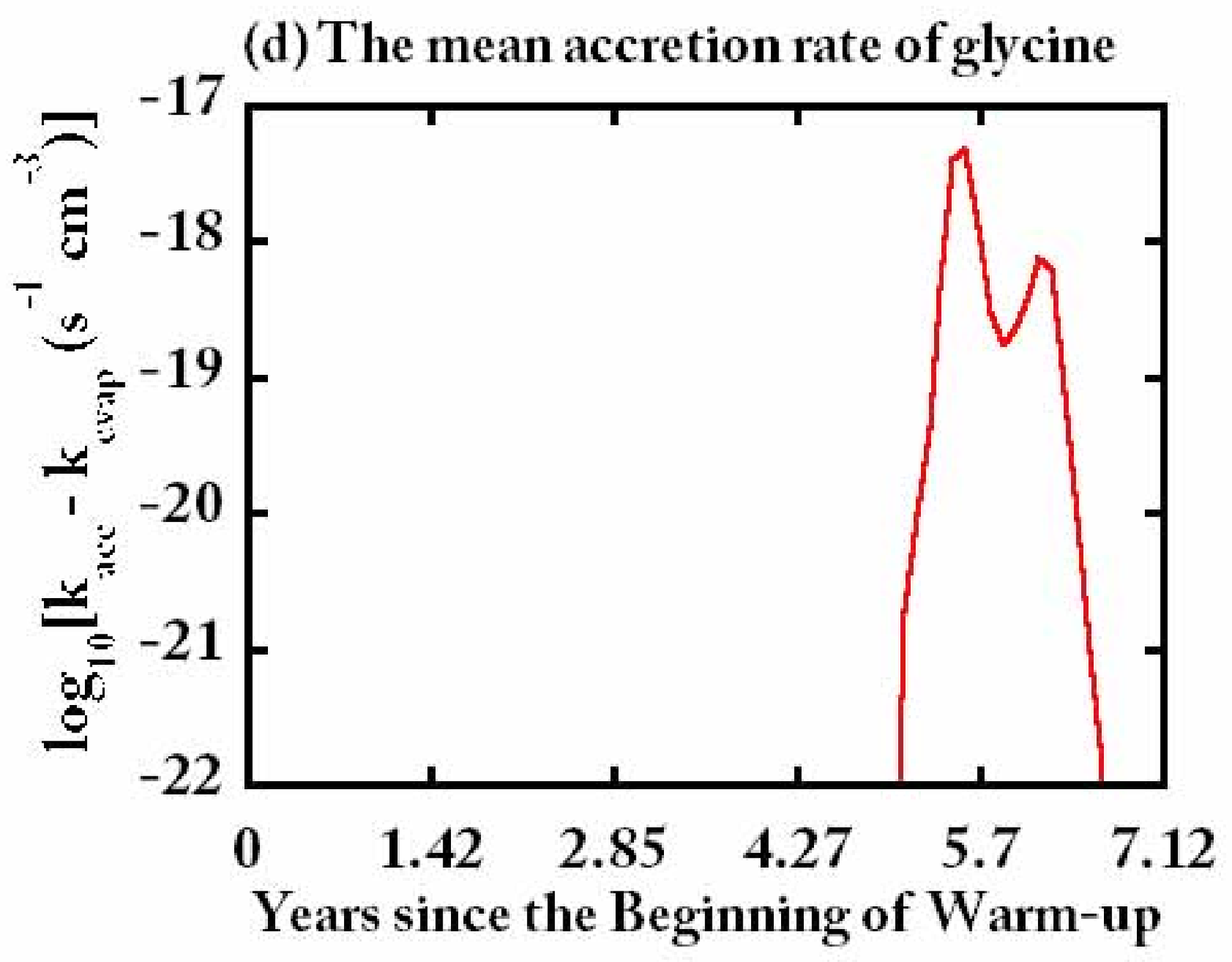}&
\includegraphics[scale=.4]{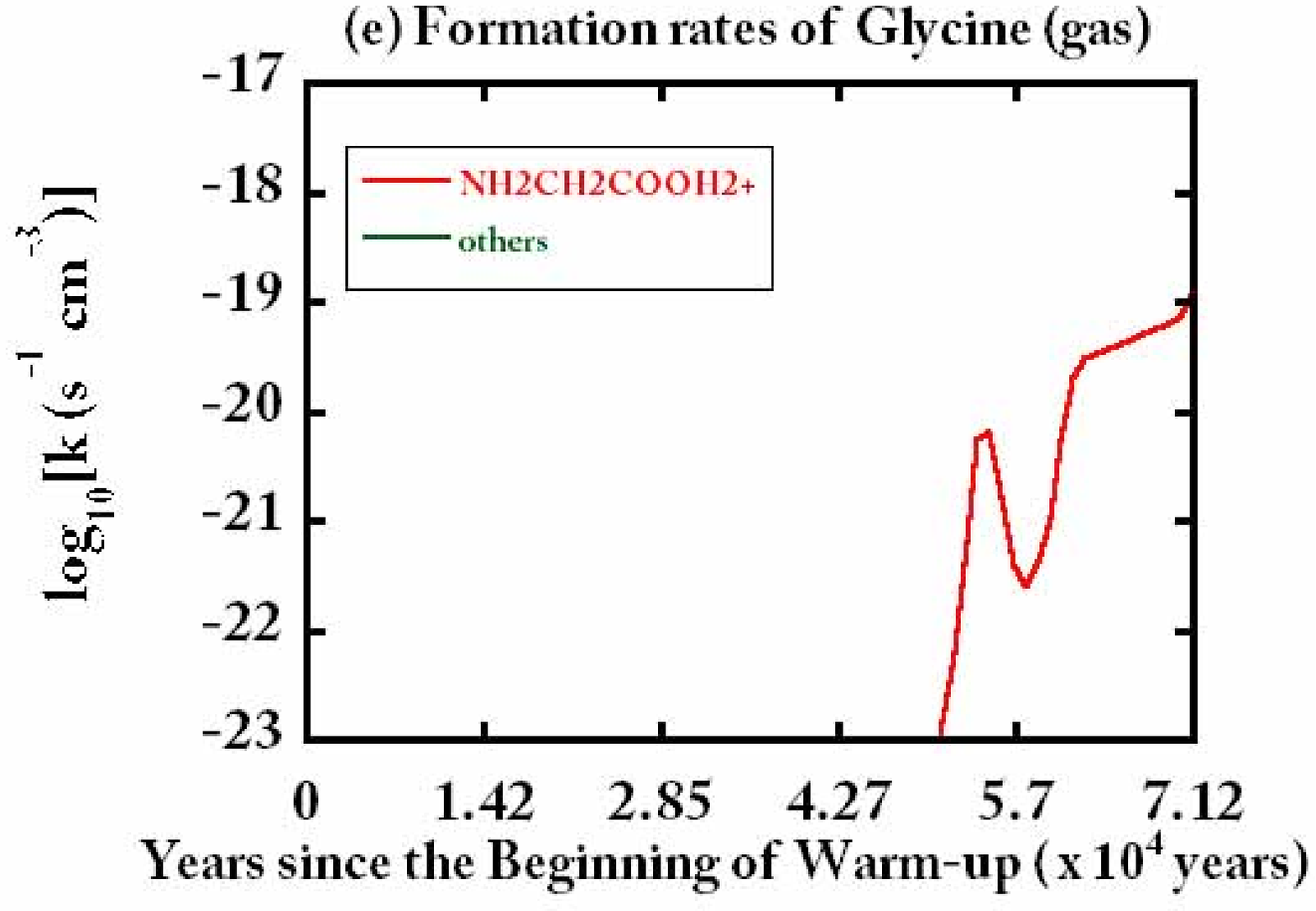}\\
\includegraphics[scale=.4]{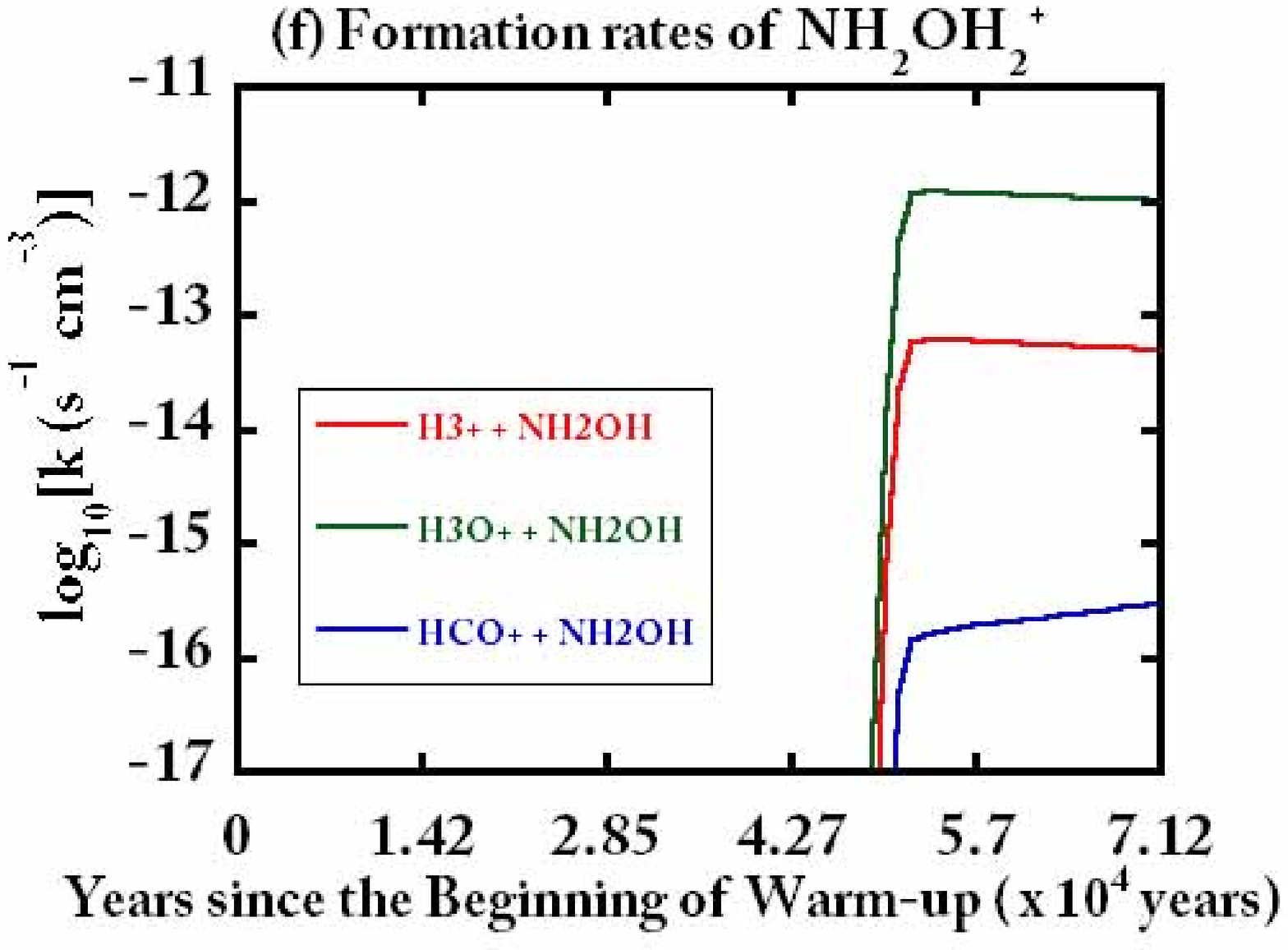}&
\includegraphics[scale=.4]{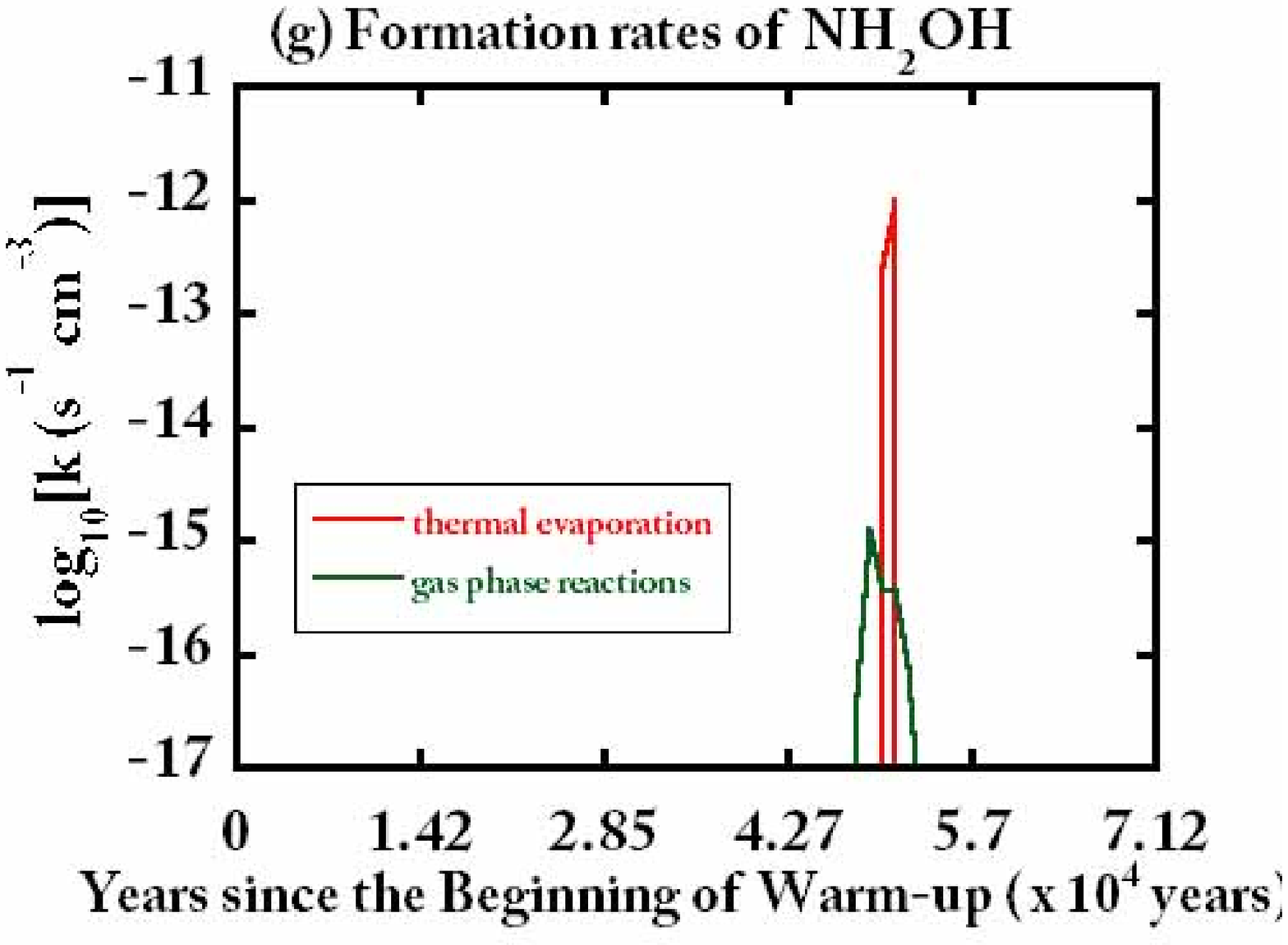}\\
\includegraphics[scale=.4]{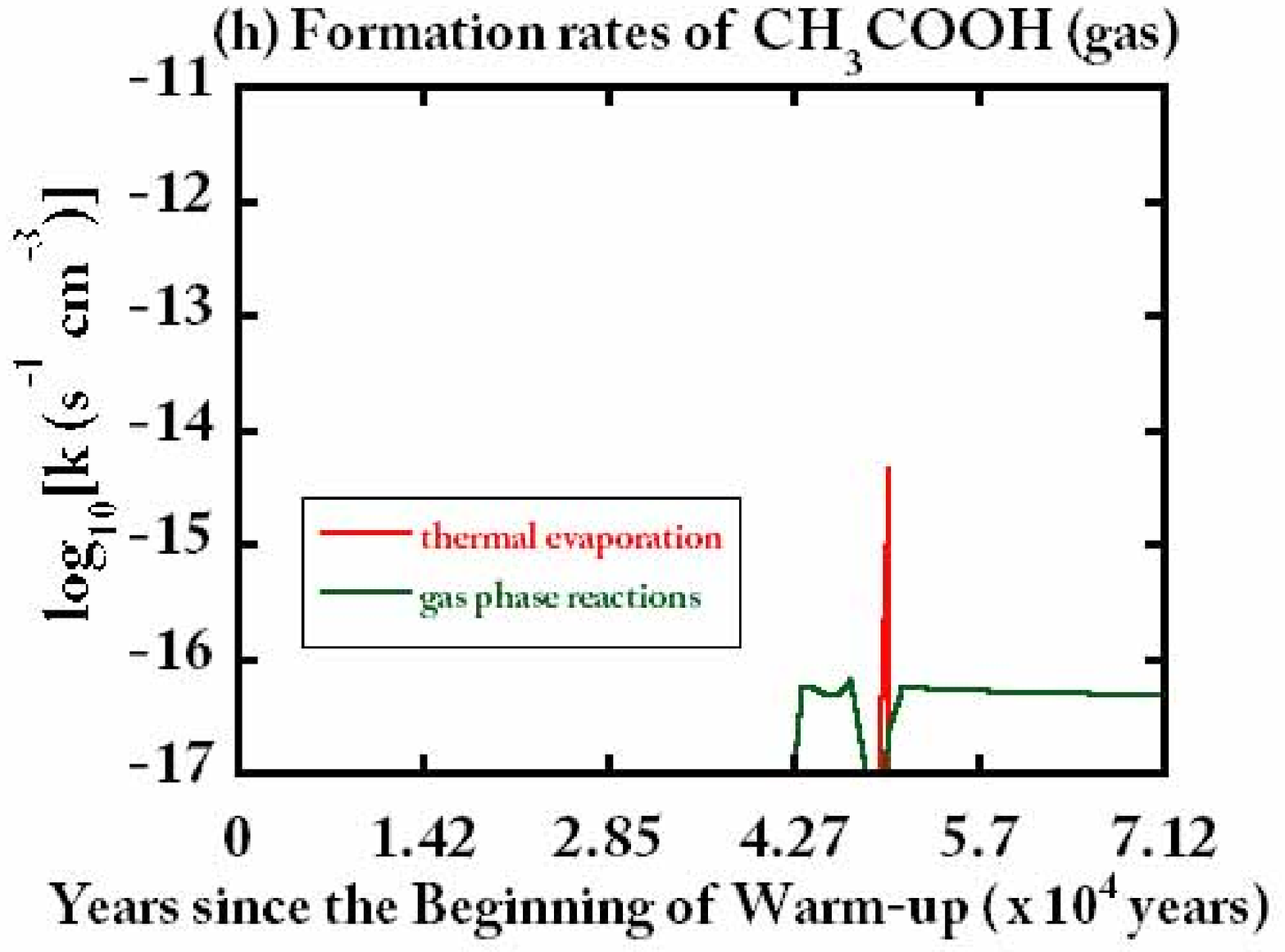}&\\
  \end{tabular}
\caption{
(continued) The formation rates (cm$^{-3}$ s$^{-1}$) of grain surface reactions for glycine and its precursors were compared using Fast Model.
(d) The subtraction of the accretion rate of gas phase glycine from the evaporation rate of glycine on grains was plotted.
(e) The formation rates for gas phase glycine were compared.
The red line represents the formation rates of glycine via the gas phase reaction of ``NH$_2$CH$_2$COOH$_2^+$ + e$^-$", while the black line represent the sum of the other reactions.
(f) The formation rates of NH$_2$OH$_2^+$ were compared.
The red, green, and the blue lines, respectively, represent the reaction of NH$_2$OH with H$_3^+$, H$_3$O$^+$, and HCO$^+$.
(g) The formation rates of NH$_2$OH were compared.
The red line represents the subtraction of accretion rates of NH$_2$OH from that of evaporation rate, while the green line represents the sum of the formation rates of NH$_2$OH through the gas phase reactions.
(h) The formation rates of CH$_3$COOH were compared.
The red line represents the subtraction of accretion rates of CH$_3$COOH from that of evaporation rate, while the green line represents the sum of the formation rates of CH$_3$COOH through the gas phase reactions.
\label{fig:Formation_Rate1_fast}
}
\end{figure}
\clearpage

\begin{figure}
 \begin{tabular}{ll}
\includegraphics[scale=.4]{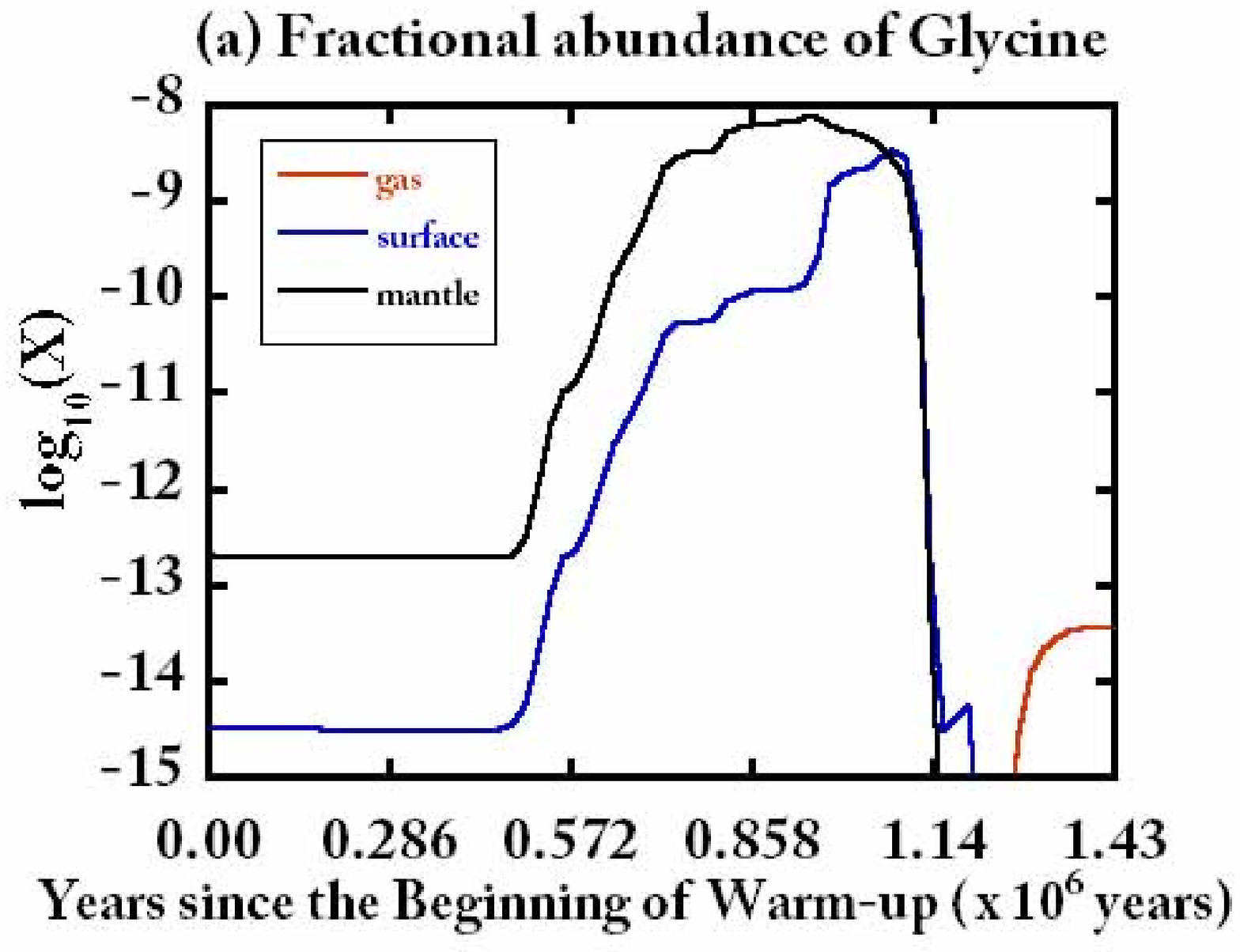}&
\includegraphics[scale=.4]{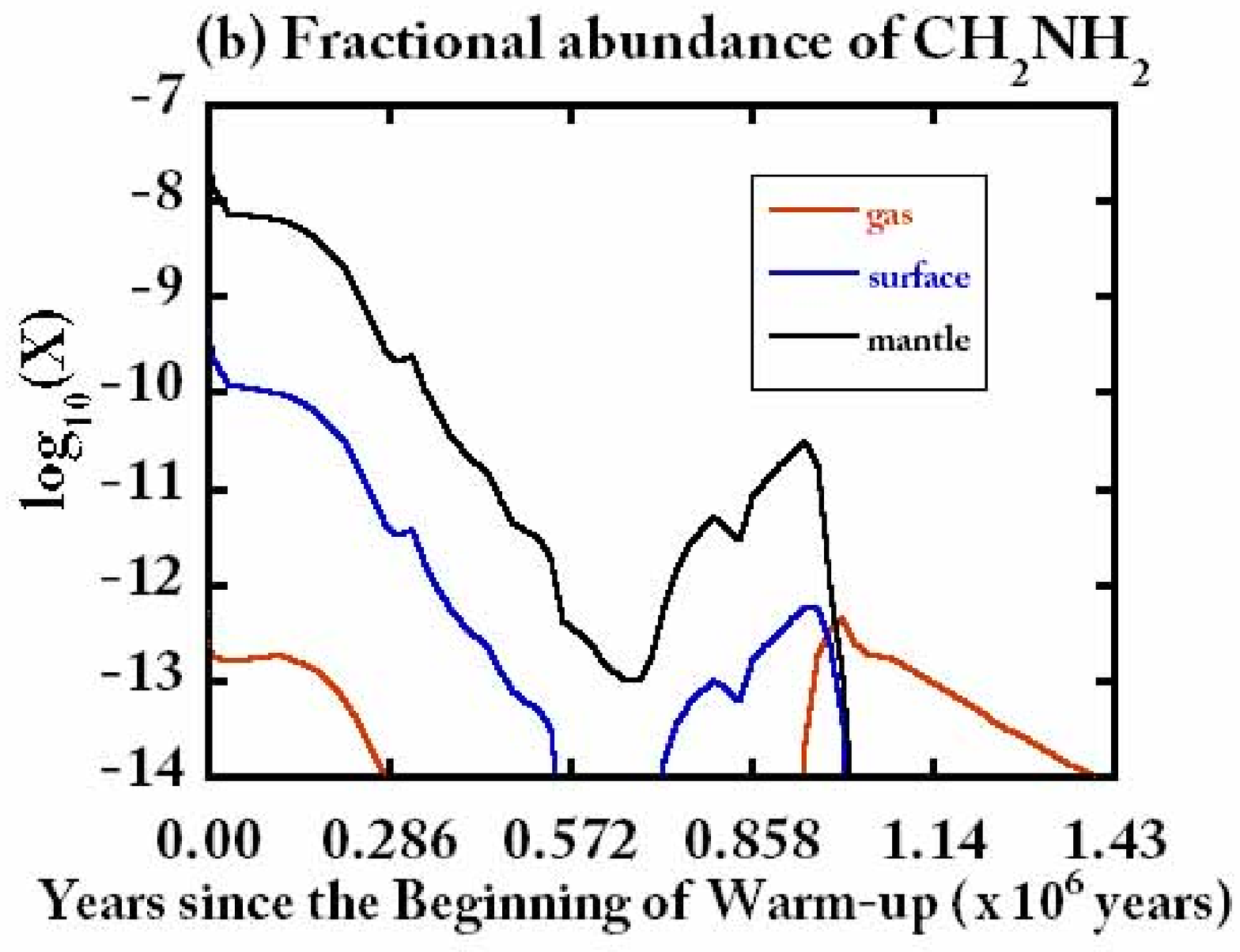}\\
\includegraphics[scale=.4]{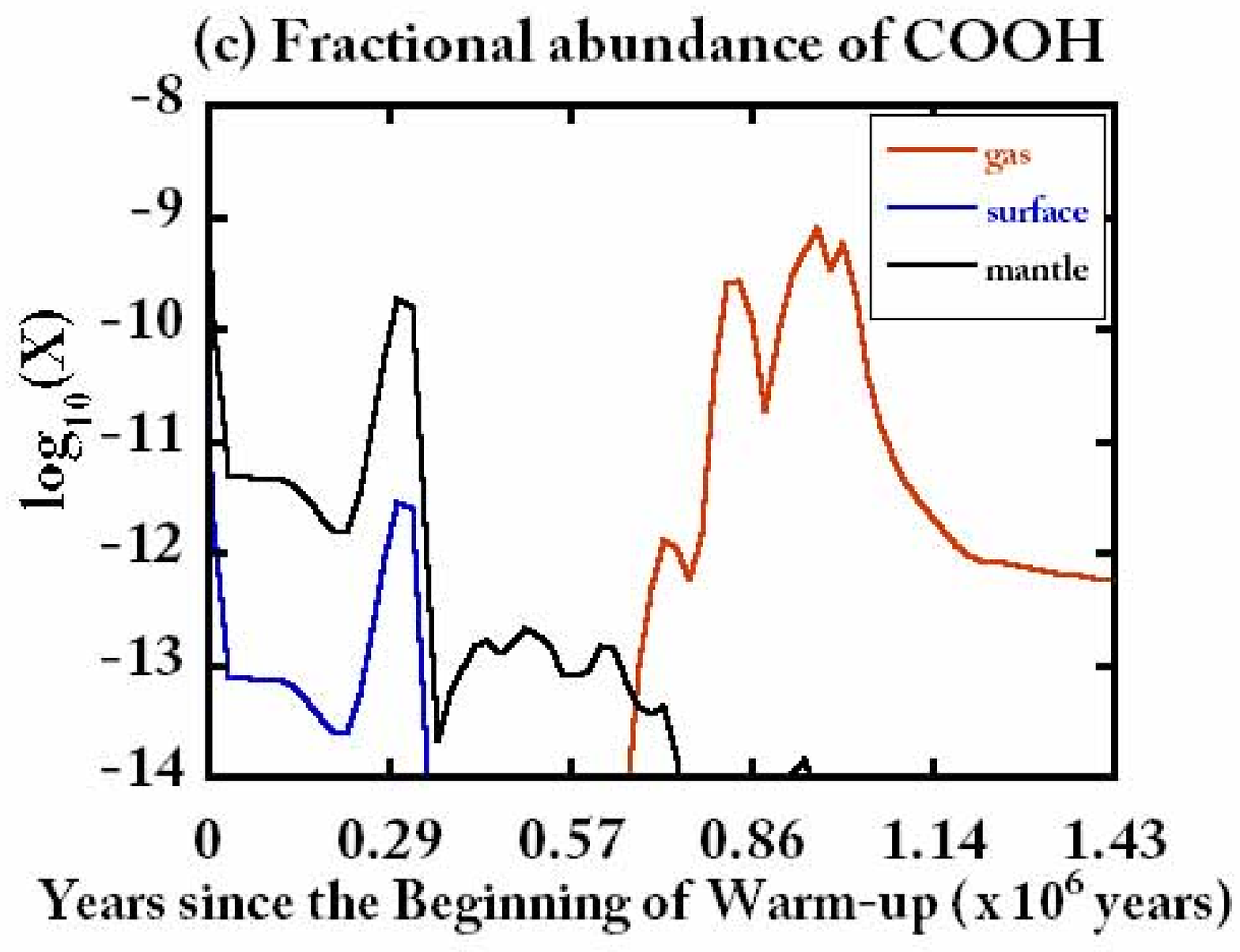}&
\includegraphics[scale=.4]{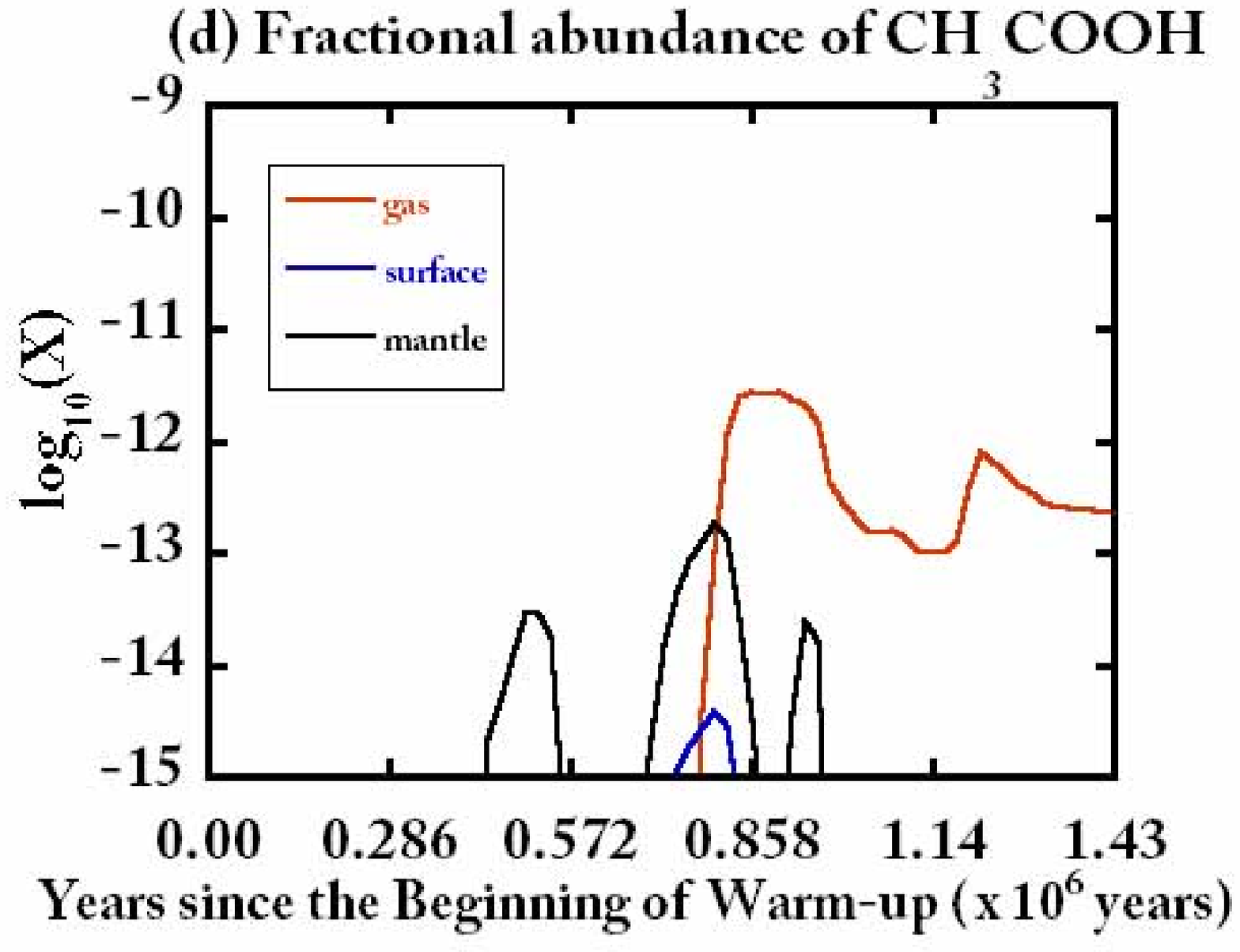}\\
\includegraphics[scale=.4]{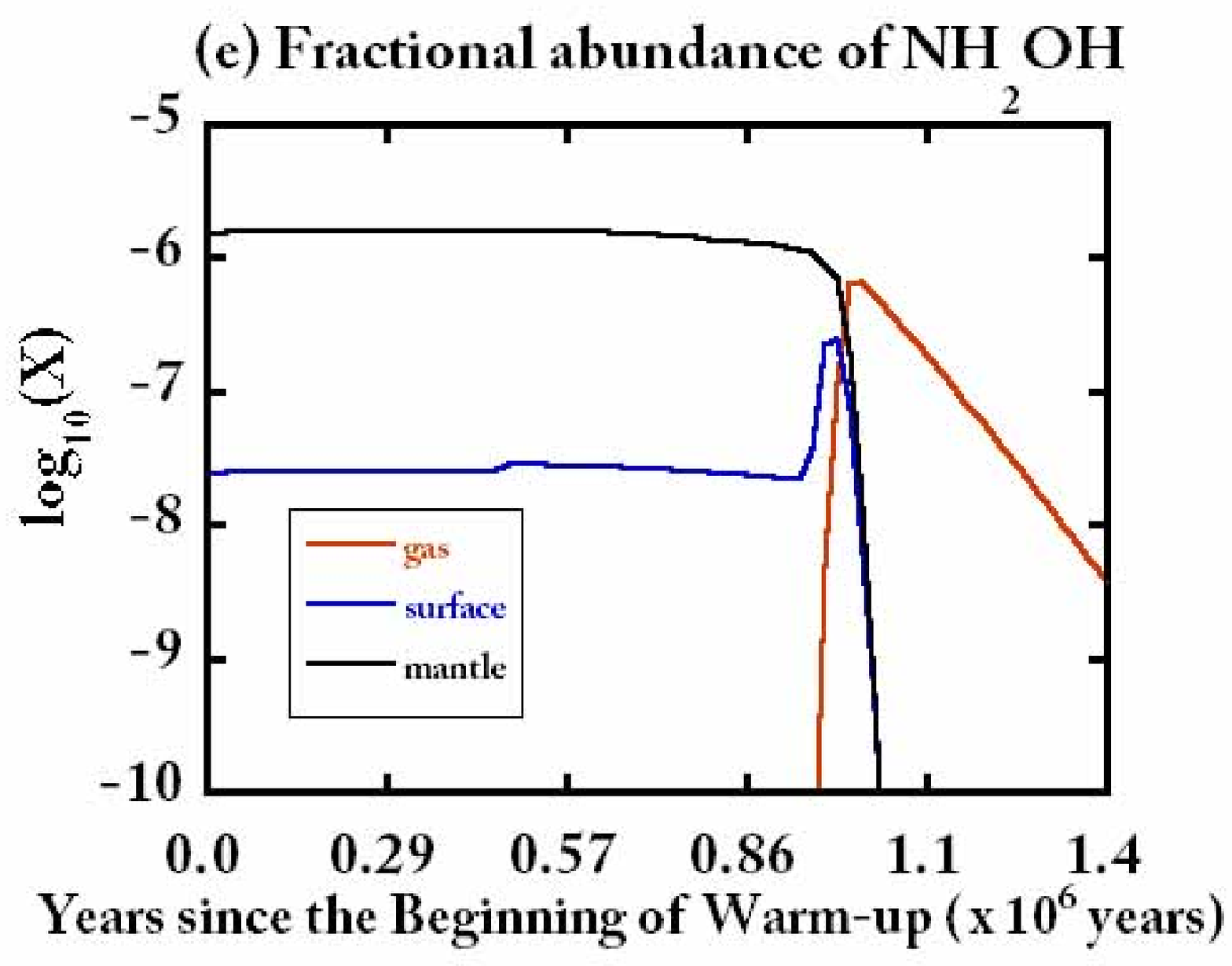}&\\
  \end{tabular}
\caption{
The simulated abundances for the Slow Model were shown in the same way as Figure~\ref{fig:Glycine_Abundances_fast}.
\label{fig:Glycine_Abundances_slow}
}
\end{figure}
\clearpage

\begin{figure}
 \begin{tabular}{ll}
\includegraphics[scale=.4]{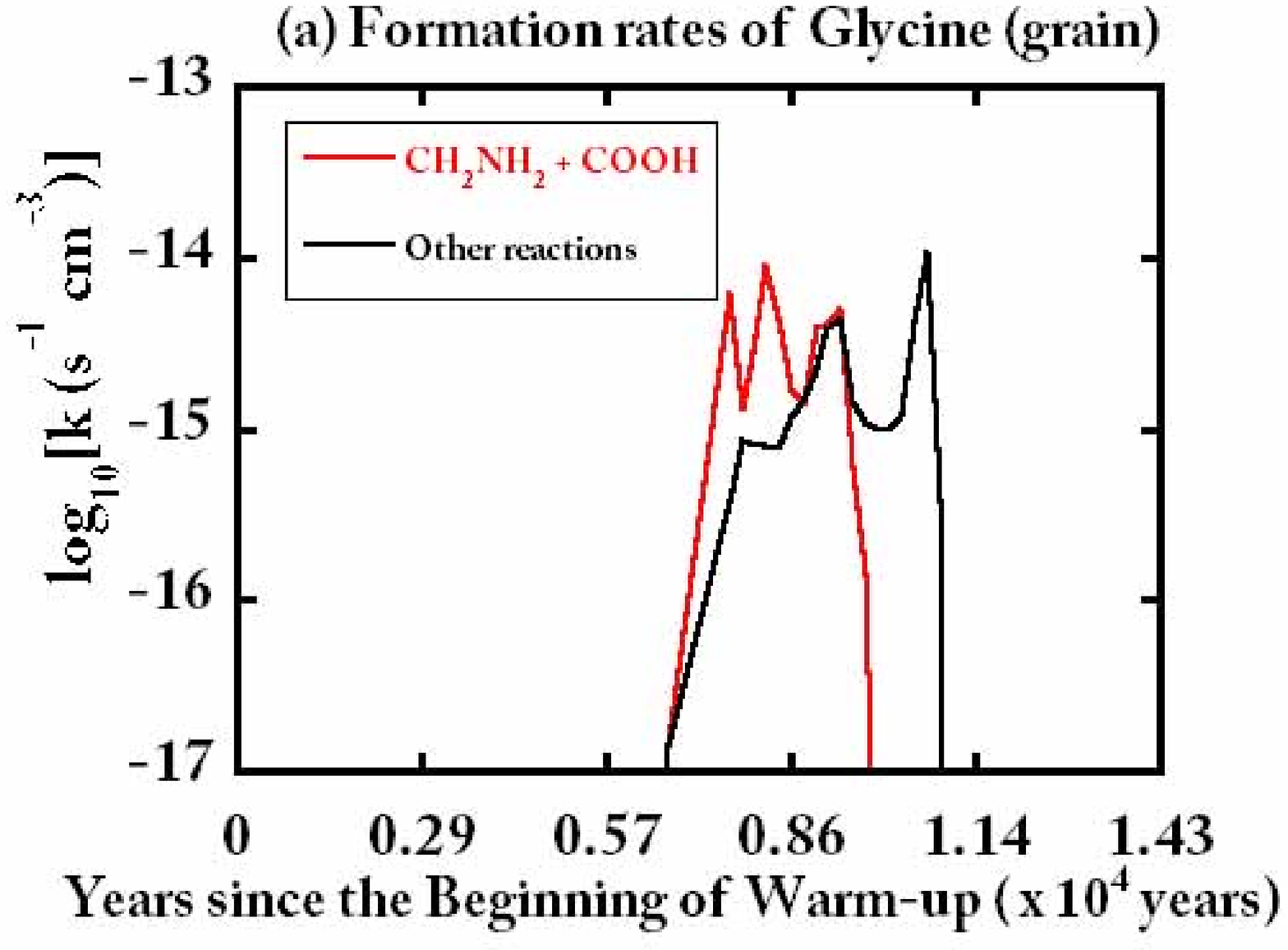}&
\includegraphics[scale=.4]{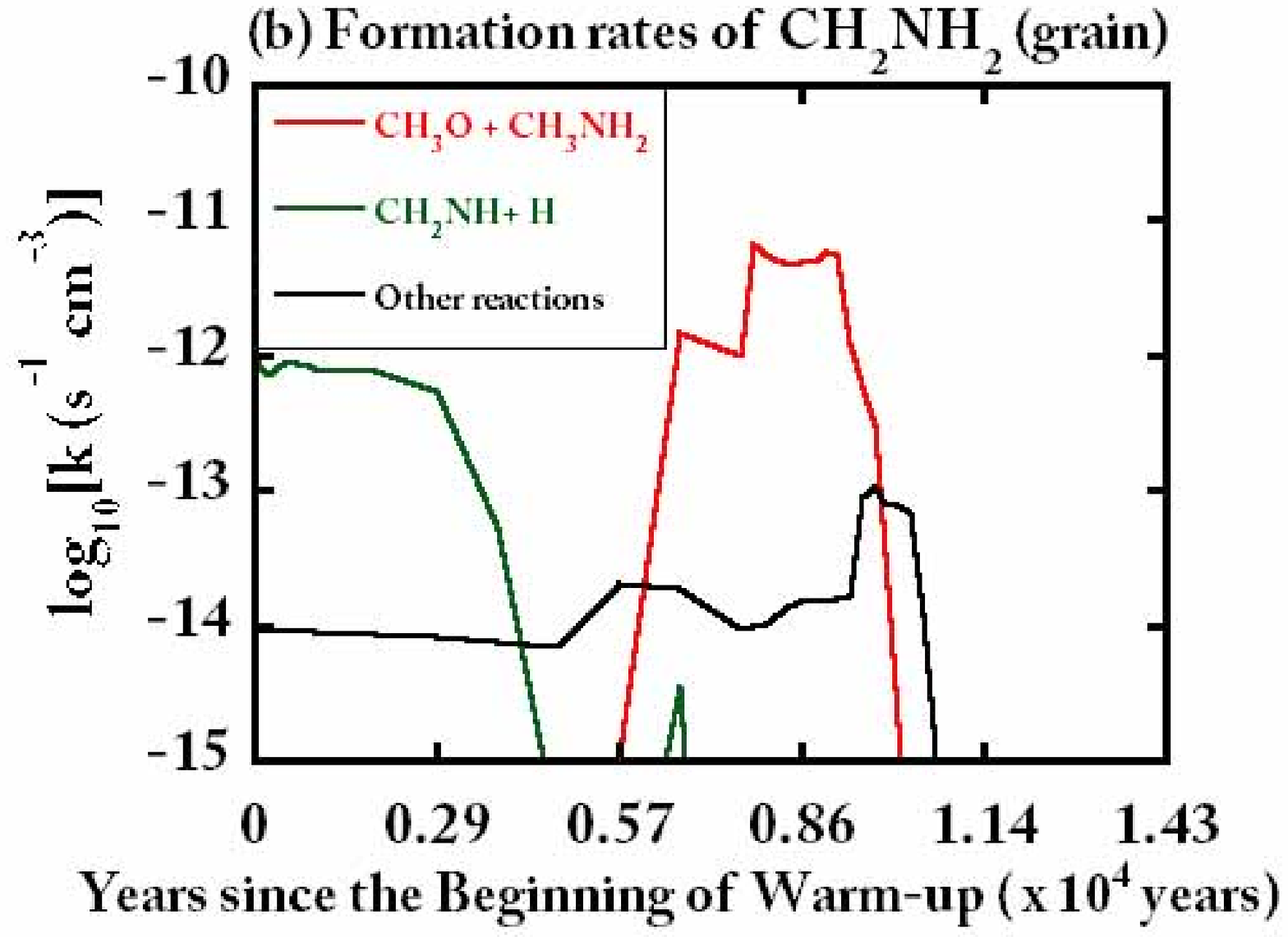}\\
\includegraphics[scale=.4]{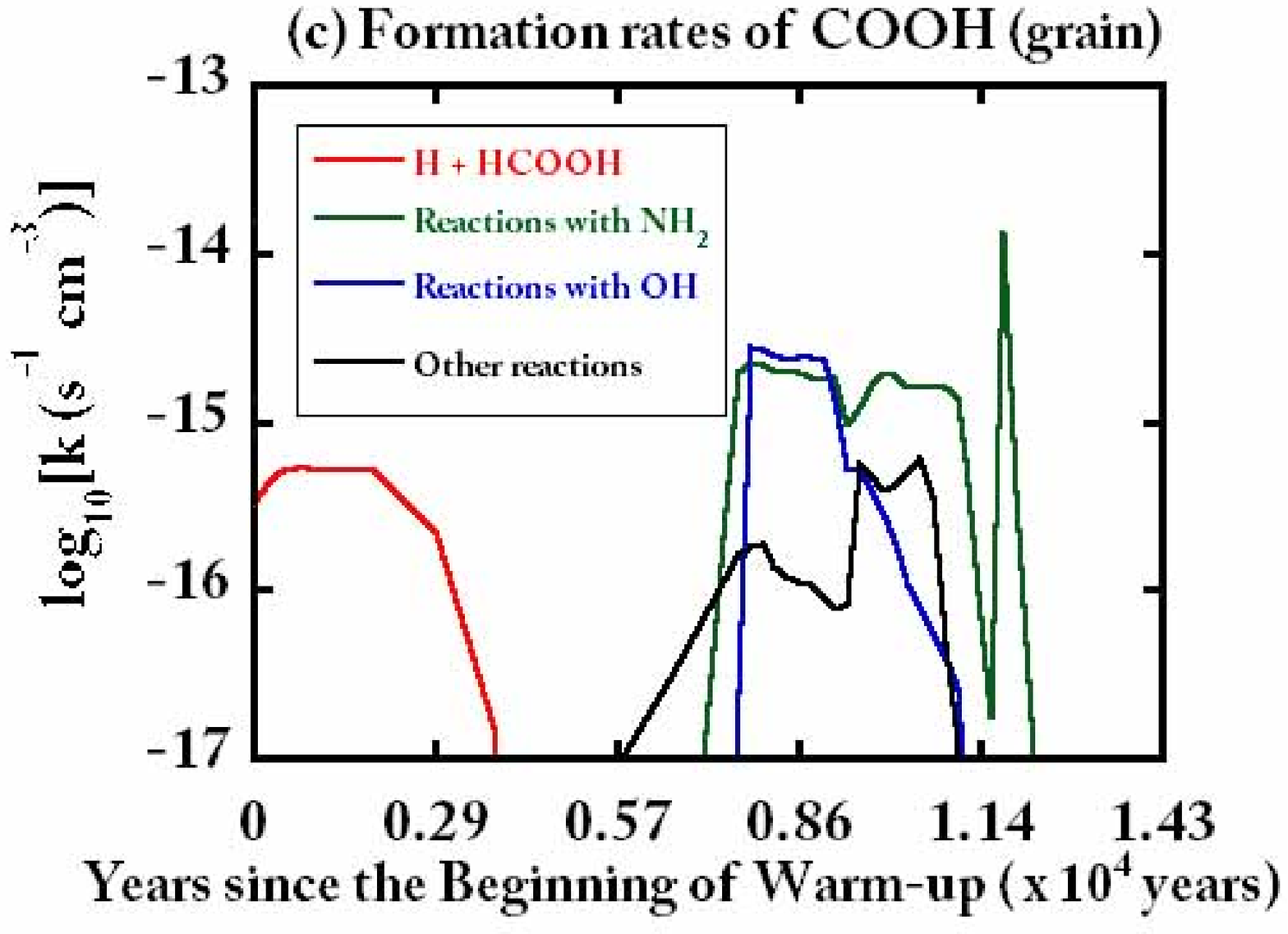}&
\includegraphics[scale=.4]{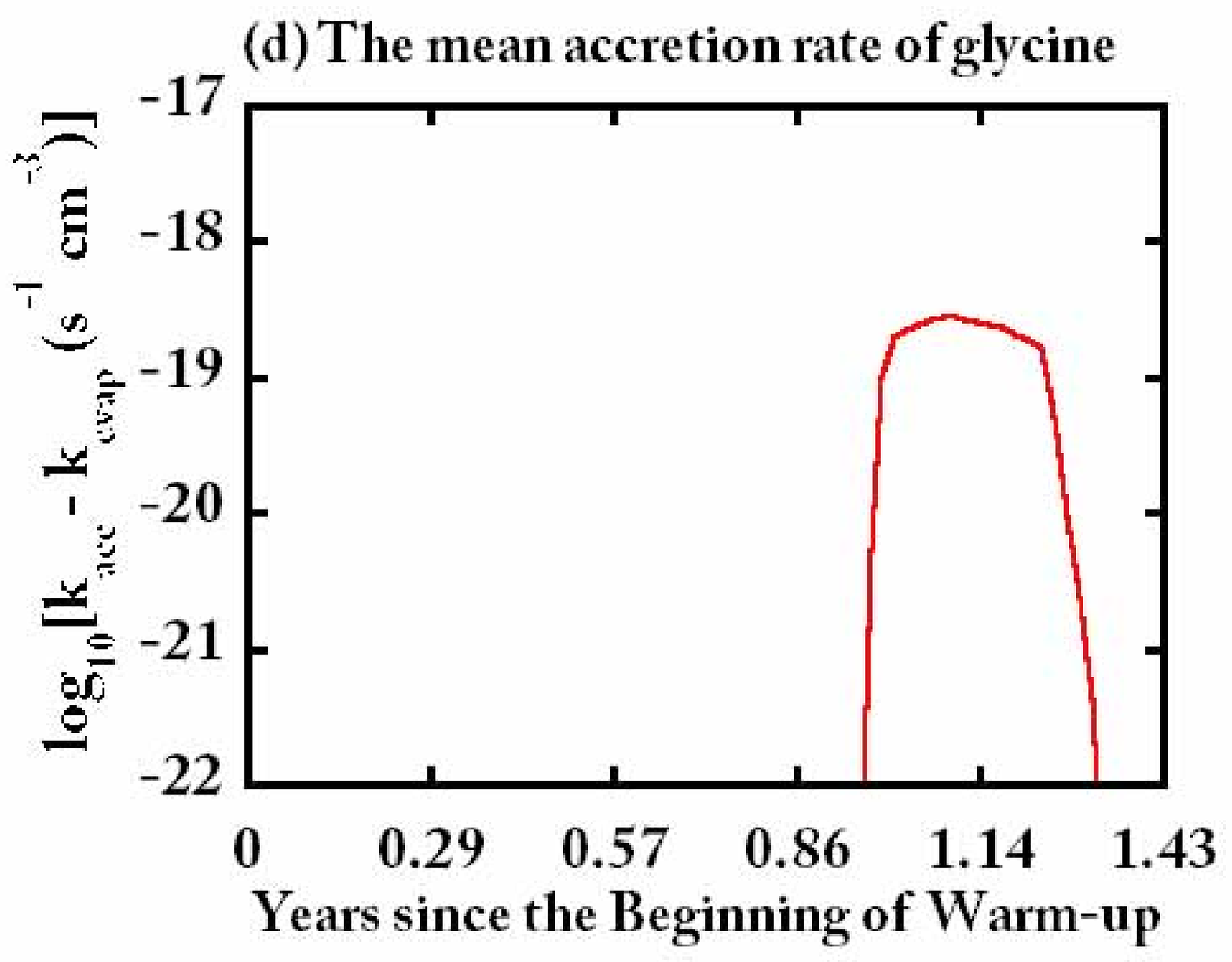}\\
\includegraphics[scale=.4]{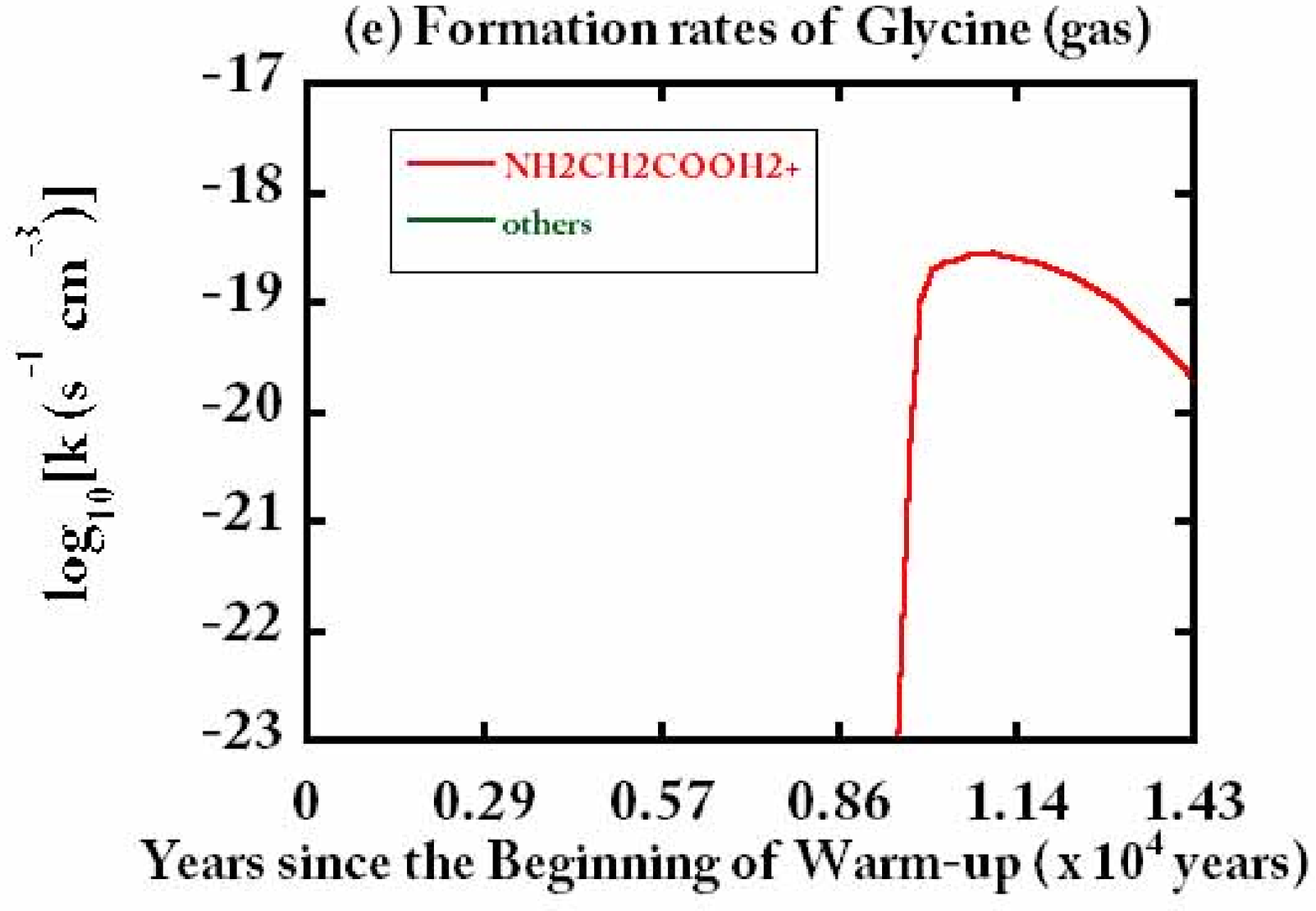}&
\includegraphics[scale=.4]{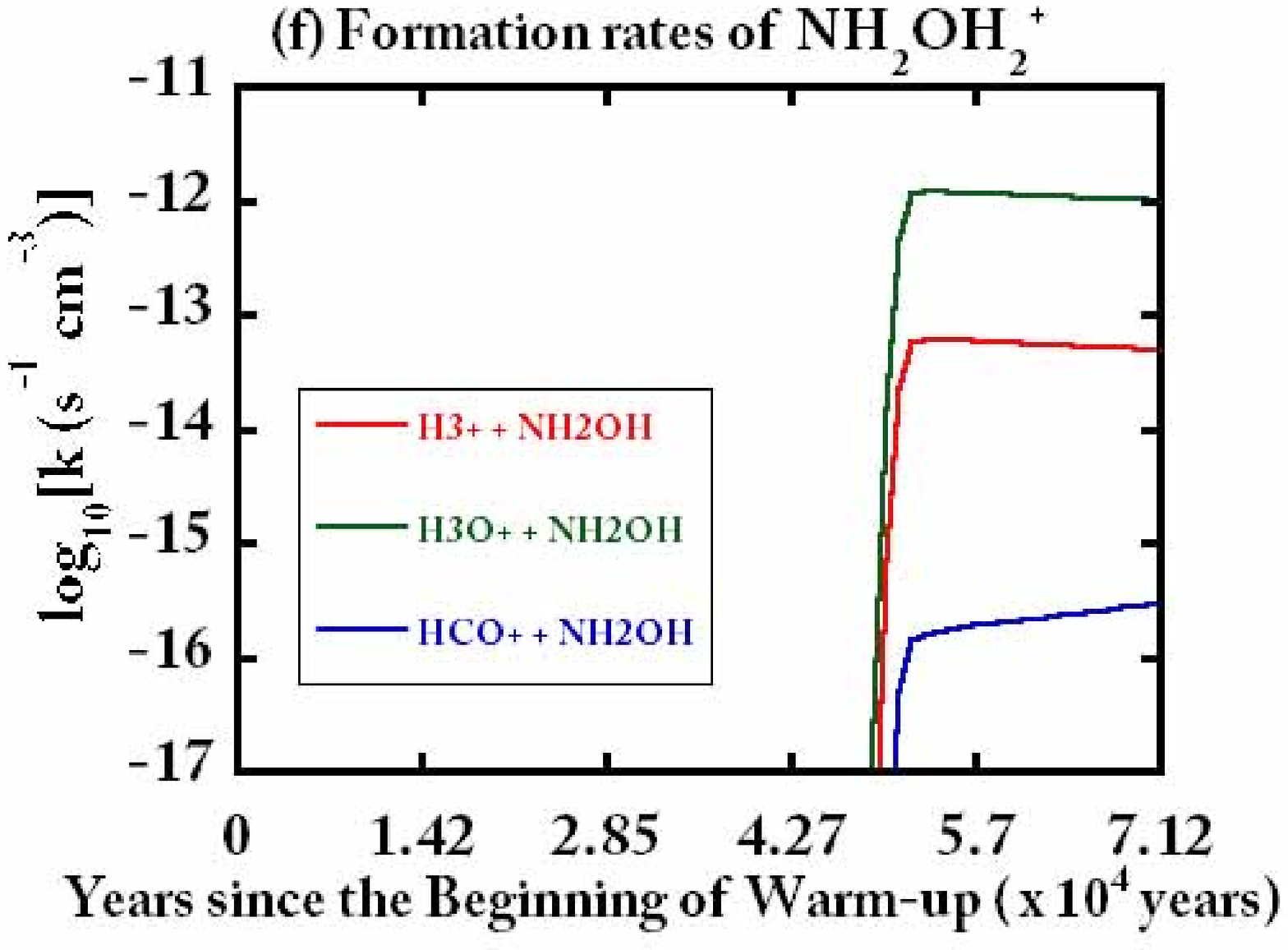}\\
\includegraphics[scale=.4]{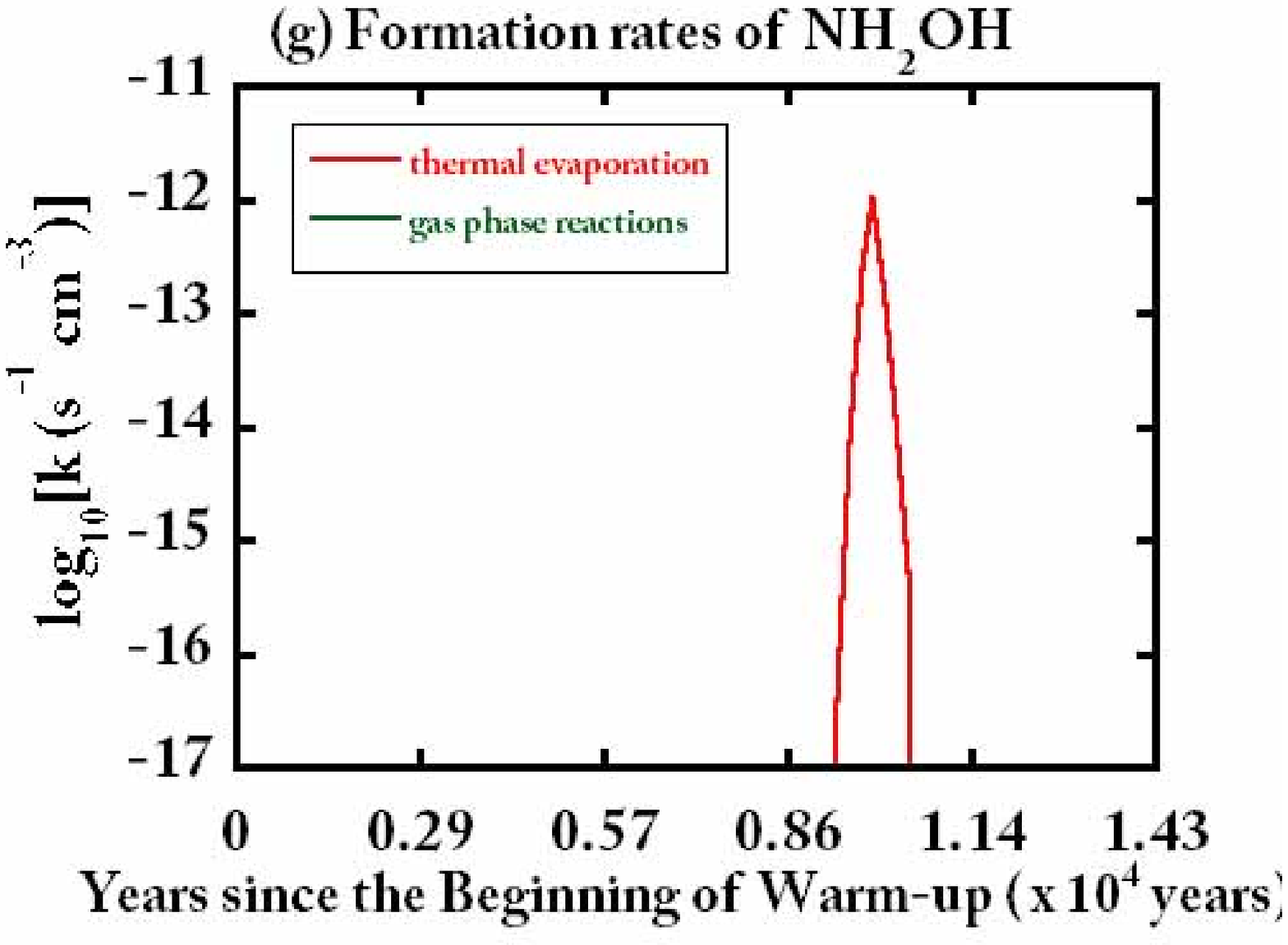}&
\includegraphics[scale=.4]{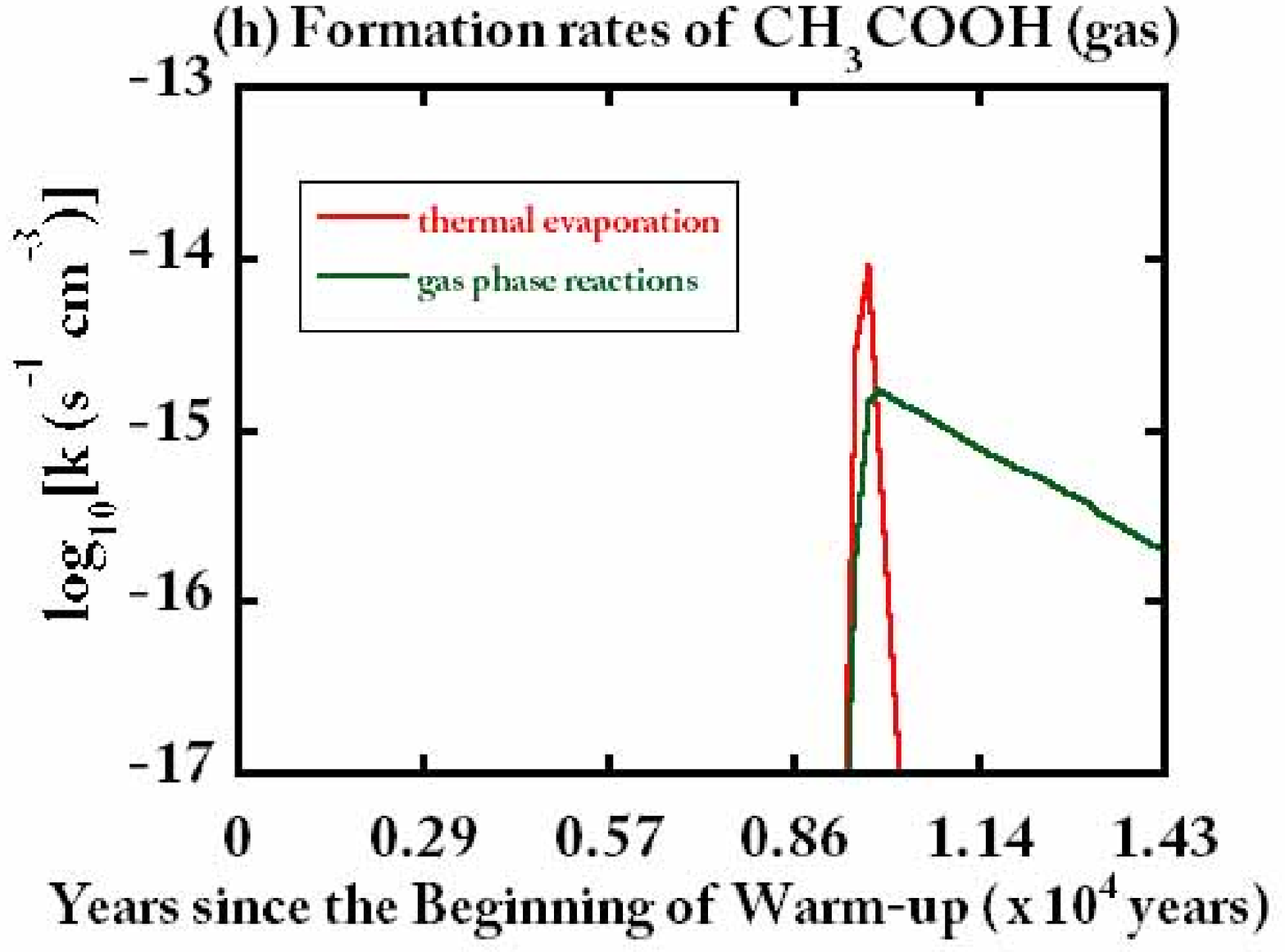}\\
  \end{tabular}
\caption{
The formation rates (cm$^{-3}$ s$^{-1}$) of grain surface reactions for glycine and its precursors were compared using Slow Model.
The formation rates (cm$^{-3}$ s$^{-1}$) of grain surface reactions for glycine and its precursors were compared with Fast Model in the same way as Figure~\ref{fig:Formation_Rate1_fast}.
The labels denote the formation rates of  (a) grain surface glycine, (b) grain surface CH$_2$NH$_2$, and (c) grain surface COOH, (e) gas phase glycine, (f) gas phase NH$_2$OH$_2^+$, (g) gas phase NH$_2$OH, and (h) gas phase CH$_3$COOH.
(d) represent the subtraction of the accretion rate of gas phase glycine from the evaporation rate of glycine on grains.
\label{fig:Formation_Rate1_slow}
}
\end{figure}
\clearpage

\begin{figure}
 \begin{tabular}{ll}
\includegraphics[scale=.4]{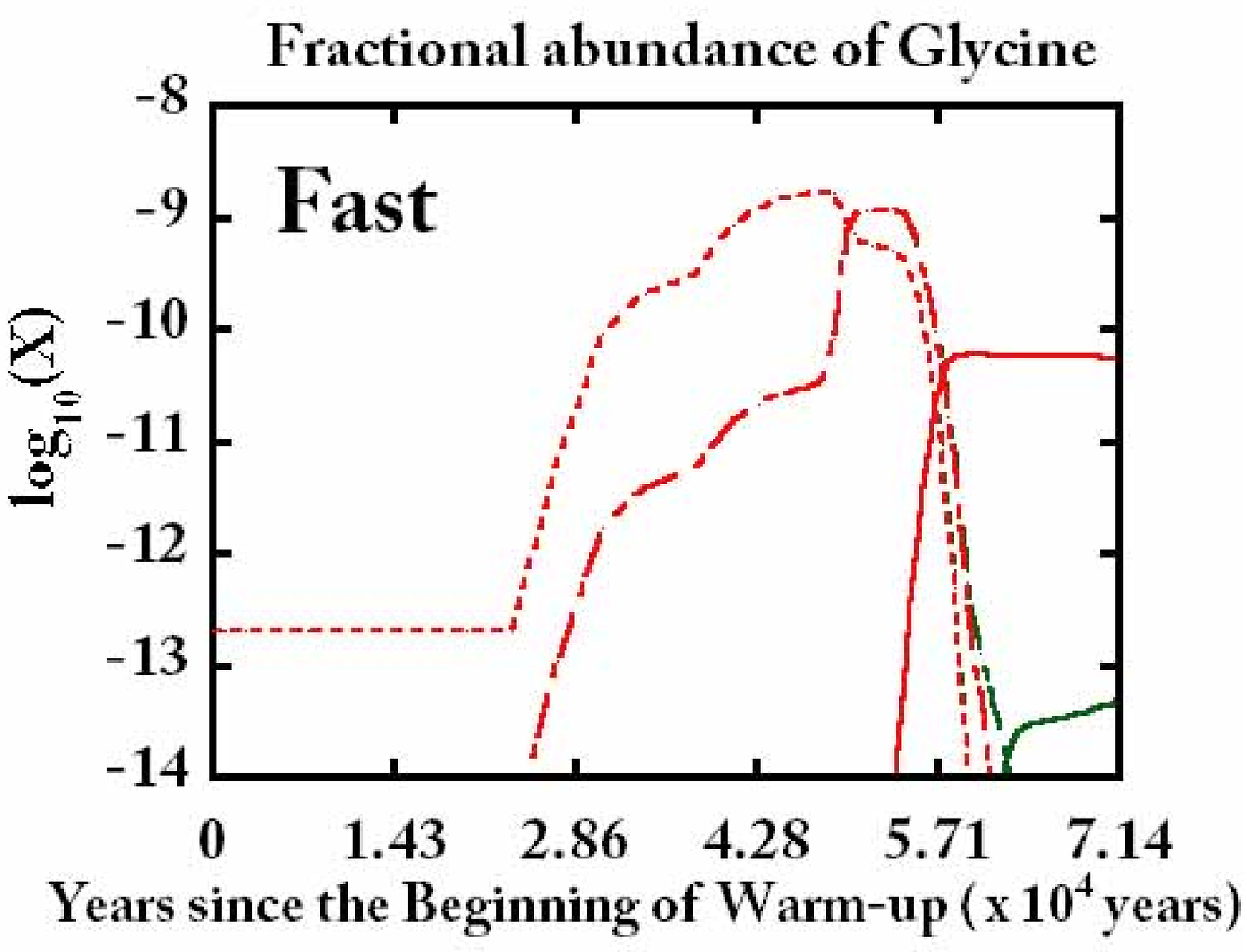}&
\includegraphics[scale=.4]{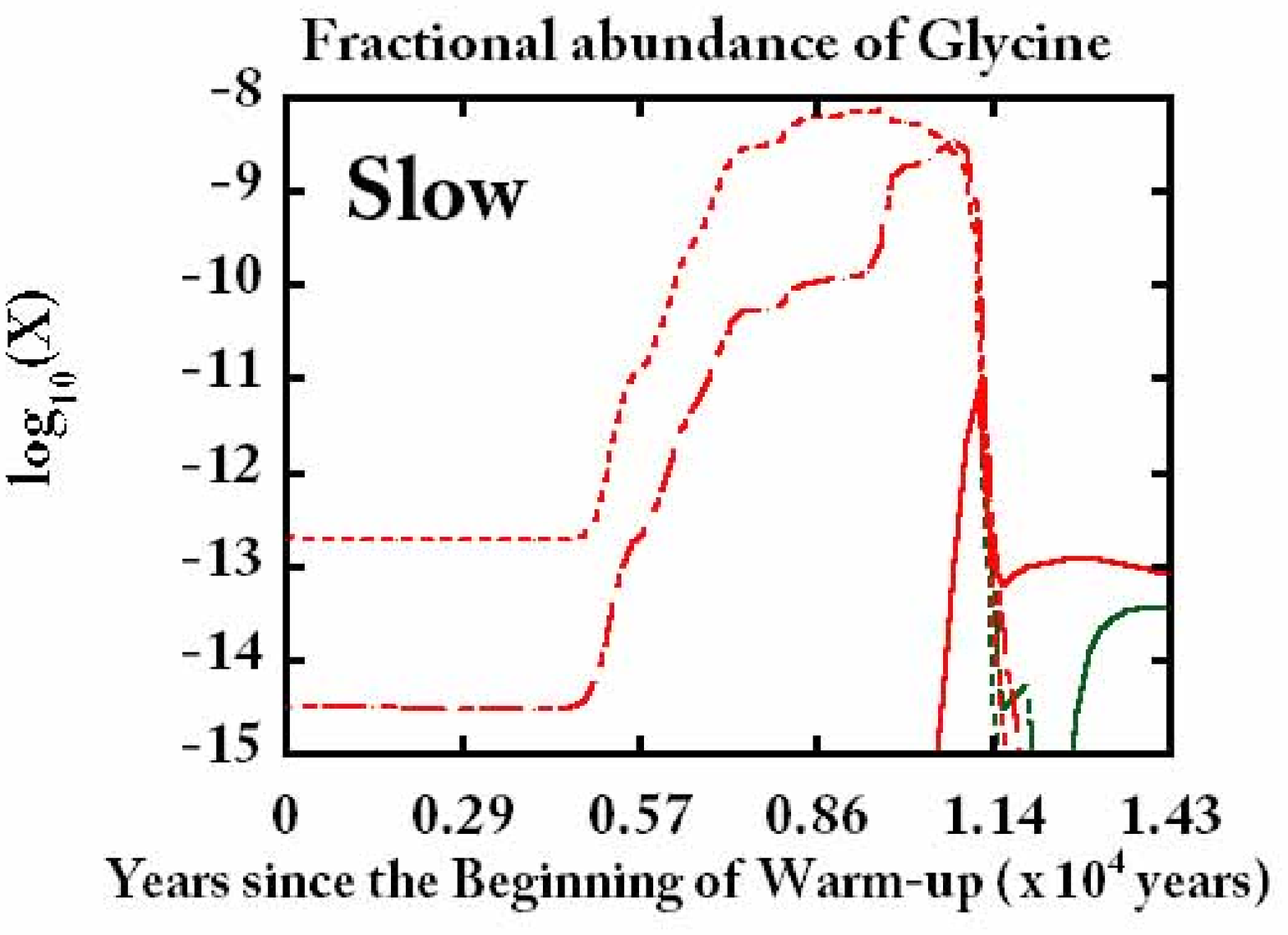}\\
  \end{tabular}
\caption{
The simulated abundances of glycine under the different desorption energies.
The solid, solid-dotted, and dotted lines, respectively, represent the abundances in the gas phase, on the grain surface, and in the grain mantle.
The green and red lines are, respectively, corresponding to the cases of the desorption energy of 13000 and 10100~K.
\label{fig:Abundances_Glycine_ED}
}
\end{figure}
\clearpage

\begin{figure}
 \begin{tabular}{ll}
\includegraphics[scale=.4]{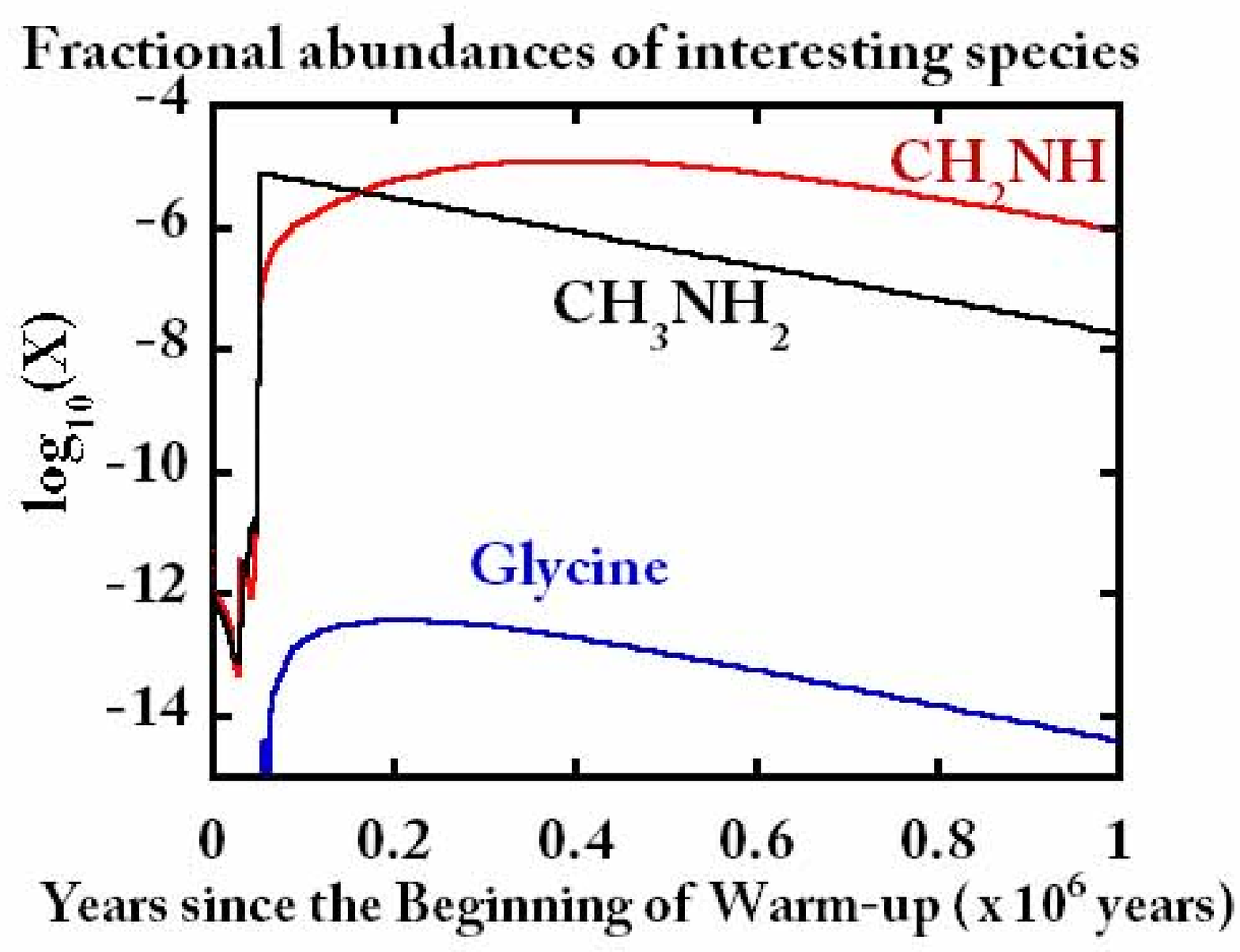}&
\includegraphics[scale=.4]{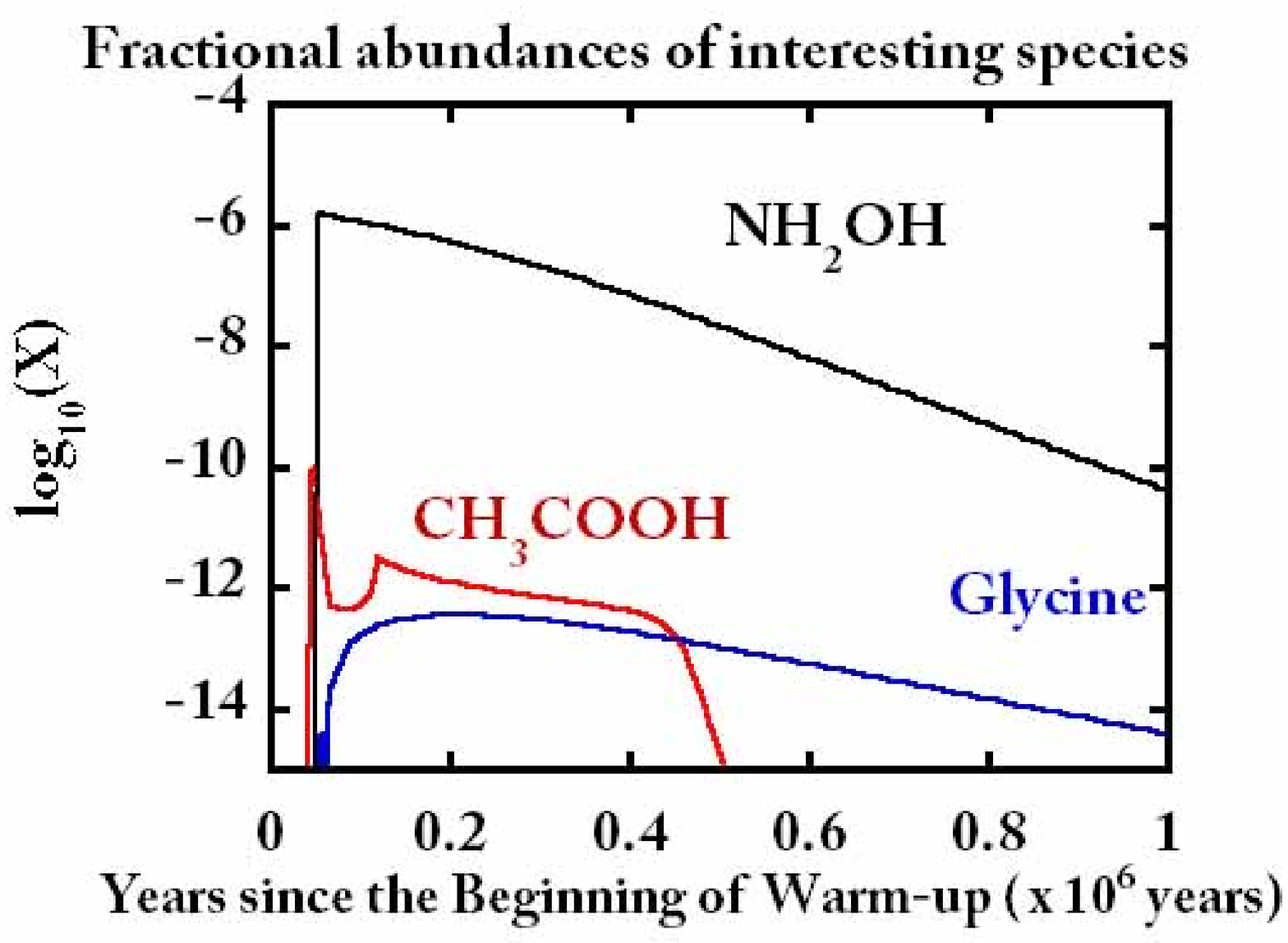}\\
  \end{tabular}
\caption{
The time evolution of gas phase abundances of glycine, CH$_3$NH$_2$, CH$_2$NH, NH$_2$OH, and CH$_3$COOH were plotted with Fast Model.
The gas phase abundance of glycine, CH$_3$NH$_2$, NH$_2$OH, and CH$_3$COOH are  decreased among the gas phase reactions with positive ions and radicals whereas CH$_2$NH, is still kept in high abundance.
Our observations of COMs indicated that G10.47+0.03 and NGC6334F would be in the age of between 6.5 and 8$\times$10$^{5}$~years \citep{Suzuki17}.
\label{fig:Glycine_Suzuki17}
}
\end{figure}
\clearpage

\begin{figure}
 \begin{tabular}{ll}
\includegraphics[scale=.4]{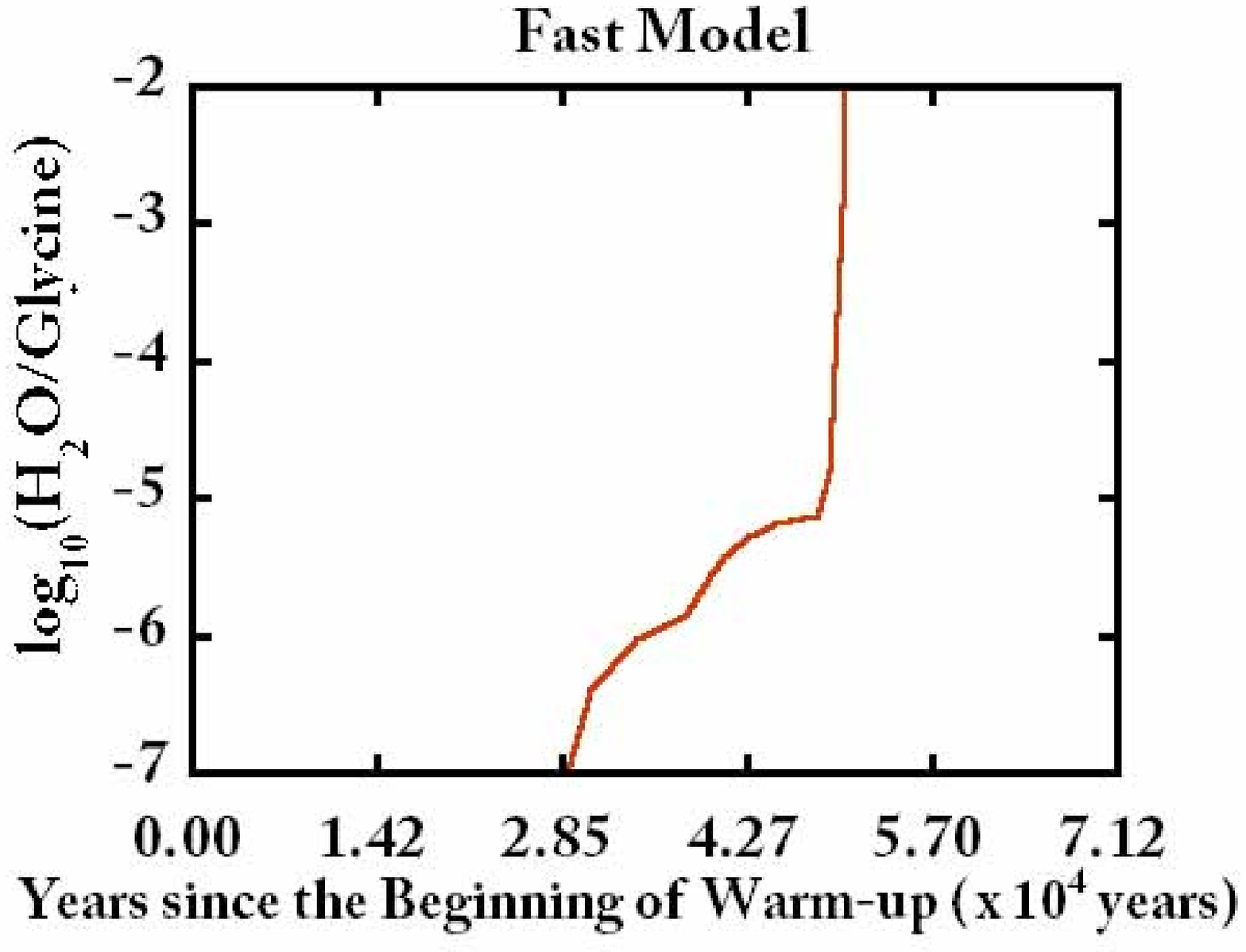}&
\includegraphics[scale=.4]{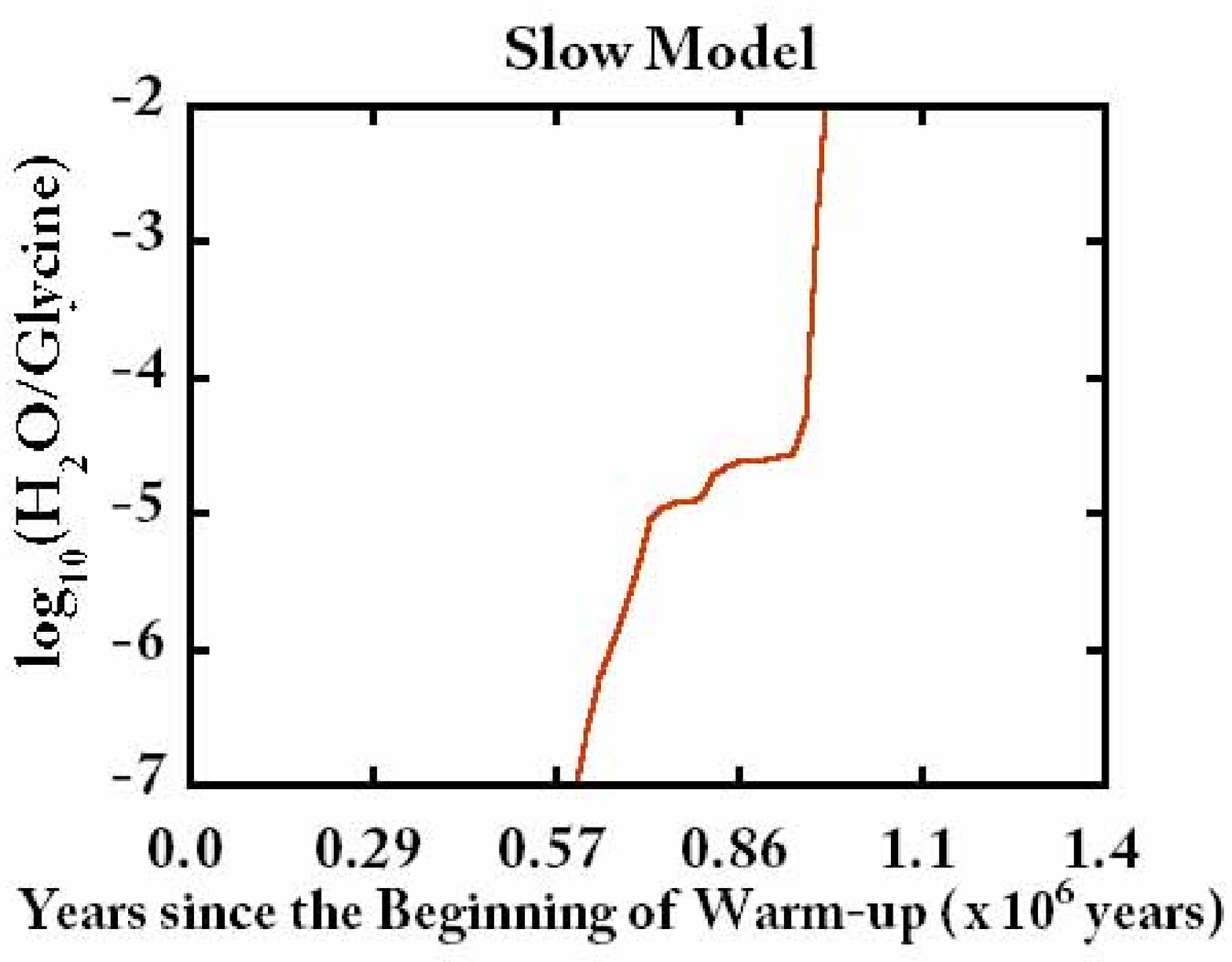}\\
  \end{tabular}
\caption{
The time evolution of the abundance ratio ``Glycine/H$_2$O'' on the grain mantle using fast and slow warm-up models. 
For both cases, the ratio was increased along with the warm-up of the core, since the evaporation of H$_2$O began at much lower temperature than glycine.
\label{fig:Glycine_Comet}
}
\end{figure}
\clearpage

\begin{figure}
\includegraphics[scale=.6]{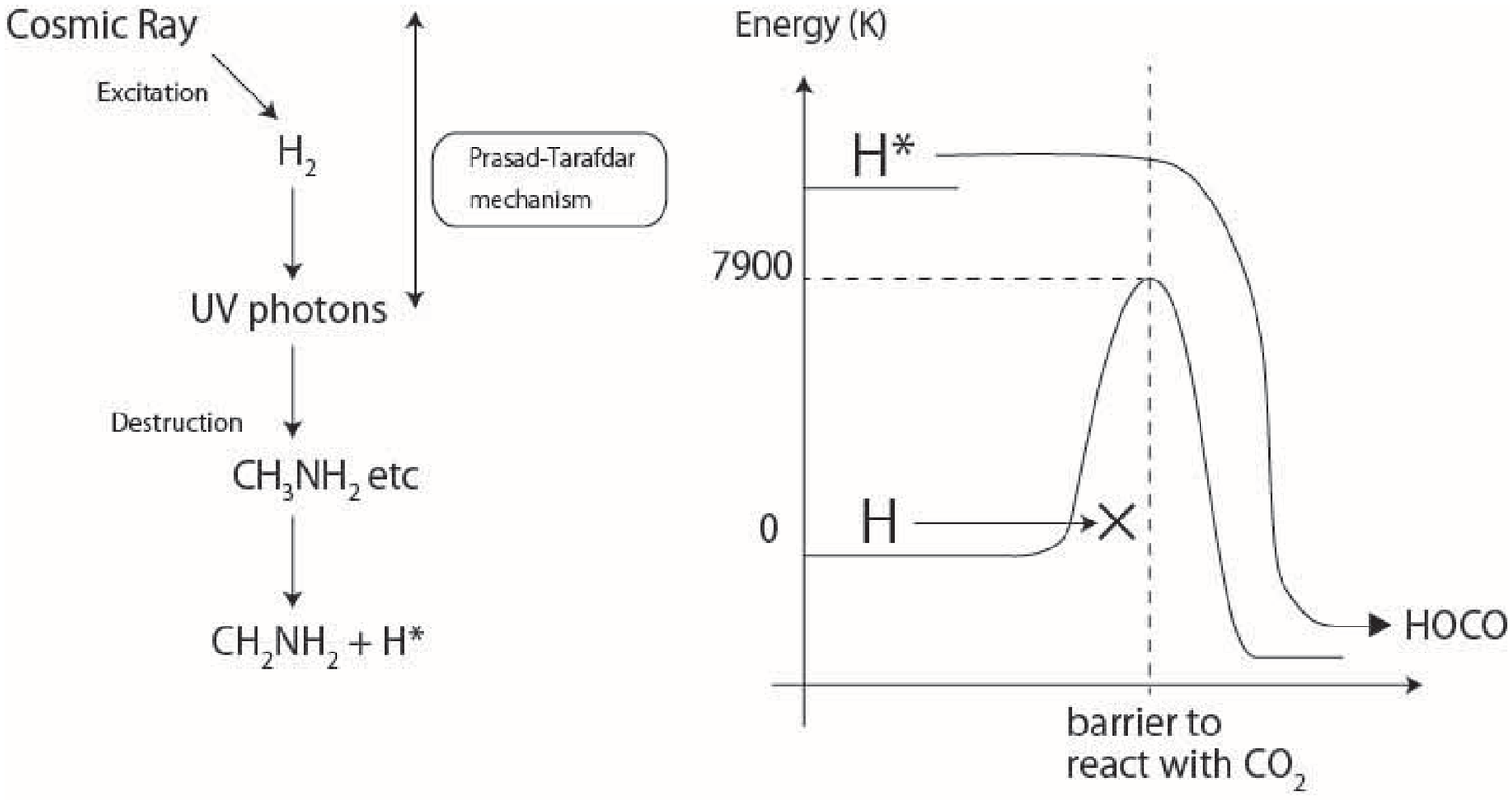}
\caption{
It was assumed that suprathermal hydrogen is formed via UV radiation field.
In dense cores, UV radiation field would be formed by Prasad-Tarafdar mechanism, starting from the destruction of H$_2$ by cosmic rays.
When such UV radiation field destroy molecules, dissociated hydrogen atoms would have extra energy and referred to as ``suprathermal hydrogen'' H*.
Although it is unlikely for usual H atoms to react with CO$_2$, the extra energy of H* make it possible to overcome the activation barrier. 
\label{fig:suprathermal_hydrogen}
}
\end{figure}
\clearpage

\begin{figure}
 \begin{tabular}{ll}
\includegraphics[scale=.4]{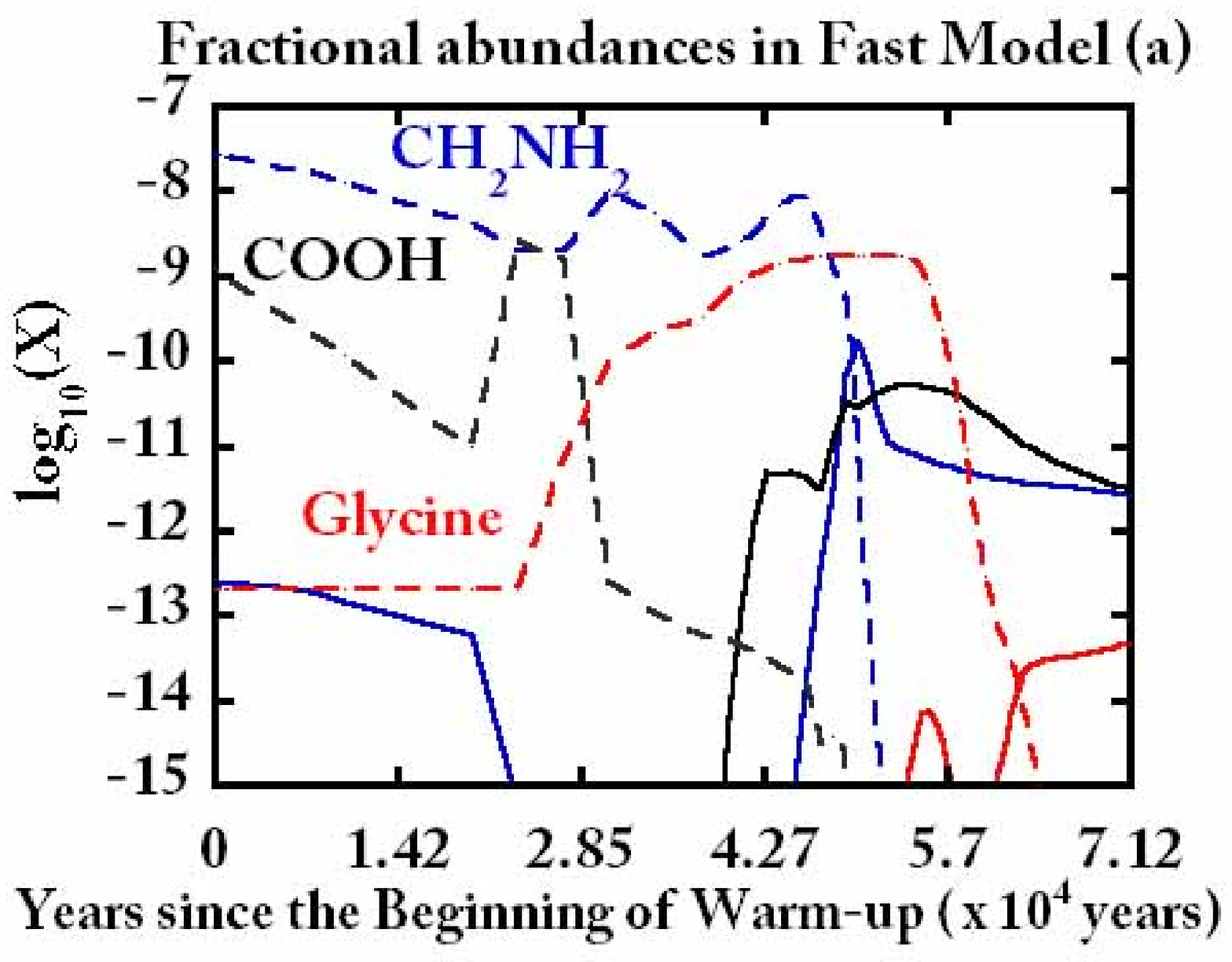}&
\includegraphics[scale=.4]{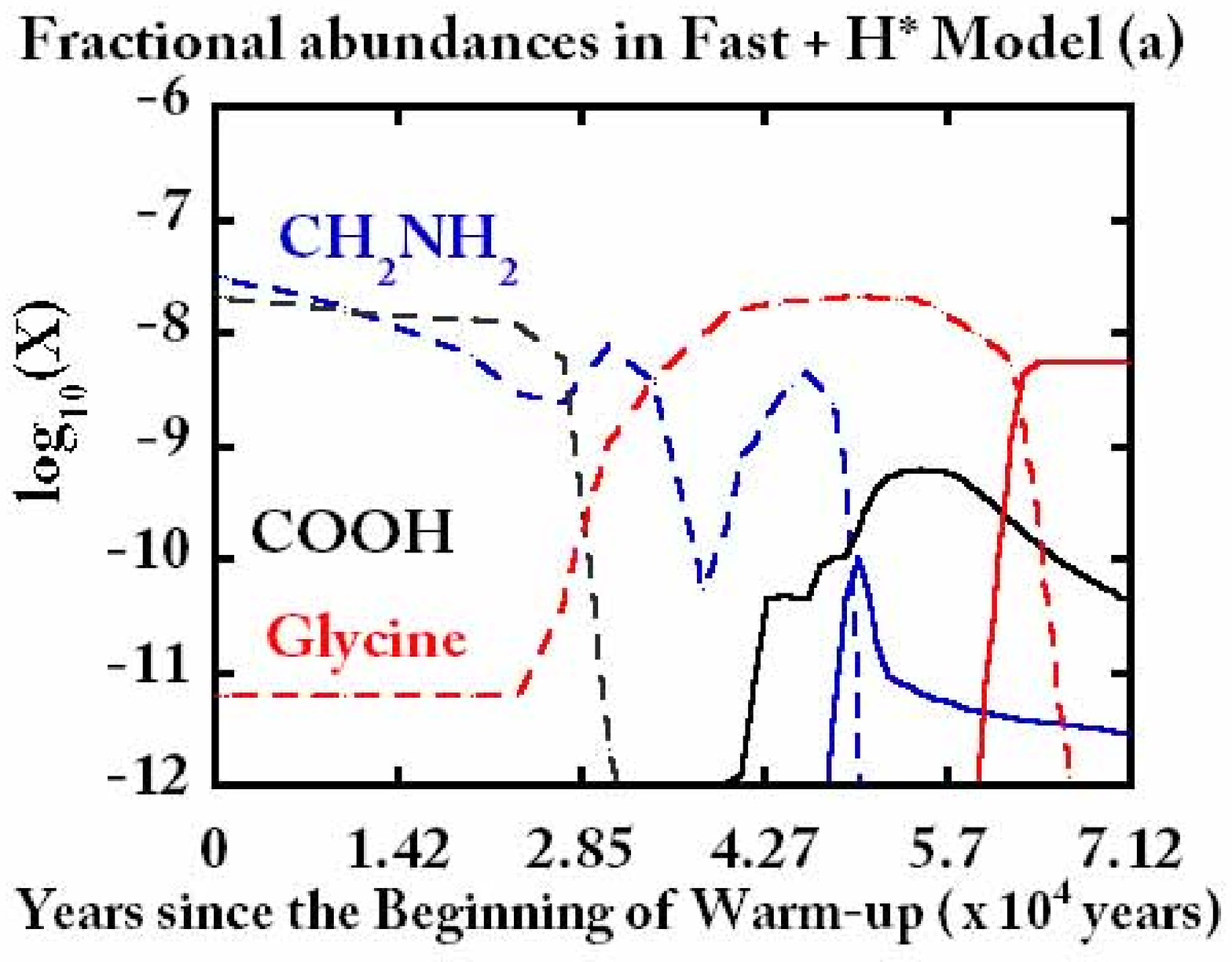}\\
\includegraphics[scale=.4]{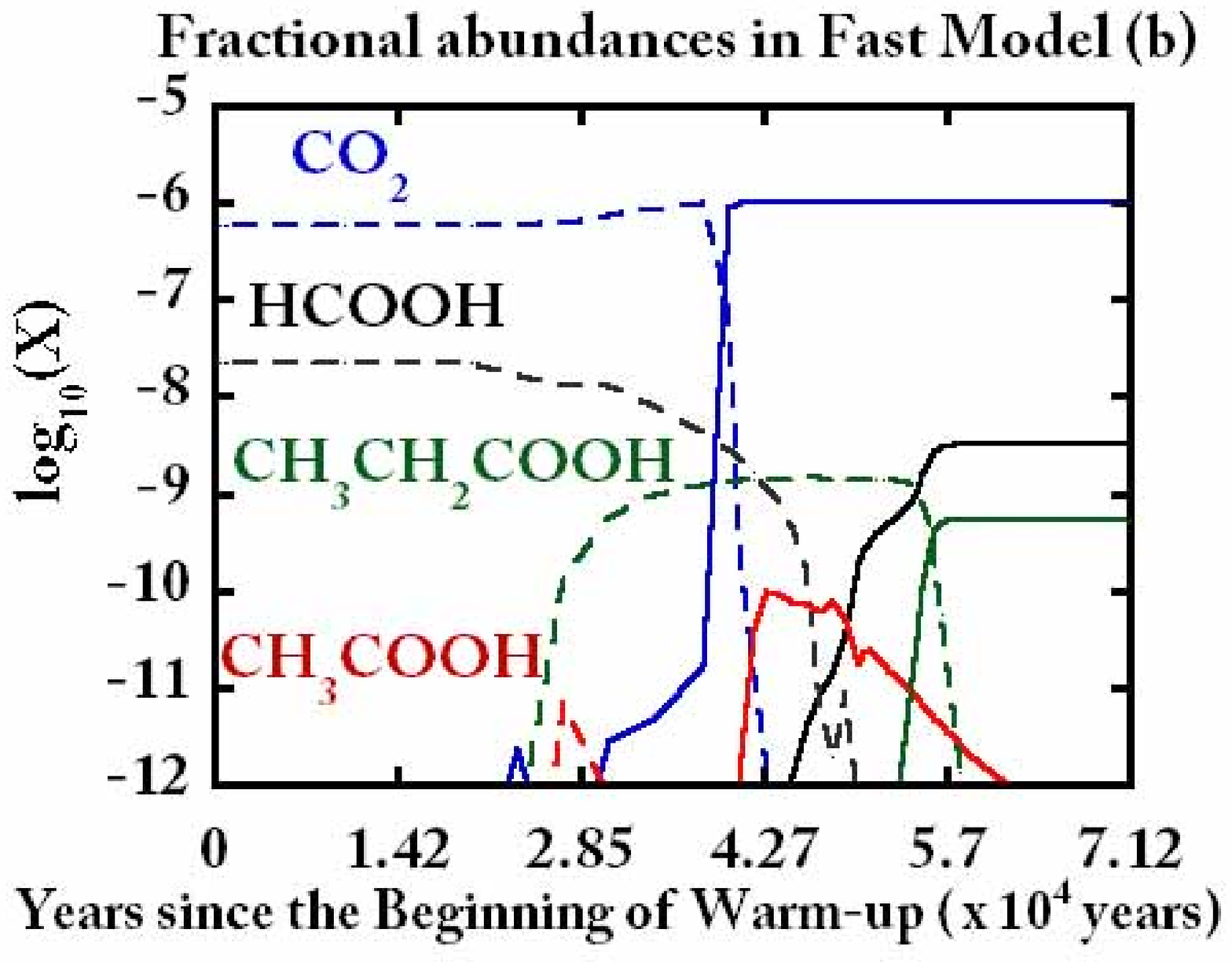}&
\includegraphics[scale=.4]{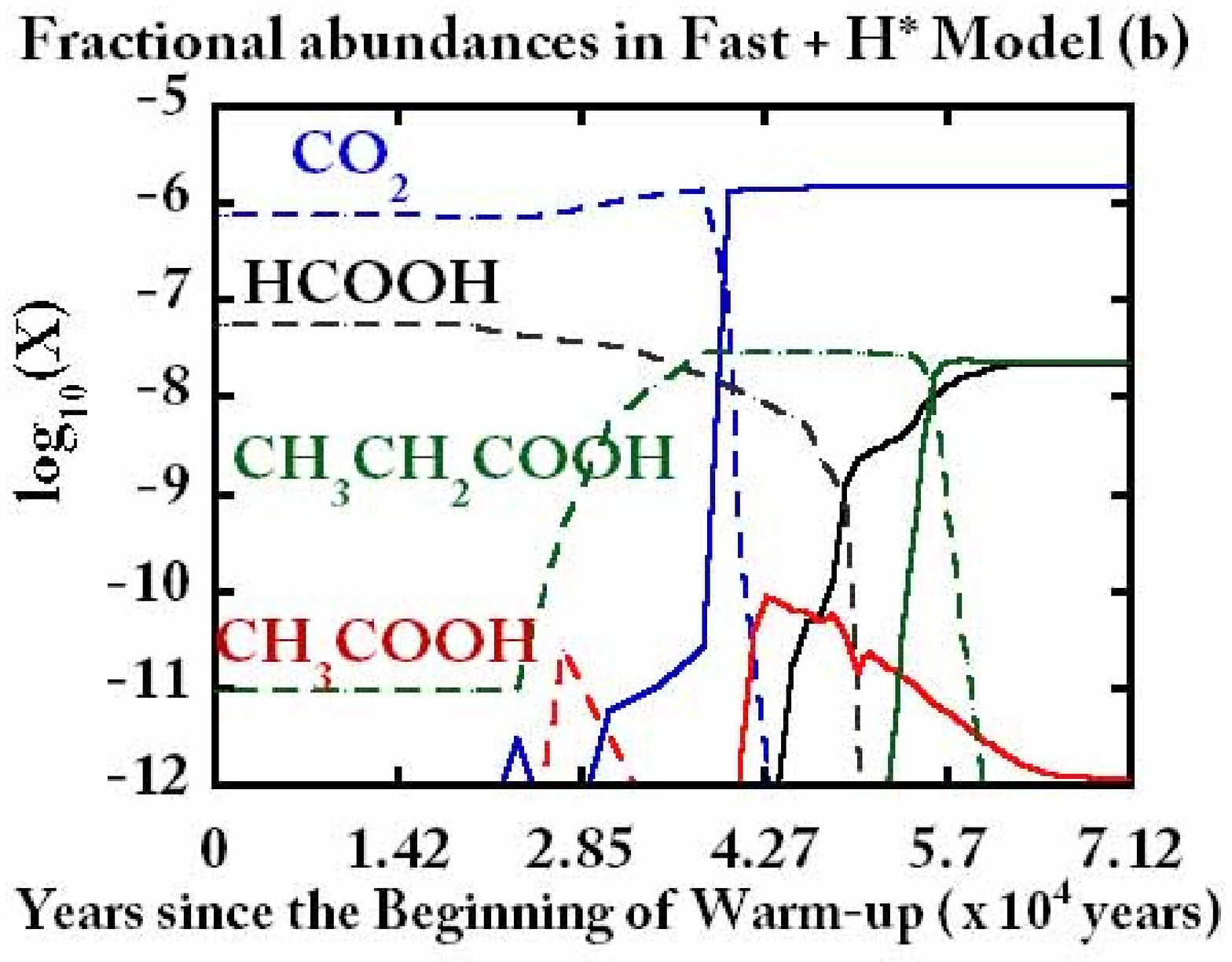}\\
\includegraphics[scale=.4]{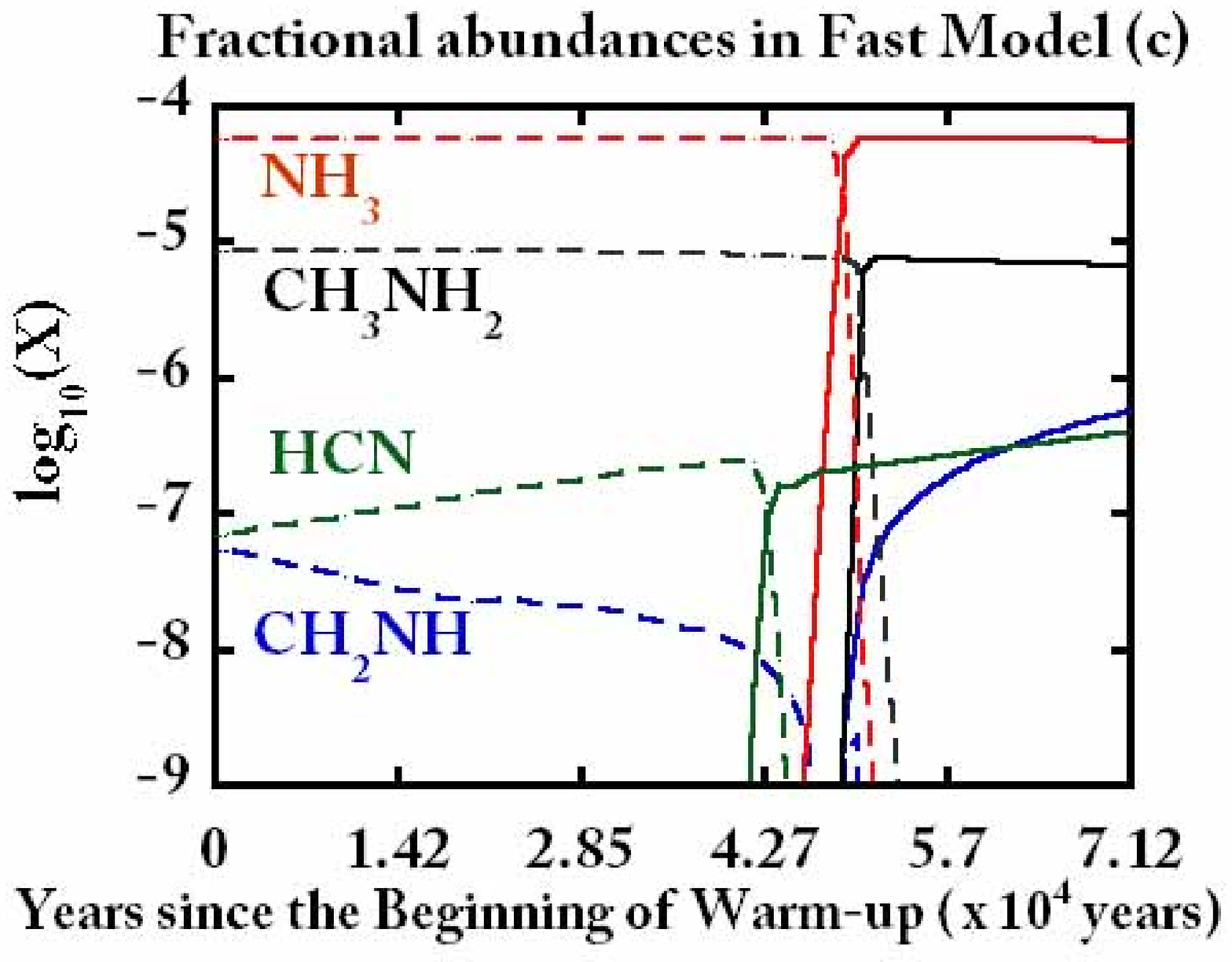}&
\includegraphics[scale=.4]{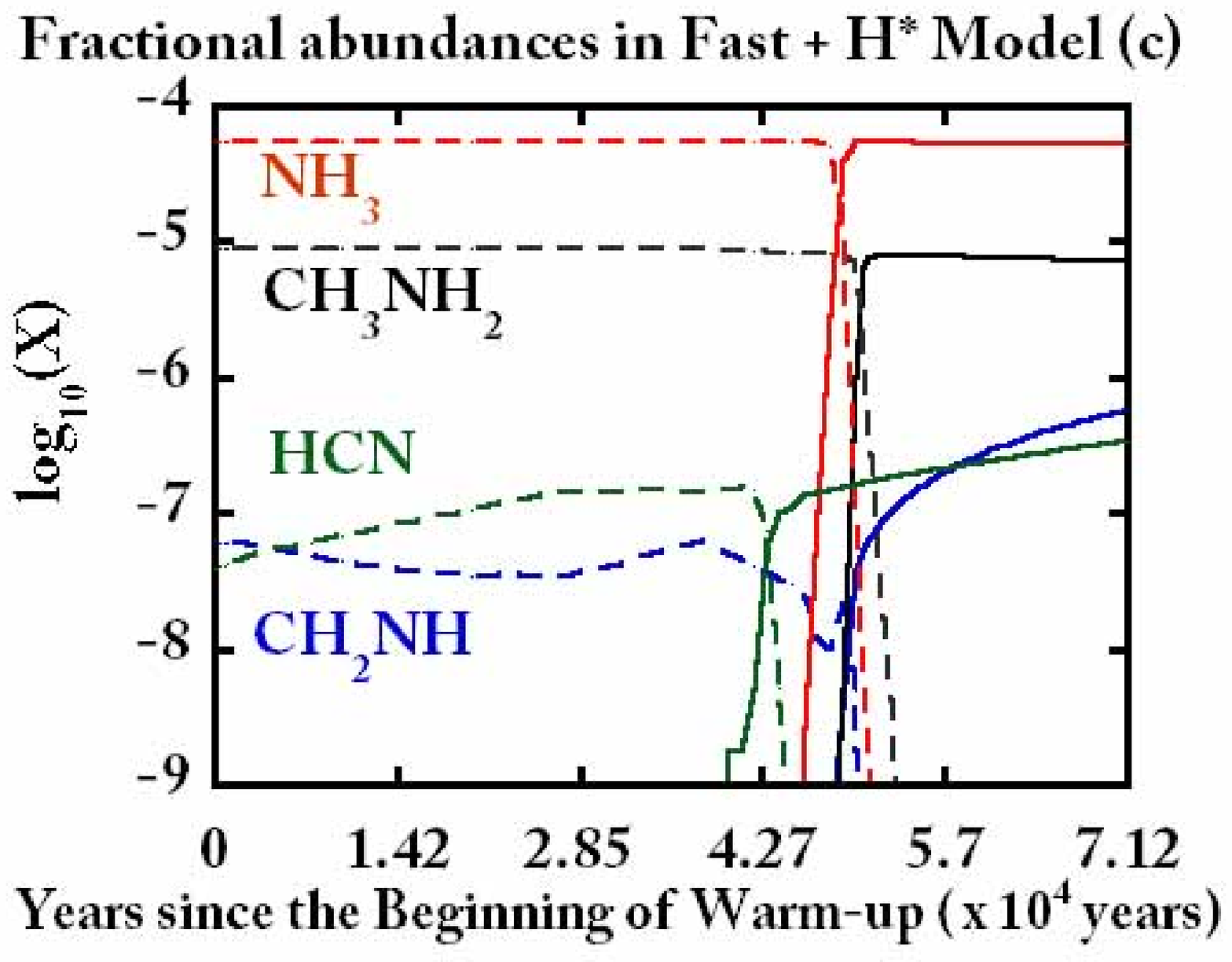}\\
\includegraphics[scale=.4]{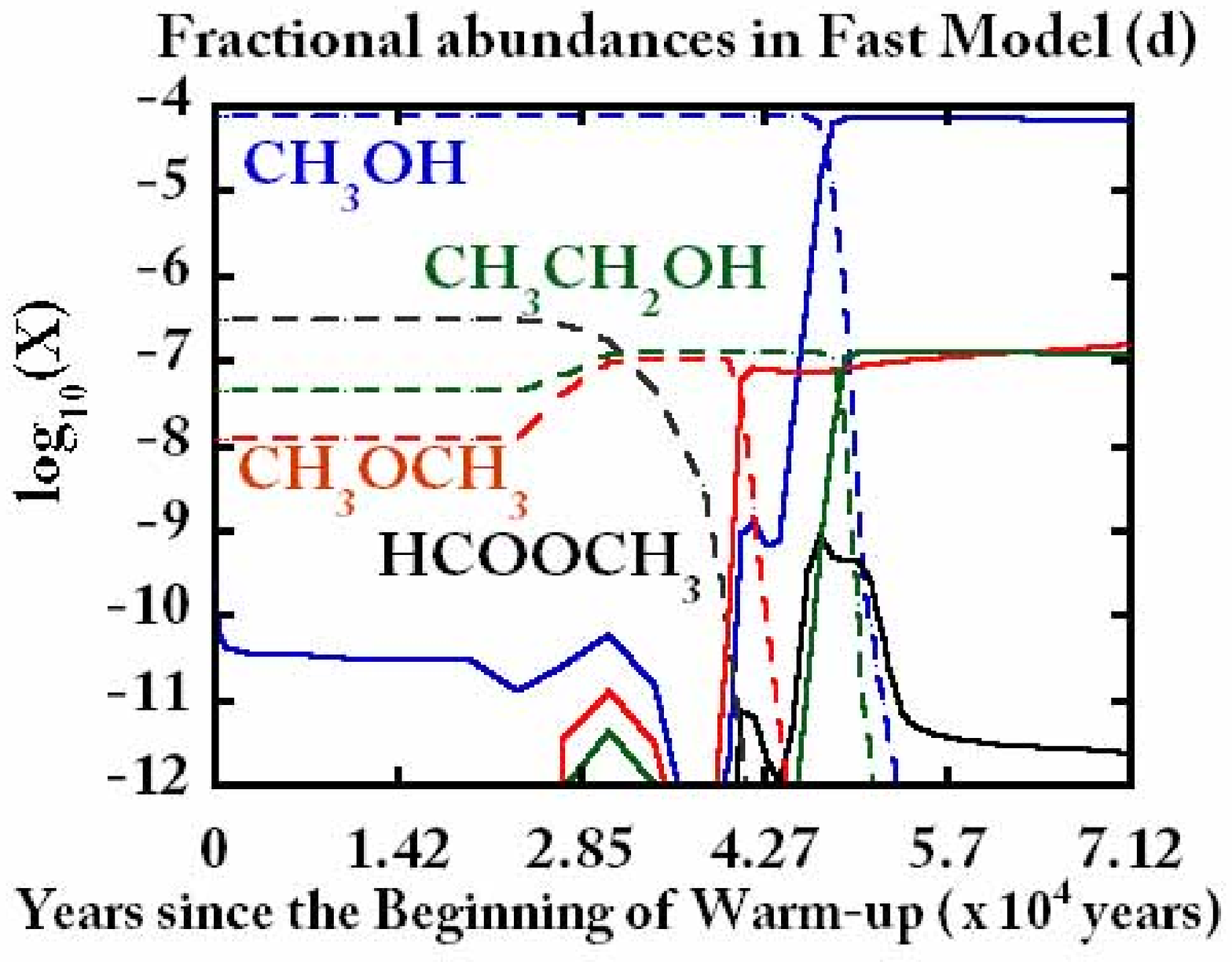}&
\includegraphics[scale=.4]{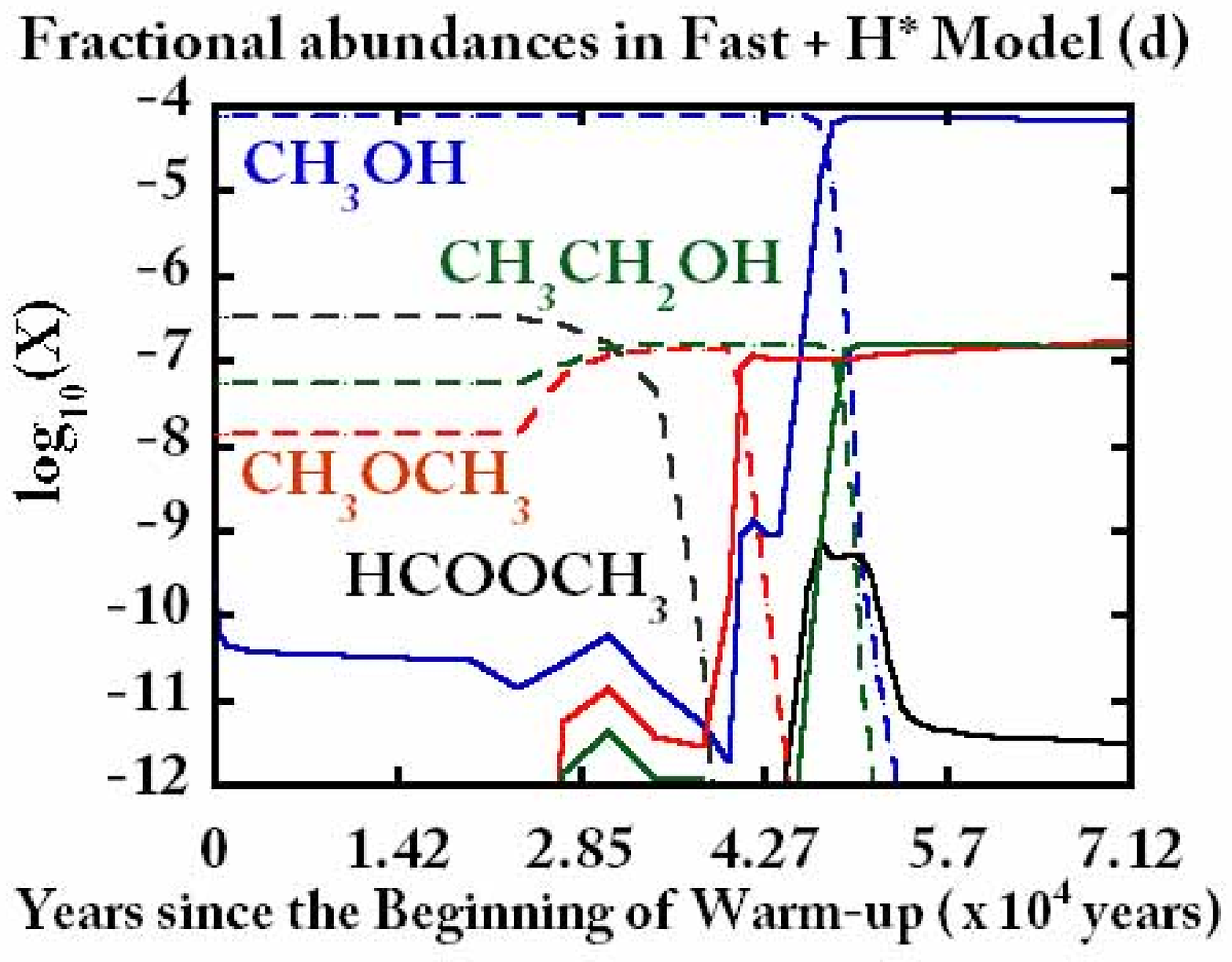}\\
  \end{tabular}
\caption{
The time evolution of the fractional abundances of important species for Fast Model and ``Fast + H*~Model" were shown with different colors
The solid and dotted lines, respectively, represent the abundances in gas phase and on grains (surface and mantle).
The time of zero corresponds to the beginning of warm-up phase.
\label{fig:COM_abundances}
}
\end{figure}
\clearpage

\begin{figure}
 \begin{tabular}{ll}
\includegraphics[scale=.4]{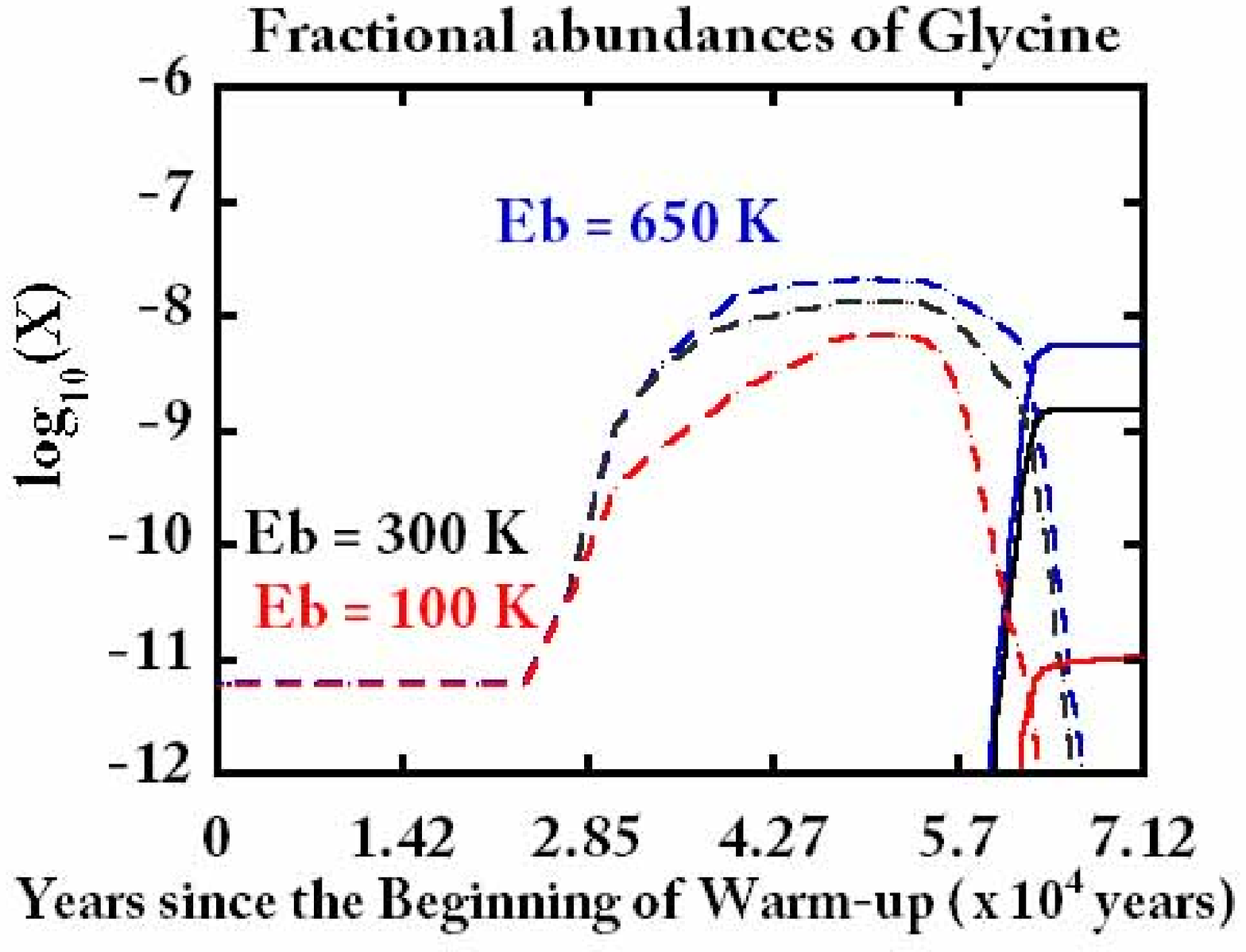}&\\
  \end{tabular}
\caption{
The abundance of glycine in the gas phase and on grains, with different binding energies.
The solid and dotted lines, respectively, represent the abundances in the gas phase and on grains (surface and mantle).
\label{fig:abundance_glycine_HotH_binding_energy}
}
\end{figure}
\clearpage

\begin{deluxetable}{cccc}
\tablecaption{Initial Elemental Abundances Compared to Total Proton Density}
\tablewidth{0pt}
\tablehead{
\colhead{Element} & \colhead{Abundance} & \colhead{Element} & \colhead{Abundance}
}
\startdata
H$_2$&0.5&Si$^+$&8.0 (-9)\\
He&0.09&Fe$^+$&3.0 (-9)\\
N&7.5 (-5)&Na$^+$&2.0 (-9)\\
O&3.2 (-4)&Mg$^+$&7.0 (-9)\\
C$^+$&1.4 (-4)&P$^+$&3.0 (-9)\\
S$^+$&8.0 (-8)&Cl$^+$&4.0 (-9)\\
F&6.7(-9)&&\\ 
\enddata
\tablecomments{Elemental abundance used in our chemical reaction model.
}
\label{table:initial_abundance}
\end{deluxetable}

\begin{deluxetable}{lrlr}
\tablecaption{Binding Energies of Key Species}
\tablewidth{0pt}
\tablehead{
\colhead{Species} & \colhead{E$_{D}$ (K)} & \colhead{Species} & \colhead{E$_{D}$ (K)}
}
\startdata
\hline
\multicolumn{4}{l}{(1) From \cite{Wakelam17}}\\
\hline
H&650&C$_2$H$_3$&2800\\
H$_2$&440&C$_2$H$_4$&2500\\
C&10000&C$_2$H$_5$&3100\\
N&720&l-C$_3$H&4000\\
O&1600&c-C$_3$H&5200\\
CH$_2$&1400&HCO&1300\\
CH$_3$&1600&H$_2$CO&4500\\
CH$_4$&960&CN&2800\\
OH&4600&HCN&3700\\
H$_2$O&5600&HNC&3800\\
NH&2600&CH$_3$CN&4680\\
NH$_2$&3200&CH$_3$OH&5000\\
NH$_3$&5500&CH$_3$CH$_2$OH&5400\\
HOOH&6000&HNCO&4400\\
CO&1300&NH$_2$CHO&6300\\
CO$_2$&2600&CH$_2$OH&4400\\
CCH&3000&C$_2$H6&1600\\
C$_2$H$_2$&2587&CH$_3$O&4400\\
\hline
\multicolumn{4}{l}{(2) Uncertain Values}\\
\hline
COOH&2000&CH$_2$NH$_2$&5530\\
NH$_2$OH&6810&CH$_3$CHO&5400\\
CH$_3$NH$_2$&6500&HNOH&5228\\
CH$_3$COOH&6300&CH$_3$NH&3553\\
HCOOCH$_3$&6295&CH$_3$OCH$_3$&3150\\
\hline
\multicolumn{4}{l}{(3) Experimental Value}\\
\hline
NH$_2$CH$_2$COOH&13000&&\\
\enddata
\tablecomments{The binding energies for key species.
(1) The binding energies for some species were given by theoretical prediction described in \cite{Wakelam17}.
(2) The species whose binding energy was guessed \citep[e.g.,][]{Ruaud15} due to lack of both theoretical or experimental studies.
(3) The binding energy of glycine was estimated based on the experiment by \cite{Tzvetkov04}.
}
\label{table:binding_energy}
\end{deluxetable}

\begin{deluxetable}{ll}
\tablecaption{Hydrogenation Processes to HCN discussed in \cite{Suzuki16}}
\tablewidth{0pt}
\tablehead{
\colhead{Reaction}
& \colhead{\shortstack {E$_A$ \\ (K)}}
}
\startdata
N + CH$_3$ $\longrightarrow$ CH$_2$NH& E$_A$= 0~K\\
NH + CH$_2$ $\longrightarrow$ CH$_2$NH& E$_A$= 0~K\\
NH$_2$ + CH $\longrightarrow$ CH$_2$NH& E$_A$= 0~K\\
HCN + H $\longrightarrow$ H$_2$CN& E$_A$= 3647~K \\
HCN + H $\longrightarrow$ HCNH& E$_A$= 6440~K \\
H$_2$CN + H $\longrightarrow$ CH$_2$NH& E$_A$= 0~K \\
HCNH + H $\longrightarrow$ CH$_2$NH& E$_A$= 0~K \\
CH$_2$NH + H $\longrightarrow$ CH$_3$NH& E$_A$= 2134~K \\
CH$_2$NH + H $\longrightarrow$ CH$_2$NH$_2$& E$_A$= 3170~K \\
CH$_3$NH + H $\longrightarrow$ CH$_3$NH$_2$& E$_A$= 0~K \\
CH$_2$NH$_2$ + H $\longrightarrow$ CH$_3$NH$_2$& E$_A$= 0~K \\
\enddata
\tablecomments{
The dust surface reactions related to CH$_2$NH and CH$_3$NH$_2$ are shown.
E$_A$ represents the value of the activation barrier.
Since radical species are so reactive, radical-radical reactions would have no activation barriers.
The activation barriers for HCN and CH$_2$NH were cited from the theoretical study by \cite{Woon02}.
}
\label{table:HCN_hydrogenation}
\end{deluxetable}

\begin{deluxetable}{ccccccccc}
\tabletypesize{\scriptsize}
\tablecaption{Comparison of Molecular Abundances on Icy Grains}
\tablewidth{0pt}
\tablehead{
\colhead{} & \colhead{best-fitting time\shortstack {years}} & \colhead{H$_2$CO} & \colhead{CO} & \colhead{CO$_2$} & \colhead{CH$_3$OH} & \colhead{NH$_3$} & \colhead{CH$_4$} & \colhead{H$_2$CO}
}
\startdata
This work&2.026$\times$10$^{8}$&100&98&0.2&32&26&9&4\\
LYSO&&100&4-14&12-25&5-23&$\sim$7&1-3&$\sim$2-7\\
MYSO&&100&12-35&23-37&5-12&4-8&3-6&$\sim$6\\
\enddata
\tablecomments{
We show the best-fitted icy molecular abundances of basic species in our modeling at the end of the collapsing phase, compared to actual low-mass and massive young stellar object (LYSO and MYSO) obtained by \cite{Boogert15}.
}
\label{table:ice_abundances}
\end{deluxetable}

\begin{deluxetable}{cccc}
\tabletypesize{\scriptsize}
\tablecaption{The Peak Abundances of Gas Phase Glycine in Different Models}
\tablewidth{0pt}
\tablehead{
\colhead{Model} & \colhead{ \shortstack {Eb \\ (K)}} & \colhead{X} & \colhead{ \shortstack {T \\ (K)}}
}
\startdata
\cite{Garrod13}&10100&8.4 (-11)&216\\
Fast Model&13000&4.3 (-14)&400\\
Fast Model (b)&10100&3.1 (-11)&232\\
\enddata
\tablecomments{
The symbols, Eb, X, T, respectively, represent the binding energy of glycine used in the models, the peak of gas phase fractional abundance of glycine, and the temperature when the peak abundance of glycine was achieved.
}
\label{table:glycine_abundances}
\end{deluxetable}

\begin{deluxetable}{cccccccccc}
\tabletypesize{\scriptsize}
\tablecaption{The Peaks of the Simulated Abundances for the Fast Warm-up Model  in ``Garrod (2013) Model'' and Fast Model}
\tablewidth{0pt}
\tablehead{
\multicolumn{1}{c}{ }  &
\multicolumn{2}{c}{Garrod Model}  &
\multicolumn{1}{c}{ }  &
\multicolumn{6}{c}{Fast~Model}  \\ \cline{2-3}  \cline{5-10}\\
\multicolumn{1}{c}{ }  &
\multicolumn{2}{c}{Gas}  &
\multicolumn{1}{c}{ }  &
\multicolumn{2}{c}{Gas}  &
\multicolumn{2}{c}{Surface}  &
\multicolumn{2}{c}{Mantle}  \\
\cline{2-3} \cline{5-6} \cline{7-8} \cline{9-10}\\
\colhead{Species}
& \colhead{X} & \colhead{ \shortstack {T \\ (K)}}&
& \colhead{X} & \colhead{ \shortstack {T \\ (K)}}
& \colhead{X} & \colhead{ \shortstack {T \\ (K)}}
& \colhead{X} & \colhead{ \shortstack {T \\ (K)}}
}
\startdata
NH$_2$OH&1.6  (-10)&145&&1.5  (-6)&158&6.8  (-7)&132&1.5  (-6)&25\\
CH$_2$NH&1.1  (-8)&400&&5.6  (-7)&400&1.0  (-9)&132&5.3  (-8)&10\\
CH$_3$NH$_2$&8.0  (-8)&114&&7.4  (-6)&149&3.0  (-6)&132&8.3  (-6)&11\\
NH$_2$CH$_2$COOH&4.3  (-9)&123&&4.6  (-14)&400&1.2  (-9)&163&1.6  (-9)&117\\
HCOOH&4.8  (-8)&123&&3.4  (-9)&314&3.6  (-10)&10&2.2  (-8)&10\\
CH$_3$COOH&1.0  (-10)&134&&1.0  (-10)&84&1.1  (-13)&25&6.7  (-12)&25\\
\enddata
\tablecomments{
The peak of the fractional abundances and the temperature of grains at that moment in Fast Model were compared with Garrod (2013) Model.
The notation of ``a (b)" represents a $\times$10$^b$.
}
\label{table:Garrod_vs_Model1}
\end{deluxetable}

\begin{deluxetable}{cccccccccc}
\tabletypesize{\scriptsize}
\tablecaption{The Peaks of the Simulated Abundances for the Slow Warm-up Model in ``Garrod (2013) Model'' and Slow Model}
\tablewidth{0pt}
\tablehead{
\multicolumn{1}{c}{ }  &
\multicolumn{2}{c}{Garrod Model}  &
\multicolumn{1}{c}{ }  &
\multicolumn{6}{c}{Slow~Model}  \\ \cline{2-3}  \cline{5-10}\\
\multicolumn{1}{c}{ }  &
\multicolumn{2}{c}{Gas}  &
\multicolumn{1}{c}{ }  &
\multicolumn{2}{c}{Gas}  &
\multicolumn{2}{c}{Surface}  &
\multicolumn{2}{c}{Mantle}  \\
\cline{2-3} \cline{5-6} \cline{7-8} \cline{9-10}\\
\colhead{Species}
& \colhead{X} & \colhead{ \shortstack {T \\ (K)}}&
& \colhead{X} & \colhead{ \shortstack {T \\ (K)}}
& \colhead{X} & \colhead{ \shortstack {T \\ (K)}}
& \colhead{X} & \colhead{ \shortstack {T \\ (K)}}
}
\startdata
NH$_2$OH&1.6  (-10)&145&&6.8  (-7)&156&2.5  (-7)&139&1.6  (-6)&10\\
CH$_2$NH&1.1  (-8)&400&&6.6  (-6)&338&1.0  (-8)&132&5.3  (-8)&10\\
CH$_3$NH$_2$&8.0  (-8)&114&&2.6  (-6)&147&6.3  (-7)&132&8.3  (-6)&10\\
NH$_2$CH$_2$COOH&4.3  (-9)&123&&3.6  (-14)&368&3.5  (-9)&173&7.3  (-9)&124\\
HCOOH&4.8  (-8)&123&&2.6  (-8)&202&3.6  (-10)&10&2.2  (-8)&10\\
CH$_3$COOH&1.0  (-10)&134&&2.8  (-12)&104&3.8  (-15)&76&1.8  (-13)&76\\
\enddata
\tablecomments{
The peak of the fractional abundances and the temperature of grains at that moment in Slow Model were compared with Garrod (2013) Model.
The notation of ``a (b)" represents a $\times$10$^b$.
}
\label{table:Garrod_vs_Model2}
\end{deluxetable}

\begin{deluxetable}{ccccc}
\tabletypesize{\scriptsize}
\tablecaption{Comparison of Abundances between Our Model and Actual Abundances}
\tablewidth{0pt}
\tablehead{
\colhead{Source} & \colhead{ NH$_2$OH} & \colhead{CH$_3$COOH} & \colhead{HCOOH} & \colhead{CH$_3$NH$_2$}
}
\startdata
Orion KL&$<$ 3 (-11)$^a$&3-10 (-10)$^b$&3 (-9)$^c$&1 (-9)$^d$\\
Sgr B2&$<$ 8 (-12)$^a$&0.9-7 (-10)$^e$&1 (-11)$^f$&1.7 (-9)$^g$\\
Fast Model&1.5 (-6)&1.0 (-10)&3.4 (-9)&7.4 (-6)\\
\enddata
\tablecomments{
We compared the peak gas phase abundances of glycine's precursors, NH$_2$OH, CH$_3$COOH, HCOOH, and CH$_3$NH$_2$ in our model, with actual observed abundances towards star-forming regions.
Since observations of these species has been performed towards high-mass stars, we used Fast Model for this comparison.
The notation of ``a (b)'' represents ``a$\times$10$^{b}$''.
Reference: (a) \cite{Pulliam12}, (b) \cite{Favre17}, (c) \cite{Liu02}, (d) \cite{Pagani17}, (e) \cite{Mehringer97}, (f) \cite{Ikeda01}, (g) \cite{Halfen13}
}
\label{table:compare_abundances}
\end{deluxetable}

\begin{deluxetable}{cccccccccccccc}
\tabletypesize{\scriptsize}
\tablecaption{The Peaks of the Simulated Abundances for Fast Model and ``Fast + H* Model''}
\tablewidth{0pt}
\tablehead{
\multicolumn{1}{c}{ }  &
\multicolumn{6}{c}{Fast Model}  &
\multicolumn{1}{c}{ }  &
\multicolumn{6}{c}{Fast + H* Model}  \\ \cline{2-7}  \cline{9-14}\\
\multicolumn{1}{c}{ }  &
\multicolumn{2}{c}{Gas}  &
\multicolumn{2}{c}{Surface}  &
\multicolumn{2}{c}{Mantle}  &
\multicolumn{1}{c}{ }  &
\multicolumn{2}{c}{Gas}  &
\multicolumn{2}{c}{Surface}  &
\multicolumn{2}{c}{Mantle}  \\
\cline{2-3} \cline{4-5} \cline{6-7} \cline{9-10} \cline{11-12} \cline{13-14} \\
\colhead{Species}
& \colhead{X} & \colhead{ \shortstack {T \\ (K)}}
& \colhead{X} & \colhead{ \shortstack {T \\ (K)}}
& \colhead{X} & \colhead{ \shortstack {T \\ (K)}}&
& \colhead{X} & \colhead{ \shortstack {T \\ (K)}}
& \colhead{X} & \colhead{ \shortstack {T \\ (K)}}
& \colhead{X} & \colhead{ \shortstack {T \\ (K)}}
}
\startdata
CH$_3$&2.5  (-6)&400&7.5  (-13)&10&4.0  (-11)&10&&2.5  (-6)&400&1.4  (-12)&10&7.8  (-11)&10\\
NH$_2$&4.1  (-9)&400&7.4  (-13)&10&3.6  (-11)&10&&4.0  (-9)&400&7.0  (-13)&10&3.4  (-11)&10\\
CN&3.2  (-10)&90&1.5  (-16)&44&9.2  (-15)&10&&2.1  (-10)&90&2.3  (-16)&33&1.8  (-14)&44\\
OCH$_3$&2.4  (-9)&125&1.5  (-8)&10&9.2  (-7)&10&&1.6  (-9)&125&1.8  (-8)&10&1.1  (-6)&10\\
CH$_2$OH&5.1  (-11)&132&4.8  (-9)&10&2.9  (-7)&10&&3.4  (-11)&132&6.0  (-9)&10&3.7  (-7)&10\\
HOCO&5.2  (-11)&167&4.2  (-11)&19&2.6  (-9)&19&&6.0  (-10)&177&3.6  (-10)&10&2.2  (-8)&10\\
CH$_2$NH$_2$&1.7  (-10)&132&4.6  (-10)&10&2.8  (-8)&10&&9.9  (-11)&132&5.4  (-10)&10&3.3  (-8)&10\\
H$_2$O&2.4  (-4)&400&5.3  (-6)&125&2.3  (-4)&103&&2.4  (-4)&400&5.3  (-6)&125&2.3  (-4)&103\\
CO&4.4  (-6)&400&7.3  (-9)&19&4.5  (-7)&19&&4.9  (-6)&400&8.2  (-9)&19&5.1  (-7)&19\\
H$_2$CO&1.7  (-6)&224&3.9  (-8)&10&2.4  (-6)&10&&2.3  (-6)&207&4.7  (-8)&10&2.9  (-6)&10\\
CH$_3$OH&7.3  (-5)&125&1.3  (-6)&33&7.7  (-5)&11&&7.3  (-5)&125&1.4  (-6)&33&7.6  (-5)&11\\
CO$_2$&1.0  (-6)&132&1.7  (-8)&59&9.7  (-7)&59&&1.4  (-6)&132&2.4  (-8)&59&1.2  (-6)&44\\
O$_2$&5.3  (-9)&68&3.0  (-13)&10&1.8  (-11)&10&&9.1  (-9)&68&1.1  (-12)&10&6.9  (-11)&10\\
O$_3$&7.9  (-11)&68&4.4  (-12)&25&2.7  (-10)&25&&3.3  (-10)&68&1.3  (-11)&33&8.1  (-10)&33\\
HCN&3.9  (-7)&400&4.3  (-9)&73&2.4  (-7)&73&&3.4  (-7)&400&2.7  (-9)&73&2.1  (-7)&68\\
HNC&4.9  (-8)&97&3.2  (-10)&84&1.8  (-8)&84&&3.3  (-8)&90&3.2  (-10)&84&3.1  (-8)&84\\
CH$_4$&3.5  (-5)&33&5.7  (-7)&19&3.5  (-5)&19&&3.4  (-5)&33&5.7  (-7)&19&3.5  (-5)&19\\
NH$_3$&5.7  (-5)&141&1.2  (-6)&117&5.6  (-5)&14&&5.4  (-5)&141&1.1  (-6)&117&5.3  (-5)&14\\
C$_2$H$_2$&1.3  (-7)&400&7.9  (-10)&59&4.4  (-8)&59&&1.3  (-7)&400&9.8  (-10)&59&4.2  (-8)&44\\
C$_2$H$_4$&5.0  (-7)&400&3.2  (-9)&59&1.8  (-7)&59&&5.8  (-7)&400&3.9  (-9)&59&1.7  (-7)&44\\
C$_2$H$_6$&1.8  (-7)&59&5.0  (-9)&14&3.1  (-7)&14&&1.6  (-7)&59&5.0  (-9)&14&3.1  (-7)&14\\
HNO&5.8  (-9)&79&1.0  (-9)&10&6.3  (-8)&10&&3.9  (-9)&73&1.6  (-9)&10&1.0  (-7)&10\\
NO&3.6  (-8)&125&6.0  (-10)&10&3.7  (-8)&10&&3.3  (-8)&125&9.6  (-10)&10&5.9  (-8)&10\\
OCN&1.9  (-10)&103&7.1  (-14)&10&4.4  (-12)&10&&1.9  (-10)&90&6.2  (-13)&10&3.7  (-11)&10\\
SO&1.1  (-8)&73&2.9  (-10)&33&1.6  (-8)&33&&1.4  (-8)&73&3.9  (-10)&33&2.2  (-8)&33\\
SO$_2$&3.7  (-9)&400&6.5  (-12)&33&3.6  (-10)&33&&4.2  (-9)&400&1.6  (-11)&33&7.3  (-10)&33\\
NH$_2$CHO&1.8  (-6)&400&4.0  (-10)&136&2.3  (-8)&19&&1.9  (-6)&400&4.5  (-10)&19&2.8  (-8)&19\\
NH$_2$OH&1.5  (-6)&158&6.8  (-7)&132&1.5  (-6)&25&&2.0  (-6)&158&8.9  (-7)&136&2.1  (-6)&33\\
CH$_3$CN&6.2  (-9)&307&8.9  (-11)&97&4.9  (-9)&97&&4.7  (-9)&307&7.0  (-11)&97&4.1  (-9)&97\\
CH$_2$NH&5.6  (-7)&400&1.0  (-9)&132&5.3  (-8)&10&&5.7  (-7)&400&2.1  (-9)&132&6.3  (-8)&90\\
CH$_3$NH$_2$&7.4  (-6)&149&3.0  (-6)&132&8.3  (-6)&11&&8.0  (-6)&149&3.0  (-6)&132&8.9  (-6)&11\\
NH$_2$CH$_2$COOH&4.6  (-14)&400&1.2  (-9)&163&1.6  (-9)&117&&5.5  (-9)&293&1.4  (-8)&167&1.4  (-7)&117\\
HC$_3$N&2.6  (-9)&400&1.1  (-11)&97&5.8  (-10)&97&&1.5  (-9)&400&4.0  (-12)&90&1.7  (-10)&90\\
CH$_2$CHCN&3.9  (-9)&90&3.9  (-10)&97&2.2  (-8)&97&&4.8  (-9)&400&4.3  (-10)&97&2.9  (-8)&97\\
CH$_3$CH$_2$CN&4.8  (-8)&125&6.3  (-10)&103&3.5  (-8)&103&&2.5  (-8)&125&2.8  (-10)&33&1.6  (-8)&103\\
CH$_3$CH$_2$OH&1.2  (-7)&136&2.4  (-9)&117&1.2  (-7)&44&&1.5  (-7)&136&3.0  (-9)&117&2.5  (-7)&68\\
CH$_3$CHO&4.2  (-9)&400&1.8  (-11)&10&1.1  (-9)&10&&4.3  (-9)&400&2.6  (-11)&10&1.6  (-9)&10\\
CH$_3$COCH$_3$&2.8  (-9)&400&3.5  (-13)&68&2.0  (-11)&59&&2.4  (-9)&400&5.9  (-13)&68&1.2  (-8)&68\\
HCOOCH$_3$&8.7  (-10)&110&4.9  (-9)&10&3.0  (-7)&14&&7.4  (-10)&110&5.2  (-9)&19&3.2  (-7)&19\\
CH$_3$OCH$_3$&1.5  (-7)&400&1.8  (-9)&44&1.0  (-7)&44&&1.7  (-7)&400&2.4  (-9)&68&2.3  (-7)&68\\
(CH$_2$OH)$_2$&5.4  (-14)&248&4.1  (-14)&192&6.0  (-14)&79&&2.6  (-13)&248&2.0  (-13)&192&4.7  (-14)&73\\
HCOOH&3.4  (-9)&314&3.6  (-10)&10&2.2  (-8)&10&&2.2  (-8)&286&9.0  (-10)&14&5.5  (-8)&14\\
CH$_3$COOH&1.0  (-10)&84&1.1  (-13)&25&6.7  (-12)&25&&8.5  (-11)&84&3.9  (-13)&25&1.2  (-11)&25\\
NH$_2$COOH&2.1  (-11)&207&3.9  (-11)&163&5.8  (-11)&117&&1.4  (-10)&207&1.2  (-10)&163&1.9  (-10)&117\\
CH$_3$OCOOH&1.0  (-9)&182&6.6  (-10)&145&1.1  (-9)&44&&1.3  (-8)&182&8.9  (-9)&145&5.4  (-8)&68\\
CH$_2$OHCOOH&9.1  (-10)&242&7.0  (-10)&192&9.8  (-10)&44&&1.2  (-8)&242&9.5  (-9)&192&7.3  (-8)&68\\
C$_2$H$_5$COOH&5.4  (-10)&207&9.5  (-10)&163&1.4  (-9)&103&&2.3  (-8)&213&2.0  (-8)&163&1.7  (-7)&84\\
\enddata
\tablecomments{
The peak of the fractional abundances and the temperature of grains at that moment were summarized.
The notation of ``a (b)" represents a $\times$10$^b$.
}
\label{table:Model1_vs_Model3}
\end{deluxetable}




\end{document}